\documentclass[a4paper,10pt]{amsart}
\usepackage[foot]{amsaddr}
\usepackage{geometry}
\theoremstyle{remark}

\usepackage{graphicx}
\usepackage{lineno}
\usepackage{bm}
\usepackage{amsfonts} 
\usepackage{amsmath}
\usepackage{framed} 
\usepackage{multicol} 
\usepackage{tabularx}
\usepackage{tabulary}
\usepackage{siunitx}
\usepackage{caption}
\usepackage{float}
\usepackage{url} 
\usepackage{multicol}
\usepackage{breakurl} 
\usepackage[breaklinks]{hyperref} 

\usepackage{subcaption}
\usepackage{xpatch}
\usepackage{nomencl} 
\makenomenclature
\makenomenclature 
\RequirePackage{ifthen}
\let\oldref\ref
\renewcommand{\ref}[1]{(\oldref{#1})}

 \graphicspath{
    {figures_ok/}
    {cyl/}
    {cavity/}
}

\usepackage{amssymb}
\providecommand{\norm}[1]{\lVert#1\rVert}

\newcommand{\overbar}[1]{\mkern 1.5mu\overline{\mkern-1.5mu#1\mkern-1.5mu}\mkern 1.5mu}

\DeclareMathAlphabet\mathbfcal{OMS}{cmsy}{b}{n}

\xpatchcmd{\thenomenclature}{%
  \section*{\nomname}
}{
}{\typeout{Success}}{\typeout{Failure}}

\usepackage{ifthen}
\renewcommand{\nomgroup}[1]{%
  \ifthenelse{\equal{#1}{A}}{\item[\textbf{Abbreviations}]}{%
    \ifthenelse{\equal{#1}{G}}{\item[\textbf{Symbols}]}{%
      \ifthenelse{\equal{#1}{C}}{\item[\textbf{Abbreviations}]}{%
        \ifthenelse{\equal{#1}{S}}{\item[\textbf{Subscripts}]}{%
          \ifthenelse{\equal{#1}{Z}}{\item[\textbf{Mathematical Symbols}]}{}
        }
      }
    }
  }
}

\ifpdf
\usepackage{pdfsync}
\fi

\usepackage{mathabx}

\begin{document}

\newcommand{\dif}{\mbox{d}}

\title[]{Data-Driven POD-Galerkin reduced order model
for turbulent flows}

\author{Saddam Hijazi\textsuperscript{1,*}}
\thanks{\textsuperscript{*}Corresponding Author.}
\email{saddam.hijazi@uni-potsdam.de}

\author{Melina Freitag\textsuperscript{1}}
\email{melina.freitag@uni-potsdam.de}
\author{Niels Landwehr\textsuperscript{2}}
\email{landwehr@cs.uni-potsdam.de}

\address{\textsuperscript{1}University of Potsdam.}
\address{\textsuperscript{2}University of Hildesheim}




\date{}

\dedicatory{}


\begin{abstract} 
We present a Reduced Order Model (ROM) which exploits recent developments in Physics Informed Neural Networks (PINNs) for solving inverse problems for the Navier--Stokes equations (NSE). In the proposed approach, the presence of simulated data for the fluid dynamics fields is assumed. A POD-Galerkin ROM is then constructed by applying POD on the snapshots matrices of the fluid fields and performing a Galerkin projection of the NSE (or the modified equations in case of turbulence modeling) onto the POD reduced basis. A \textit{POD-Galerkin PINN ROM} is then derived by introducing deep neural networks which approximate the reduced outputs with the input being time and/or parameters of the model. The neural networks incorporate the physical equations (the POD-Galerkin reduced equations) into their structure as part of the loss function. Using this approach, the reduced model is able to approximate unknown parameters such as physical constants or the boundary conditions. A demonstration of the applicability of the proposed ROM is illustrated by three cases which are the steady flow around a backward step, the flow around a circular cylinder and the unsteady turbulent flow around a surface mounted cubic obstacle.
\end{abstract}

\maketitle

%
%
%
%
\section{Introduction and literature overview}\label{sec:intro}

In recent decades, research into numerical methods for solving systems of Partial Differential Equations (PDEs) has been growing rapidly. Popular methods include the finite difference (FDM), the finite element (FEM), the finite volume (FVM), and the spectral element method (SEM). However, running computational simulations using those numerical methods can be very expensive, especially in high dimensions. The situation becomes worse when simulations have to be run several times with several different input configurations (as in repetitive computational environment). These common settings can be observed in various fields such as Uncertainty Quantification (UQ), sensitivity analysis, real-time control problems, optimization, prediction and parameter estimation/inference. In such circumstances, running simulations using the classical numerical methods for each different input value could be deemed prohibitive. Therefore, numerical techniques which could bring a reduction in the computational cost are  needed. Reduced Order Methods (ROMs) represent a suitable tool for achieving the goal of having computational speed-up and providing accurate solutions to the problems of interest. These methods have been applied to a variety of mathematical problems, for greater details on ROMs we refer the reader to \cite{hesthaven2015certified,quarteroniRB2016,bennerParSys,Benner2015,Bader2016}. In this article we focus on ROMs for fluid dynamics problems in the context of the reduction of the Navier--Stokes Equations (NSE).\par

%

In reduced order modeling, projection-based ROMs \cite{Balajewicz2012,Amsallem2012} represent a popular technique for the construction of surrogate reduced models, these ROMs have been applied in several fields such as civil engineering, aerospace engineering and nuclear engineering. The reduction in the projection-based ROMs is achieved by finding the reduced solution which lies in a subspace of a much smaller dimension $N_r$, $N_r \ll N_h$, where $N_h$ is the dimension of the original space constructed by the Full Order Model (FOM). The dimension of the FOM (i.e. $N_h$) represents the number of unknowns or degrees of freedom in the discretized problem. In a projection-based ROM, there are two main ingredients: (i) low dimensional spaces called the reduced spaces which are often generated using a set of snapshots (which are FOM solutions obtained for different values of time and/or parameter) and (ii) a Galerkin or a Petrov--Galerkin projection for the construction of a low dimensional $N_r \times N_r$ problem whose solution is the ROM one. In the context of Parameterized PDEs (PPDEs), projection-based ROMs have been exploited for achieving a solution-space reduction by relying on greedy algorithms \cite{DeVore2013,Binev2011} or Proper Orthogonal Decomposition (POD) \cite{volkwein2013proper,Bergmann2009516,Baiges2014189,Burkardt2006,Ballarin2016} to generate the reduced space. The application of the POD method together with a Galerkin projection technique results in the so-called POD-Galerkin ROM. This type of ROMs has been used extensively for the reduction of PPDEs, for more details on POD-Galerkin ROMs we refer the reader to \cite{Noack1994,Akhtar2009,Bergmann2009516,Kunisch2002492,Burkardt2006,Baiges2014189}.\par


There are several challenges for the construction of efficient POD-Galerkin ROMs for the Navier--Stokes equations. The treatment of the turbulence phenomenon at both the FOM and the ROM levels is one of them. In this work, turbulence is tackled at the full order level through modeling strategies. In other words, turbulent flows are not solved using the Direct Numerical Simulations (DNS) approach because of the enormous computational resources needed to simulate these flows for the problems of interest. In particular, turbulence modeling is done with the help of the Reynolds Averaging Navier--Stokes (RANS) \cite{wilcox2006turbulence} equations and the Large Eddy Simulations (LES) \cite{berselli2005mathematics,sagaut} approaches. The RANS approach is based on solving the NSE for the time-averaged part of the fluid fields, where it basically assumes that time fluctuations are of no significant interest. On the other hand, the LES approach is based on filtering the Navier--Stokes equations to some scale, then the large scales are simulated while the small scales are modeled.\par

Beside the issue of turbulence, the treatment at the reduced order level of the non-linear convective term in the NSE is important and might affect the efficiency of the ROM. In several contributions, the ROM formulation approximates the non-linear term using a third order tensor \cite{Lorenzi2016,Stabile2017,Stabile2018,Xie2018}, this tensor has a dimension of $N_u \times N_u \times N_u$, where $N_u$ is the number of reduced velocity unknowns. However, this approach of dealing with the non-linear term could raise the computational burden when the number of reduced unknowns increases. Hyper-reduction techniques could be used for the approximation of the non-linear term such as the EIM or DEIM methods \cite{Xiao20141,Barrault2004}. In order to implement EIM or DEIM, an expensive pre-processing is required to obtain a version of parameterized operators \cite{Bonomi2017}. The Gappy-POD approach also can be employed \cite{Carlberg2013}.\par

Reduced order models have been also constructed using data-driven techniques as in \cite{Ionita2014,Peherstorfer2015,Kaiser2014,Guo2018,Hesthaven2018,NOACK2003,Guo2019}. In these ROMs, the identification of the reduced solutions is done using data-driven approaches such as regression-based methods, interpolation techniques or Neural Networks (NNs). In addition, hybrid reduced order models which merge projection-based ROMs and data-driven ROMs have been proposed \cite{GALLETTI2004,Couplet2005,noack2005,Xie2018,Hijazi2020,Hijazi2020JCP,Mou2021}. The latter ROMs include the use of calibration methods, the introduction of correction terms which can be approximated by the snapshots data, and the employment of data-driven techniques for the approximation of only the turbulent/eddy viscosity in the case of turbulent flows. Deep learning approaches have also been used in ROMs in order to perform non-linear dimensionality reduction \cite{Fresca_2021,Fresca2022,Romor2022}.\par 

In recent years, several contributions have aimed at using machine learning techniques for solving PDEs arising from physical problems. This field of scientific computing is often termed as physics-based machine learning.  We give a brief overview of related works in this field. Early work in \cite{Lee1990} presents neural minimization algorithms for solving differential equations. The work in \cite{Lagaris1998} presents an approach for solving ODEs and PDEs using feedforward neural networks which is based on approximating the solution function by a trial function that has two parts. The first part which has no tunable parameters satisfies the initial/boundary conditions, while the second part contains all the adjustable parameters which are determined by training the feedforward neural network. The construction of the second term is made in a way that guarantees no contribution to the initial/boundary conditions.\par 

Physics-Informed Neural Networks (PINNs) have been proposed in \cite{Raissi2019} for solving general non-linear PDEs as well as inverse problems which involve PDEs. The approach in \cite{Raissi2019} consists of approximating the solution function of the general PDEs by deep neural networks, the trainable parameters of these NNs are then learned by minimizing a loss function that takes into consideration initial/boundary data and at the same time penalizes the departure from the equations which model the physical problem. The PINNs are based on two different approaches, namely continuous time and discrete time models. The continuous model PINNs allow to infer the solution of the PDE across all time and space. On the other hand, PINNs with the discrete time model approach employ implicit Runge--Kutta time stepping schemes with unlimited number of stages for the prediction of the solution at large time steps without compromising the accuracy of the approximation. 


The PINNs presented in \cite{Raissi2019} were extended to coupled multi-physics problems in  \cite{Raissi2018}, where the latter work presents another application of PINNs for solving inverse problems for a Fluid Structure Interaction (FSI) problem. 
In \cite{Sirignano2018}, the authors present a Deep Galerkin Method (DGM) for the approximation of high-dimensional PDEs with a deep neural network, where they solve high-dimensional free boundary PDEs in $200$ dimensions. 
\par 

A recent work \cite{Chen2021} proposes a reduced basis method based on the use of PINNs for solving PPDEs. This work shows that training the PINNs by only minimizing the loss function that corresponds to the reduced equations does not give as accurate results as the ones obtained by the projection of FOM solution onto the reduced space. The authors indicate that for complex nonlinear problems, the PINNs trained only on the reduced equations are not accurate in approximating the original high fidelity solution. This is justified by the fact that the reduced equations do not take into account the impact of the truncated modes on the resolved ones. On the other hand, the authors demonstrate that the PINNs trained on both the output data labels and the reduced equations are more accurate.\par

The work in this paper aims at employing reduced order modeling coupled with PINNs for solving inverse problems. By solving an inverse problem, we attempt to infer unknown inputs or parameters from a given set of observations of the output (output data). For a review on inverse problems, we refer the reader to \cite{Banks1989,Kirsch2011}. Research into inverse problems in a Bayesian setting has been conducted extensively, for example see \cite{Stuart2010,Kaipio2005,Matthies2016,Dashti2017,Cotter2010}. Reduced order methods and dimension reduction techniques have been used previously for the estimation of unknown parameters in inverse problems. The work \cite{Marzouk2009} presents a Bayesian approach for solving nonlinear inverse problems with the help of a Galerkin projection. In \cite{Warner2015}, stochastic reduced order models were proposed for solving inverse problems. Active subspaces were utilized for the reduction of the parameter space in a UQ problem for turbulent combustion simulations \cite{Ji2019}. In \cite{Tiangang2014}, a hybrid data-driven/projection-based reduced order model is proposed for the Bayesian solution of inverse problems. The work in \cite{Galbally2009} proposes a non-linear reduced order model for large-scale inverse problems in a Bayesian inference setting. Another approach \cite{Garmatter2016} combines the nonlinear Landweber method with adaptive online reduced basis updates for solving the inverse problem related to the construction of the conductivity in the steady heat equation. The authors in \cite{Himpe2015} present a reduction approach of a parameterized forward model, which is utilized for obtaining a surrogate model in a Bayesian inverse problem setting, the inversion is done during the online stage by using the surrogate model constructed via the projection of the forward model onto the reduced spaces.\par 

Here, we present a model for solving inverse problems in a reduced order setting. The approach is based on integrating the structure of the POD-Galerkin ROMs into physics-informed neural networks (PINNs). In particular, we propose to incorporate the POD-Galerkin reduced order equations into the loss function of the PINNs. Consequently, the task of inferring any parameter or physical unknown which is present in the FOM equations will become feasible, thanks to the possibility of introducing additional parameters in the neural networks and making them trainable at low computational cost. The unknown parameters could be physical constants such as the physical viscosity or boundary/initial conditions or the velocity at the inlet in inlet/outlet fluid problems. The approach developed in this work is termed \textit{POD-Galerkin PINN ROM} and is based on assimilating reduced simulated data into the physical model represented by the POD-Galerkin differential algebraic system. The latter reduced simulated data is obtained from the Galerkin projection of the available FOM data onto the POD reduced spaces.\par 

This new methodology introduces a significant reduction of the computational cost associated with solving inverse problems for the NSE. In fact, the use of the PINNs directly at the full order level for inferring unknown parameters in the mathematical fluid problem could be of significant computational cost. This is due to the number of degrees of freedom in the available full order data, in turbulent $3D$ problems this number is of the order $10^6$ or higher. On the other hand, the proposed approach deals with the inference problem by introducing two levels of approximation. The first level is represented by the dimensional reduction performed by the POD and the Galerkin projection which results in a POD-Galerkin ROM, while in the second approximation level neural networks for the approximation of the reduced fluid variables are utilized. By doing so, one may leverage the power of the PINNs in inferring unknown parameters by solving the optimization problem (which has reduced number of unknowns and hence low computational cost) in which the goal is to minimize the error committed in approximating the (reduced) data and the error caused by the violation of the physics (the reduced POD-Galerkin equations). It is worth mentioning that after the training of the PINNs in the offline stage, the POD-Galerkin PINN ROM will still be able to do fast online forward computations without the need to re-train the neural networks.

This article is organized as follows: \autoref{sec:NSE} introduces the full order model and addresses the incompressible Navier--Stokes equations. In \autoref{sec:ROM}, the reduced order model structure and methodology is presented. Firstly, the proper orthogonal decomposition method is recalled, then we present the non-intrusive reduced order model developed in this work to treat inverse problems in a reduced setting in the context of the NSE. \autoref{sec:results} gives three numerical examples that illustrate the results of the parameter identification using the reduction approach proposed in this work. The first example is the steady case of the flow past a backward step in a laminar setting, while the second one is the turbulent case of the flow around a circular cylinder and the last example deals with a more complex $3D$ turbulent case which is the flow around a surface mounted cubic box.

\section{The problem setup : parameterized Navier-Stokes equations}\label{sec:NSE}
 
The NSE are ubiquitous in science and engineering where they describe the physics of many phenomena such as modeling the air flow around an airfoil, the flow in boat wakes and the motion of bluff bodies inserted in fluid flows. In this work, the focus is on the parameterized unsteady NSE, the mathematical formulation of the problem reads as follows: Given the fluid spatial domain $\Omega \in \mathbb{R}^d$, with $d=2$ or $3$ and the time window $[0,T]$ under consideration, find the vectorial velocity field $\bm{u} : \Omega \times [0,T] \mapsto \mathbb{R}^d$ and the scalar pressure field $p : \Omega \times [0,T] \mapsto \mathbb{R}$ such that:
\begin{equation}\label{eq:navstokes}
\begin{cases}
\frac{\partial\bm{u}}{\partial t}+ \bm{\nabla} \cdot (\bm{u} \otimes \bm{u}) - \bm{\nabla} \cdot \nu \left(\bm{\nabla}\bm{u}+\left(\bm{\nabla}\bm{u}\right)^T\right)=-\bm{\nabla}p &\mbox{ in } \Omega \times [0,T],\\
\bm{\nabla} \cdot \bm{u}=\bm{0} &\mbox{ in } \Omega \times [0,T],\\
\bm{u} (t,\bm{x};\bm{\mu}) = \bm{f}(\bm{x},\bm{\mu}) &\mbox{ on } \Gamma_\textrm{inlet} \times [0,T],\\
\bm{u} (t,\bm{x};\bm{\mu}) = \bm{0} &\mbox{ on } \Gamma_{0} \times [0,T],\\ 
(\nu\bm{\nabla} \bm{u} - p\bm{I})\bm{n} = \bm{0} &\mbox{ on } \Gamma_\textrm{outlet} \times [0,T],\\ 
\bm{u}(0,\bm{x})=\bm{R}(\bm{x}) &\mbox{ in } (\Omega,0),\\            
\end{cases}
\end{equation}
where $t$ is the time, $\bm{x}$ is the spatial variable vector and $\Gamma = \Gamma_\textrm{inlet} \cup \Gamma_0 \cup \Gamma_\textrm{outlet}$ is the boundary of the fluid domain $\Omega$. The three parts that form the boundary are called $\Gamma_\textrm{inlet}$, $\Gamma_\textrm{outlet}$ and $\Gamma_0$, they correspond to the inlet boundary, the outlet boundary and the physical walls, respectively. The fluid kinematic viscosity is denoted by $\nu$ and is constant across the spatial domain. The function $\bm{f}$ includes the boundary conditions for the non-homogeneous boundary. The initial velocity field is given by the function $\bm{R}(x)$. The normal unit vector is denoted by $\bm{n}$. We remark that the velocity and the pressure fields depend on time, space and the parameter $\bm{\mu} \in \mathcal{P} \subset \mathbb{R}^q$, where $\mathcal{P}$ is a $q$-dimensional parameter space, the dependencies are dropped for making the notation concise.\par

The governing equations of \eqref{eq:navstokes} are discretized using the FVM \cite{Moukalled:2015:FVM:2876154}. In this work, the numerical solver used for solving the NSE is the finite volume C\texttt{++} library OpenFOAM\textsuperscript{\textregistered} (OF) \cite{weller1998tensorial}. For more details on the finite volume discretization and the techniques used by OpenFOAM, we refer the reader to \cite{jasak1996error}.\par

The fluid dynamics problems which this work aim at tackling include turbulent problems or problems with moderate to high Reynolds number $Re =\frac{U_\infty L}{\nu}$, where $L$ and $U_\infty$ are the characteristic length and velocity of the particular fluid problem, respectively. Flows with low values of Reynolds number are called laminar flows in which fluid moves smoothly or in regular paths, laminar flows are also characterized by having high momentum diffusion and low convection. In contrast to laminar flows, turbulent flows are chaotic where sudden changes in the velocity and the pressure fields are more common. Turbulent flows can be frequently observed in real life applications, examples include external flows over airplanes or ships and oceanic and atmospheric currents. Therefore, it is important to mention how turbulence is treated at both the FOM and ROM levels.\par 

At the FOM level, turbulence is not solved directly using the so-called DNS approach, instead it is modeled using modeling strategies, namely the Reynolds Averaged Navier--Stokes equations (RANS) and the Large Eddy Simulation (LES). In both cases, one resorts to closure models which introduce an additional viscosity term known as the eddy/turbulent viscosity (denoted by $\nu_t$) which has the same unit as the physical kinematic viscosity $\nu$ \cite{boussinesq1877essa}. The estimation of the eddy viscosity requires the use of the so-called closure turbulence models. These models approximate $\nu_t$ as a function of other turbulence variables such as the turbulent kinetic energy $k$, where they resolve one or more transport-diffusion PDE for the additional turbulence variables. Examples of such closure models under the RANS approach include the one equation Spalart–Allmaras (S–A) turbulence model \cite{SPALART1992} and the two equations $k-\epsilon$ \cite{Hanjalic1972} and SST $k-\omega$ turbulence models \cite{Menter1994}, where $\epsilon$ and $\omega$ stand for the turbulent dissipation and the specific turbulent dissipation rate, respectively. As for the LES turbulence models, the Smagorinsky model \cite{SMAGORINSKY1963} is a well known LES model, other models are the dynamic eddy viscosity model proposed in \cite{Germano1991} and the one equation model named "dynamicKEqn" \cite{Kim1995} which has been utilized in this work. Closure turbulence models are also often termed as Eddy Viscosity Models (EVMs). For a comprehensive review on the issue of turbulence modeling, we refer the reader to \cite{wilcox2006turbulence,berselli2005mathematics}.\par

We report here as an example the modified equations after employing the RANS approach complemented by the $k-\omega$ turbulence model:

\begin{equation}\label{eq:RANS}
\begin{cases}
 \frac{\partial\overbar{\bm{u}}}{\partial t}+ \bm{\nabla} \cdot (\overbar{\bm{u}} \otimes \overbar{\bm{u}})  = \bm{\nabla} \cdot \left[-\overbar{p} \mathbf{I}+\left(\nu+\nu_t \right) \left(\bm{\nabla}\overbar{\bm{u}}+\left(\bm{\nabla}\overbar{\bm{u}}\right)^T\right)\right] &\mbox{ in } \Omega \times [0,T],\\
\bm{\nabla} \cdot \overbar{\bm{u}} = 0  &\mbox{ in } \Omega \times [0,T],\\
\overbar{\bm{u}} (t,\bm{x}) = \bm{f}(\bm{x},\bm{\mu}) &\mbox{ on } \Gamma_{In} \times [0,T],\\
\overbar{\bm{u}} (t,\bm{x}) = \bm{0} &\mbox{ on } \Gamma_{0} \times [0,T],\\ 
(\nu\bm{\nabla} \overbar{\bm{u}} - \overbar{p}\bm{I})\bm{n} = \bm{0} &\mbox{ on } \Gamma_{Out} \times [0,T],\\ 
\overbar{\bm{u}}(0,\bm{x})=\bm{R}(\bm{x}) &\mbox{ in } (\Omega,0),\\           
\nu_t=F(k,\omega), &\mbox{ in } \Omega, \\ 
\mbox{Transport-Diffusion equation for $k$}, \\ 
\mbox{Transport-Diffusion equation for $\omega$},\\ 
\end{cases}
\end{equation}
where $F$ is an algebraic function that relates $\nu_t$ with the turbulence variables $k$ and $\omega$. LES closure models result in a similar modified momentum equation for the NSE. We remark that in any of the turbulence modeling strategies mentioned above, there exists an additional vector field term in the momentum equation which is $\bm{\nabla} \cdot \left[ \nu_t \left(\bm{\nabla}\overbar{\bm{u}}+\left(\bm{\nabla}\overbar{\bm{u}}\right)^T\right) \right]$. The latter term will be referred to as the turbulent term in this work.\par

\section{The reduced order model (ROM)}\label{sec:ROM}
This section presents the reduced order model (ROM) constructed for the reduction of the NSE addressed in the previous section. An effective ROM is sought for the approximation of the solutions of the parameterized NSE (\eqref{eq:navstokes} and \eqref{eq:RANS}) for both laminar and turbulent flows. Therefore, the ROM will take into consideration the features of the full order model (FOM) including turbulence treatment when applicable. The ROM will then be used in parameter estimation/inference tasks.\par

The main assumption in reduced order modeling is that the dynamics of the FOM are governed by few dominant modes, and therefore, an accurate reproduction of the full order solution is possible when one combines appropriately those dominant modes. This assumption represents a cornerstone in the construction of ROMs and mathematically it implies that the FOM solution fields of the velocity and pressure can be approximated as sum of spatial modes multiplied by temporal coefficients, i.e.:
\begin{equation}\label{eq:decompose}
\bm{u}(\bm{x},t; \bm{\mu}) \approx  \sum_{i=1}^{N_u} a_i(t;\bm{\mu}) \bm{\phi}_i(\bm{x}), \quad p(\bm{x},t; \bm{\mu})\approx \sum_{i=1}^{N_p} b_i (t;\bm{\mu}) {\chi_i}(\bm{x}),
\end{equation}    
where the reduced velocity and reduced pressure modes are denoted by $\bm{\phi}_i(\bm{x})$ and ${\chi_i}(\bm{x})$, respectively. The reduced modes of both variables depend only on the spatial variables. The coefficients $a_i(t;\bm{\mu})$ and $b_i (t;\bm{\mu})$ represent the $i$-th reduced solution for velocity and pressure, respectively, they depend on both time $t$ and the parameter $\bm{\mu}$. Several methods and approaches can be applied for the generation of the reduced order spaces of the velocity and pressure defined by $\mathbb{V}_{rb}=\mbox{span}\left\{\bm{\phi}_i\right\}_{i=1}^{N_u}$ and $\mathbb{Q}_{rb}=\mbox{span}\left\{\chi_i\right\}_{i=1}^{N_p}$, respectively. The Reduced Basis (RB) method based on the greedy basis generation can be utilized \cite{quarteroniRB2016}. Proper Orthogonal Decomposition (POD) represents a popular choice for the construction of reduced spaces \cite{Lorenzi2016,Stabile2017}. The Proper Generalized Decomposition (PGD) \cite{Dumon20111387,Chinesta2011} and Dynamic Mode Decomposition (DMD) \cite{Schmid2010} may also be employed. In recent works \cite{Murata2019,Lee2020}, autoencoders have been used for the generation of non-linear reduced spaces. In the general case of unsteady parameterized problems, the generation of the reduced basis could be done using a POD-Greedy approach as in \cite{Haasdonk2008}, where POD is applied in time and the RB method with the greedy approach is used for the parameter space. Another choice is the use of a nested POD method, where POD procedures are carried out on the solutions that correspond to each parameter value separately before applying a final POD on the resulted POD modes from the first step. The method chosen in this work for the generation of the reduced space is POD applied directly on the set of all realizations of the solution fields which might correspond to different values of the parameters and/or time.\par

Efficient reduced order models rely on the notion of having two decoupled phases termed as the offline and the online phases. In the offline phase, the training procedure of the ROM is carried out. This includes sampling the parameter space and then simulating the FOM in order to generate the snapshots which are used later for the generation of the reduced order space (here the POD space). Hence, the offline stage consists of computing the POD modes and all reduced quantities, which form the reduced system and are dependent on the POD modes. The offline phase is known to have a significant computational cost due to the fact that the offline computations depend on the FOM dimension. However, the offline phase must be carried out just once for a given choice of the ROM dimension. The final result of the offline stage is the reduced order system of equations.\par

The online stage utilizes the ROM and hence results in fast computations which are dependent only on the dimension of the ROM. Ideally, the online stage should not depend on any aspect of the full order computational model such as accessing the original finite volume mesh. During the online stage, the solution of the reduced order problem is found by solving the low dimensional system produced during the offline stage.\par

In this work, POD is used for the generation of the reduced order spaces of both the velocity and pressure. After sampling the parameter space, the FOM described in \autoref{sec:NSE} is solved for each value of the parameter $\bm{\mu} \in \mathcal{P}_M = \{\bm{\mu}_1,...\bm{\mu}_M\}$ and solutions are acquired at the desired time instants $\{t_1,t_2,...,t_{N_T}\} \subset [0,T]$. This yields a total of $N_s = M * N_T$ snapshots which form the following snapshots matrices for velocity and pressure:
\begin{equation}\label{eq:snapU}
\bm{{\mathcal{S}_u}} = \{\bm{u}(\bm{x},t_1;\bm{\mu}_1),...,\bm{u}(\bm{x},t_{N_T};\bm{\mu}_M)\} ~\in~\mathbb{R}^{N_u^h\times N_s},
\end{equation}
\begin{equation}\label{eq:snapP}
\bm{{\mathcal{S}_p}} = \{p(\bm{x},t_1;\bm{\mu}_1),...,p(\bm{x},t_{N_T};\bm{\mu}_M)\} ~\in~\mathbb{R}^{N_p^h\times N_s}.
\end{equation}

The POD velocity and pressure modes are then computed using the method of snapshots \cite{sirovich1987snap}. As for the identification of the reduced coefficients of the velocity and pressure in \eqref{eq:decompose}, we utilize feedforward neural networks to achieve this task in the online stage.\par 

In more details, we use Physics Informed Neural Networks (PINNs) to solve the reduced problem. Firstly, the reduced equations are obtained by performing a Galerkin projection of the FOM equations onto the POD spaces of the velocity and pressure. Then, one may encode these reduced equations as a part of the loss function which has to be minimized by the neural network optimizer. The resulted non-intrusive reduced order model merges aspects of POD-Galerkin ROMs with Physics-Informed Neural Networks (PINNs), therefore, it is termed here as {\it{POD-Galerkin PINN ROM}}. This reduced order model is designed to solve inverse problems for the Navier--Stokes equations. The goal is to identify/infer unknown parameters or inputs in a mathematical models by comparing the predictions of these models with real or simulated measurements or outputs. The inference task is carried out by leveraging the features of neural networks which allow for the introduction of additional trainable weights. These new weights are present in the loss function through the reduced equations which makes it possible to compute gradients of the loss with respect to these weights and consequently to optimize their value. In rest of this section we describe the construction of the proposed ROM.

The construction of a POD-Galerkin ROM for the Navier--Stokes equations starts by projecting the momentum equation onto the reduced space spanned by the velocity POD modes $[\bm{\phi}_i]_{i=1}^{N_u}$, i.e.:
\begin{equation}\label{eq:l2proj_vel}
\left( \bm{\phi}_i,\frac{\partial\bm{u}}{\partial t} +  \bm{\nabla} \cdot (\bm{u} \otimes \bm{u}) - \bm{\nabla} \cdot \nu \left(\bm{\nabla}\bm{u}+\left(\bm{\nabla}\bm{u}\right)^T\right) +\bm{\nabla} p \right)_{L^2(\Omega)} = 0 .
\end{equation}

After inserting the reduced approximation of the velocity and pressure, we obtain the following ODEs which represent the reduced momentum equation:

\begin{equation}\label{eq:momentum_POD_G}
\bm{\dot{a}}= \nu (\bm{B}+\bm{B_T})\bm{a} - \bm{a}^T \bm{C} \bm{a}-\bm{H}\bm{b},
\end{equation}

where each of $\bm{B}, \bm{B_T}, \bm{C}$ and $\bm H$ is either a reduced order matrix or tensor. These terms are computed as follows:
\begin{align}
& (\bm{B})_{ij}=\left( \bm{\phi}_i ,\bm{\nabla} \cdot \bm{\nabla}\bm{\phi}_j\right)_{L^2(\Omega)}, \\
& (\bm{B_T})_{ij}=\left( \bm{\phi}_i ,\bm{\nabla} \cdot (\bm{\nabla} \bm{\phi}_j^T)\right)_{L^2(\Omega)},\\
& (\bm{C})_{ijk}=\left( \bm{\phi}_i , \bm{\nabla} \cdot (\bm{\phi}_j \otimes \bm{\phi}_k)\right)_{L^2(\Omega)} , \label{eq:div_phi}\\
& (\bm{H})_{ij} = \left( \bm{\phi}_i , \bm{\nabla} \chi_j \right)_{L^2(\Omega)}.
\end{align}

We note that the treatment of non-convective term in the reduced momentum equation above is done by the use of the third-order tensor $\bm C$. This approach might lead to a substantial increase in the computational cost of solving the reduced problem when the number of the reduced velocity unknowns $N_u$ grows. This approach approximates the projection of the non-linear term $\bm{\nabla} \cdot (\bm{u} \otimes \bm{u})$ onto the velocity POD mode $\bm{\phi}_i$ as follows:

\begin{equation}\label{eq:3rd_order_tensor_pod_g}
\left( \bm{\phi}_i,\bm{\nabla} \cdot (\bm{u} \otimes \bm{u}) \right)_{L^2(\Omega)} \approx \bm{a}^T \bm{C_{i\bullet \bullet}} \bm{a}.
\end{equation} 

However, we propose to utilize a different approach in which we add a variable for the approximation of the convective term in the reduced momentum equation named $\bm{c}$, where:

\begin{equation}\label{eq:3rd_order_tensor_pod_g}
\left( \bm{\phi}_i,\bm{\nabla} \cdot (\bm{u} \otimes \bm{u}) \right)_{L^2(\Omega)} = \bm{c}_i,
\end{equation} 
the additional variable $\bm{c}$ represents the projection of the non-linear vector field $\bm{\nabla} \cdot (\bm{u} \otimes \bm{u})$ (which can be retrieved/isolated from any velocity snapshot) onto the velocity POD modes $[\bm{\phi}_i]_{i=1}^{N_u}$. The final form of the reduced momentum equation is then given by
\begin{equation}\label{eq:momentum_POD_G_new}
\bm{\dot{a}}= \nu (\bm{B}+\bm{B_T})\bm{a} - \bm{c} -\bm{H}\bm{b}.
\end{equation}
An additional set of reduced equations can be obtained by the employment of either the supremizer enrichment approach \cite{Ballarin2014,Rozza2007} or by considering a reduced version of the Poisson Equation for Pressure (PPE) \cite{JOHNSTON2004221,Liu2010,Stabile2017}. The supremizer approach computes artificial velocity-like modes which are termed the supremizers and then it enriches the original velocity POD modes with the newly computed supremizers in a way that ensures the fulfillment of a reduced version of the inf-sup condition. The additional velocity-like fields or the supremizers are computed by solving the following problems:
\begin{equation}
\begin{cases}
\Delta \bm{s}_i = - \bm{\nabla} \chi_i &\mbox{ in } \Omega, \mbox{ }   \forall {\chi_i} \in \mathbb{V}^p_{POD},\\
\bm{s}_i=\bm{0} &\mbox{ on } \partial \Omega.
\end{cases}
\end{equation}

After that the velocity POD space is enriched with the supremizer modes:
\begin{equation}
\tilde{\mathbb{V}}^u_{POD} = [\bm{\phi}_1,\dots,\bm{\phi}_{N_u}] \oplus [\bm{s}_1,\dots,\bm{s}_{N_S}] \in \mathbb{R}^{N_u^h \times (N_u+N_S)}.
\end{equation}

The original velocity POD modes are divergence free by construction since they are just a linear combination of the velocity snapshots. This implies that the projection of the continuity equation onto the pressure modes before enriching the velocity POD space would have added no reduced equations. The newly added supremizer modes are not divergence free, therefore, the continuity equation could be utilized for obtaining an additional set of scalar reduced algebraic equations, as follows:

\begin{equation}
\left(\chi_i,\bm{\nabla} \cdot \bm{u} \right)_{L^2(\Omega)} = 0.\label{eq:cont_sup_pod_g}
\end{equation}

The final POD-Galerkin ROM with the supremizer enrichment approach is given by

\begin{subequations}\label{eq:SUP_ROM}
\begin{align}
&\bm{M} \bm{\dot{a}} = \nu (\bm{B}+\bm{B_T})\bm{a}-\bm{c}-\bm{H}\bm{b},
 \label{eq:SUP_alg_mom} \\
&\bm{P} \bm{a}= \bm{0},
\label{eq:SUP_alg_PPE}
\end{align}
\end{subequations}
with two additional reduced matrices $\bm{M}$ and $\bm{P}$. The first matrix $\bm{M}$ is the mass matrix, which is not unitary anymore as a result of the additional supremizer modes. The matrix $\bm{P}$ is called the divergence reduced matrix. The entries of the additional matrices are given by:
\begin{align}
& (\bm{M})_{ij}=\left( \bm{\phi}_i , \bm{\phi}_j \right)_{L^2(\Omega)},\\
& (\bm{P})_{ij}=\left(\chi_i, \bm{\nabla} \cdot  \bm{\phi}_j \right)_{L^2(\Omega)}.
\end{align}

In the turbulent case, an additional term in the reduced momentum equation will appear. This term corresponds to the projection of the added turbulence modeling term in the momentum equation (in the RANS or the LES formulation at the FOM level) onto the velocity POD modes. The turbulent POD-Galerkin ROM with the employment of the supremizer enrichment approach is given by
\begin{subequations}\label{eq:SUP_ROM_Turb}
\begin{align}
&\bm{M} \bm{\dot{a}} = \nu (\bm{B}+\bm{B_T})\bm{a}-\bm{c}-\bm{H}\bm{b}+\bm{h},\\
&\bm{P} \bm{a}= \bm{0},
\end{align}
\end{subequations}
where $\bm{h}$ is the turbulent reduced variable.\par

At this point, we describe the structure of the PINNs which are used to approximate the solution of the POD-Galerkin ROMs. The PINNs have as input the time and the parameter. The outputs are the reduced velocity, pressure, convective and turbulent terms denoted by $\bm{a}, \bm{b}, \bm{c}$ and $\bm{h}$, respectively. The dimension of each of these output terms is $N_u$ except for the reduced pressure which is of $N_p$ dimension.\par 

The starting point of the PINN training is the computation of output label data. This data consist of the $L^2$ projection coefficients for each of the FOM variables fields onto the velocity or pressure POD basis. The velocity $L^2$ projection coefficients are computed as follows:
\begin{equation}\label{eq:vec_coeff}
\mathbb{R} \ni a_{i,L^2}^{j} = \left(\bm{{\mathcal{S}^i_u}}, \bm{\phi}_j\right)_{L^2(\Omega)}, \quad \text{for} \quad i=1,2,...,N_s, \quad j=1,2,...,N_u,
\end{equation}
similarly the pressure coefficients are given by:

\begin{equation}\label{eq:pressure_coeff}
\mathbb{R} \ni b_{i,L^2}^{j} = \left(\bm{{\mathcal{S}^i_p}}, \chi_j\right)_{L^2(\Omega)}, \quad \text{for} \quad i=1,2,...,N_s, \quad j=1,2,...,N_p.
\end{equation}

Then, the FOM vectorial fields of the convective and turbulent terms are retrieved from the original snapshots of the velocity, pressure and the turbulent eddy viscosity. Then, one may compute the projection of these vectorial fields onto the velocity POD modes $[\bm{\phi}_i]_{i=1}^{N_u}$. This yields the output data for the vectors $\bm{c}$ and $\bm{h}$ which are also needed for the training of the PINN. The additional coefficients are given by:
\begin{equation}\label{eq:conv_coeff}
\mathbb{R} \ni c_{i,L^2}^{j} = \left(\bm{\nabla} \cdot (\bm{{\mathcal{S}^i_u}} \otimes \bm{{\mathcal{S}^i_u}})
, \bm{\phi}_j\right)_{L^2(\Omega)}, \quad \text{for} \quad i=1,2,...,N_s, \quad j=1,2,...,N_u,
\end{equation}

\begin{equation}\label{eq:turb_coeff}
\mathbb{R} \ni h_{i,L^2}^{j} = \left(\bm{{\mathcal{S}^i_t}}, \bm{\phi}_j\right)_{L^2(\Omega)}, \quad \text{for} \quad i=1,2,...,N_s, \quad j=1,2,...,N_u,
\end{equation}
where $\bm{{\mathcal{S}^i_t}}$ is the $i$-th snapshot of the turbulent additional term in the FOM formulation of the RANS or the LES turbulence modeling approach.\par 

The input and output data matrices $\tilde{\bm{A}}~\in~\mathbb{R}^{N_s \times (q+1)} $ and $\tilde{\bm{G}}~\in~\mathbb{R}^{N_s \times (3N_u + N_p)}$ are given by
\begin{equation}
\tilde{\bm{A}} = \begin{bmatrix} 
    \bm{\mu}_1 & t_1  \\
    \bm{\mu}_1 & t_2  \\
    \vdots & \vdots \\
    \bm{\mu}_1 & t_{N_T} \\
    \bm{\mu}_2 & t_1 	\\
	\bm{\mu}_2 & t_2  \\  
    \vdots & \vdots \\
    \bm{\mu}_2 & t_{N_T} \\
    \vdots & \vdots \\
    \bm{\mu}_M & t_1  \\
    \bm{\mu}_M & t_2  \\
    \vdots & \vdots \\
    \bm{\mu}_M & t_{N_T}
    \end{bmatrix},
\end{equation}
\begin{equation}
 \setcounter{MaxMatrixCols}{20}
    \tilde{\bm{G}} = \begin{bmatrix} 
    a_{1,L^2}^{1} & \hdots & a_{1,L^2}^{N_u} &  b_{1,L^2}^{1} & \hdots & b_{1,L^2}^{N_p} &  h_{1,L^2}^{1} & \hdots & h_{1,L^2}^{N_u} &  c_{1,L^2}^{1} & \hdots & c_{1,L^2}^{N_u} \\
  	a_{2,L^2}^{1} & \hdots & a_{2,L^2}^{N_u} &  b_{2,L^2}^{1} & \hdots & b_{2,L^2}^{N_p} &  h_{2,L^2}^{1} & \hdots & h_{2,L^2}^{N_u} &  c_{2,L^2}^{1} & \hdots & c_{2,L^2}^{N_u} \\
  	\vdots & \hdots & \vdots & \vdots & \hdots & \vdots & \vdots & \hdots & \vdots & \vdots & \hdots & \vdots \\
  	a_{N_s,L^2}^{1} & \hdots & a_{N_s,L^2}^{N_u} &  b_{N_s,L^2}^{1} & \hdots & b_{N_s,L^2}^{N_p} &  h_{N_s,L^2}^{1} & \hdots & h_{N_s,L^2}^{N_u} &  c_{N_s,L^2}^{1} & \hdots & c_{N_s,L^2}^{N_u}
    \end{bmatrix}.
\end{equation}
We remark that the number of the PINN outputs is increased by $2N_u$ variables compared to the POD-NN case \cite{Hesthaven2018}. However, this approach ensures that the dimension of the reduced problem scales linearly with the number of reduced variables of both velocity and pressure. The POD-Galerkin PINN ROM is then constructed by training deep neural networks whose input is time and parameter and whose output is the reduced velocity, pressure, convective and turbulent terms. The loss function which has to be minimized will be a weighted loss that takes into consideration the available data and the POD-Galerkin formulation imposed by the algebraic differential system in \eqref{eq:SUP_ROM_Turb}. The loss function can be written as follows
\begin{equation}
E(\bm{w}) = E_\textrm{data}(\bm{w}) + \alpha_1 E_1(\bm{w}) + \alpha_2 E_2(\bm{w}),
\end{equation}
where
\begin{equation}\label{eq:dataLoss}
E_\textrm{data}(\bm{w}) = \sum_{n=1}^{N_s} \frac{1}{3N_u+N_p}  \sum_{k=1}^{3N_u+N_p} \{ y_k(\bm{l}^n,\bm{w}) - r_k^n \}^2,
\end{equation}

\begin{equation}\label{eq:eq1Loss}
E_1(\bm{w}) = \sum_{n=1}^{N_s} \frac{1}{N_u}  \sum_{k=1}^{N_u} \{ {\bm{R}^a_k(\bm{l}^n,\bm{y},\bm{w})} \}^2,
\end{equation}

\begin{equation}\label{eq:eq2Loss}
E_2(\bm{w}) = \sum_{n=1}^{N_s} \frac{1}{N_p}  \sum_{k=1}^{N_p} \{ {\bm{R}^b_k(\bm{l}^n,\bm{y},\bm{w})} \}^2,
\end{equation}
and
\begin{align}
& \bm{R}^a = - \bm{M} \bm{\dot{a}}_\textrm{PINN} + \nu (\bm{B}+\bm{B_T})\bm{a}_\textrm{PINN}-\bm{c}_\textrm{PINN}-\bm{H}\bm{b}_\textrm{PINN} + \bm{h}_\textrm{PINN} \in \mathbb{R}^{N_u},\label{eq:Ra}\\
& \bm{R}^b = \bm{P} \bm{a}_\textrm{PINN} \in \mathbb{R}^{N_p},\label{eq:Rb}\\
& \bm{y} = [ \bm{a}_\textrm{PINN}, \bm{b}_\textrm{PINN}, \bm{h}_\textrm{PINN}, \bm{c}_\textrm{PINN}] \in \mathbb{R}^{3N_u+N_p}.
\end{align} 
The two loss functions $E_1$ and $E_2$ enforce the reduced equations given by the POD-Galerkin model. The two weighting coefficients $\alpha_1$ and $\alpha_2$ are tuned heuristically depending on the problem but are also in general trainable. The above formulation gives the PINN the ability to estimate unknown parameters which are present in the POD-Galerkin formulation, such parameters might include for example the physical viscosity $\nu$. The approximation of the time derivative of the reduced velocity $\bm{\dot{a}}_\textrm{PINN}$ which appears in $\bm{R}^a$ is done with the help of automatic differentiation \cite{Baydin2017}. Automatic differentiation represents a crucial tool in PINNs, where it is capable of differentiating the neural networks with respect to their input coordinates and model parameters, the latter model parameters do not include only the weights and biases stacked in the vector $\bm{w}$ but also any other unknown physical quantity in the model.\par

It is worth mentioning that the POD-Galerkin PINN ROM could incorporate physical constraints related to the velocity at the boundary. In fact, it is common to have inhomogeneous Dirichlet boundary conditions for the velocity field at specific parts of the boundary. This is typical in inlet-outlet problems such as the flow around a circular cylinder or the flow past a backward step. In these circumstances, an additional effort has to be made for the treatment of the inhomogeneous velocity boundary conditions at the ROM level. The common strategies for tackling this issue are the lifting function method \cite{Graham1999,Gunzburger2007,HijaziStabileMolaRozza2020a} and the penalty method \cite{Bizon2012,Babuka1973,Barrett1986,Kalashnikova2012,Sirisup2005}. A brief description of the two methods will be given and then the strategy of incorporating them inside the PINN formulation will be addressed.\par

The lifting function method treats the non-homogeneous Dirichlet boundary condition through the introduction of a lifting function (or several lifting functions). In this method the inhomogeneity is transferred to the lifting function and a new set of velocity snapshots is created. The POD procedure is then performed on the newly created set of velocity snapshots (which have homogeneous Dirichlet boundary conditions). This results in velocity POD modes which have also homogeneous boundary conditions at the Dirichlet boundary. We remark that the lifting function must satisfy certain conditions such as being divergence free. This function is also added to the velocity POD basis. Unlike the lifting method, the penalty method does not involve any modification on the velocity snapshots. The penalty method enforces the inhomogeneous boundary condition by the introduction of an additional term in the reduced momentum equation, this term corresponds to the projection of a function (which has zero value everywhere except on the Dirichlet boundary) onto the velocity POD modes. The penalty method modifies the POD-Galerkin ROM as follows:
\begin{subequations}\label{eq:penalty_DAE}
\begin{align}
&\bm{M} \bm{\dot{a}} = \nu (\bm{B}+\bm{B_T})\bm{a}-\bm{c}-\bm{H}\bm{b}+\bm{h} + \tau \left( U_{BC} \bm{D} - \bm{E} \bm{a} \right),
 \\
&\bm{P} \bm{a}= \bm{0},
\end{align}
\end{subequations}
where $U_{BC}$ is the velocity value at the Dirichlet boundary $\Gamma_D$, and $\tau$ is the penalization factor whose value is tuned heuristically. Higher values of $\tau$ generally tend to enforce the boundary conditions in a stronger fashion. The additional reduced operators $\bm{D}$ and $\bm{E}$ are defined as follows:
\begin{align}
& (\bm{D})_{i} = \left(\bm{\phi_i} \right)_{L^2({\Gamma_D})},\\
& (\bm{E})_{ij} = \left(\bm{\phi_i}, \bm{\phi_j} \right)_{L^2({\Gamma_D})}.
\end{align} 
In equation \eqref{eq:penalty_DAE}, we assumed to have only one inhomogeneous boundary condition at the Dirichlet boundary, however, a generalization  for more than one condition can be done \cite{PhDHijazi}. The POD-Galerkin PINN model can be adopted to the penalty method by including the additional constraint as a separate loss function denoted by $E_3(\bm{w})$ defined as follows:
\begin{equation}\label{eq:eq3Loss}
E_3(\bm{w}) = \sum_{n=1}^{N_s} \frac{1}{N_u}  \sum_{k=1}^{N_u} \{ {\bm{R}^c_k(\bm{l}^n,\bm{y},\bm{w})} \}^2,
\end{equation}
with 
\begin{align}
& \bm{R}^c = U_{BC} \bm{D} - \bm{E} \bm{a}_\textrm{PINN}  \in \mathbb{R}^{N_u}.
\end{align}
As for the case of the lifting function method, the homogenization procedure leads to the transfer of the inhomogeneous Dirichlet boundary conditions from the velocity snapshots to the lifting functions. Therefore, the lifting velocity mode will have a normalized version of the velocity at the Dirichlet boundary at the inlet \footnote{the velocity POD modes including the lifting function are normalized and this causes $a_l$ to be a normalized version of the velocity at the inlet}.

The reduced approximation of the velocity field is modified to the following one:
\begin{equation}\label{eq:uROM_Lift}
\bm{u}(\bm{x},t; \bm{\mu}) \approx  a_l \bm{\phi}_l(\bm{x}) + \sum_{i=1}^{N_u} a^u_i(t;\bm{\mu}) \bm{\phi}_i(\bm{x}) + \sum_{i=1}^{N_S} a^S_i(t;\bm{\mu}) \bm{s}_i(\bm{x}),
\end{equation}
where $\bm{a} = [a_l, a^u_1,\hdots,a^u_{N_u},a^S_1,\hdots,a^S_{N_S}] \in \mathbb{R}^{1+N_u+N_S}$, here the upper-subscript in $a^u_i$ and $a^S_i$ refer to the reduced coefficients corresponding to the original velocity POD modes and the supremizer added ones, respectively. The coefficient $a_l$ is the one that corresponds to the lifting mode and represents as mentioned a normalized version of the velocity at $\Gamma_D$. In this case, the reduced operators $\bm{R}^a$ and $\bm{R}^b$ from 
\eqref{eq:Ra} and \eqref{eq:Rb} in the PINN formulation become:
\begin{align}
& \bm{R}^a = - \bm{M} [a_l, \bm{a}_\textrm{PINN}] + \nu (\bm{B}+\bm{B_T})[a_l, \bm{a}_\textrm{PINN}]-\bm{c}_\textrm{PINN}-\bm{H}\bm{b}_\textrm{PINN} + \bm{h}_\textrm{PINN} \in \mathbb{R}^{N_u},\\
& \bm{R}^b = \bm{P} [a_l, \bm{a}_\textrm{PINN}] \in \mathbb{R}^{N_p},
\end{align}
Hence, we observe that the PINNs are able to incorporate the velocity at the boundary in their formulation using both the lifting function and penalty methods, making it possible to learn these physical parameters through the training procedure and the optimization of the total loss function.

\section{Numerical results}\label{sec:results}
This section presents the application of the POD-Galerkin PINN reduced order models on three problems with unknown inputs or parameters. The first problem is the benchmark case of the flow past a backward step. In the first problem the POD-Galerkin PINN model is used for solving inverse and forward problems in the parameterized setting. The second one is the flow around a circular cylinder in a turbulent setting modeled by the RANS approach, in this problem we consider a situation of incomplete data, where the ROM will be used to infer an unknown input value and its corresponding missing output data. The last problem is the $3D$ flow around a surface mounted cube, this problem is a turbulent one with a large number of degrees of freedom, where turbulence is modeled using the LES turbulence approach. In the latter problem, the POD-Galerkin PINN ROM is used for the identification of the physical viscosity which is assumed to be unknown. This is done by assuming the presence of simulated data for the velocity, pressure and the eddy viscosity fields. The full order solver utilized is OpenFOAM\textsuperscript{\textregistered} (OF) \cite{weller1998tensorial} which is an open-source C\texttt{++}-based library for solving fluid problems with the finite volume method. At the reduced order level, the POD modes and the $L^2$ projection coefficients needed for the training of the PINNs are computed using the library ITHACA-FV \cite{RoSta17} which is also based on C\texttt{++}, while the training of the neural network is done using the Python library TensorFlow V2 \cite{TensorFlow}.\par 

\subsection{Steady case: The Flow Past a Backward Step}\label{sec:steady_lam_backStep}
\begin{figure}
\centering
\includegraphics[width=0.8\textwidth]{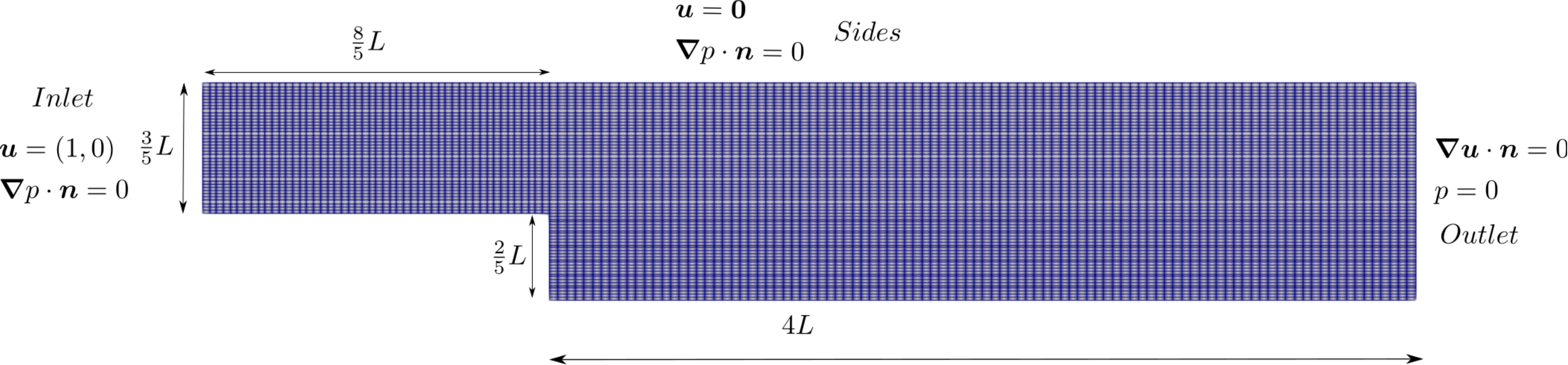}
\caption{The computational domain used in the numerical simulations of the flow past a backward step, all lengths are described in terms of the characteristic length $L$ that is equal to $1$ meter.}\label{fig:comp_domain_bs}
\end{figure}

We consider the application of the POD-Galerkin PINN based reduced order model to the benchmark case of the flow past a backward step. The problem is studied here in laminar steady setting with physical parameterization. In \autoref{fig:comp_domain_bs}, the computational domain is depicted, one may see the lengths of the different parts of the domain expressed in terms of the characteristic length $L=1$ \si{m}. The boundary conditions for both the velocity and pressure fields are reported in the latter figure. The objective is to utilize the POD-Galerkin PINN model for solving inverse problems, these inverse problems involve the estimation of certain physical parameters such as the velocity at the inlet or the physical viscosity. In the current setting, the physical viscosity $\nu$ is parameterized, where it is varied in the range of $[0.02,7.1]$ \si{m^2\per s}. The velocity at the inlet is fixed at $1$ \si{m\per s}, this gives a parameterized Reynolds number which lies in the range of $[0.1408,50]$. As for the numerical schemes used at the FOM level, a $1$-st order bounded Gauss upwind scheme is used to approximate the convective term. The gradients are approximated using a Gauss linear scheme, while a Gauss linear scheme with non-orthogonal correction has been utilized for the approximation of the Laplacian terms.\par  

\begin{figure}
\centering
\includegraphics[width=0.95\textwidth]{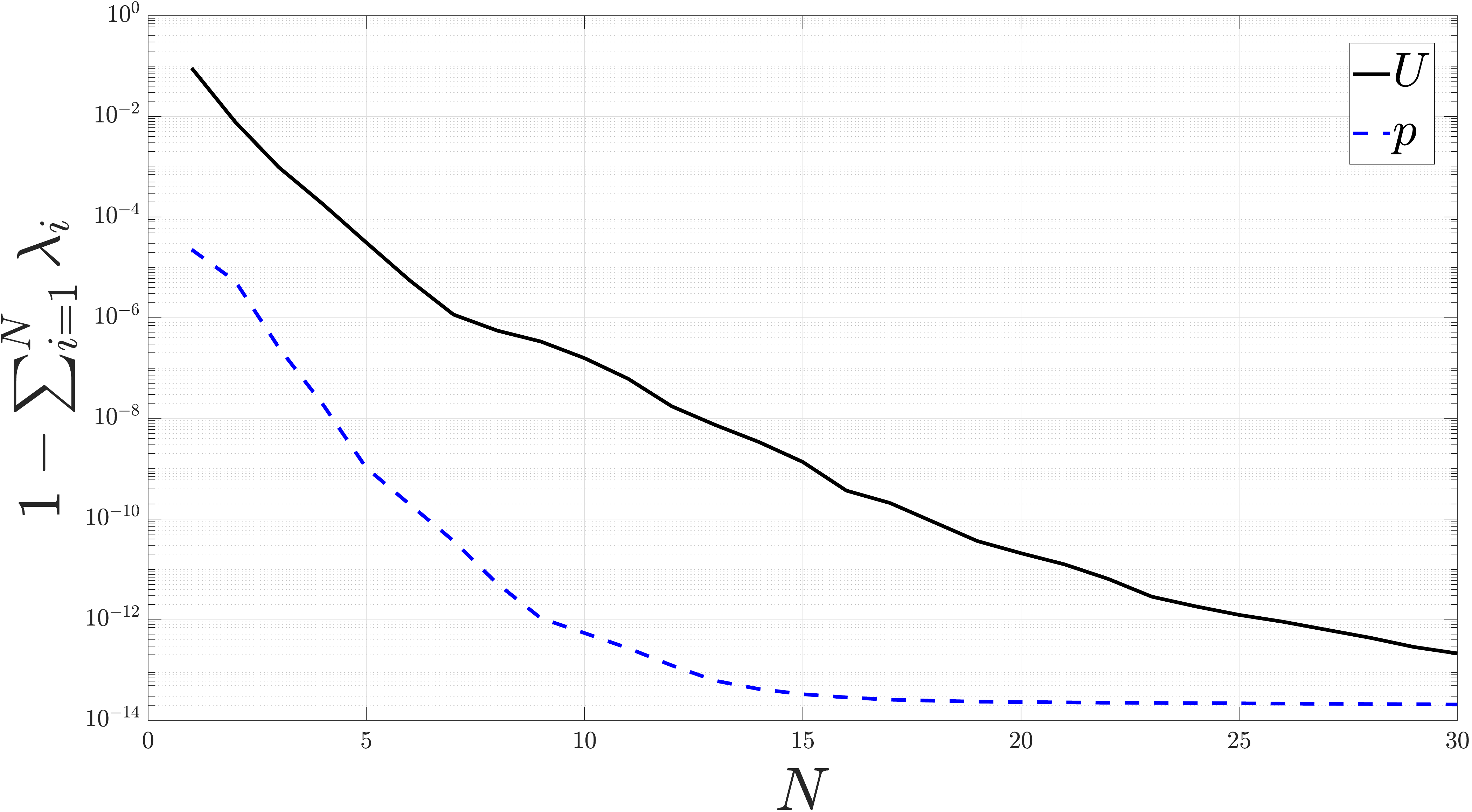}
\caption{The cumulative ignored eigenvalues decay for the first numerical case of the flow past a backward step. In this figure, the solid black line refers to the velocity eigenvalues and the dashed blue line corresponds to the pressure eigenvalues.}\label{fig:Cum_backstep}
\end{figure}

The offline stage starts by collecting snapshots for different values of the parameter $\nu$, in particular $150$ samples were drawn from the range $[0.02,7.1]$ and for each viscosity sample, the FOM was run using the SIMPLE solver. The snapshots of velocity, pressure and flux fields were stored for the computation of the POD modes and the reduced operators involved in the formulation of the POD-Galerkin PINN model. Firstly, the non-homogeneous velocity boundary condition at the inlet is treated with the help of the lifting function method. In particular, the average of the velocity snapshots computed for different values of the parameter is computed and then used as the lifting function. The lifting function is then subtracted from the original snapshots which lead to the creation of a new set of homogenized velocity snapshots (velocity snapshots which have homogeneous Dirichlet boundary condition at the inlet). At this stage one can apply the POD procedure on the snapshots matrices of both the homogenized velocity and the pressure. The cumulative eigenvalues decay can be seen in \autoref{fig:Cum_backstep}, where the cumulative eigenvalue decay for the pressure is observed to be slower. The second step involves the computation of the supremizer modes which can be done by solving the supremizer problems corresponding to each pressure POD mode. The velocity POD space is then enriched by the supremizer modes. The convective part of each snapshots is then retrieved for later use during the training stage of the neural network.\par 

At this point, we describe the numerical tests which will be done in this subsection. The first test has two objectives which are:

\begin{itemize}
\item To estimate a physical unknown which is the velocity at the boundary, where we assume for this test that this value is unknown. The identification of this physical unknown is carried out by the POD-Galerkin PINN model.

\item To approximate the velocity and pressure fields for test values of the physical viscosity $\nu$ which were not seen during the offline stage.
\end{itemize} 

The two objectives are achieved simultaneously by training only once the PINN informed by the POD-Galerkin system using the training data and then performing a prediction task for the test data.\par 

In the second test, the input test data of the physical viscosity mentioned above will be assumed to be unknown and then the POD-Galerkin PINN model will be used to solve the inverse problems associated with the latter test data while assuming that the velocity at the inlet is known.\par 

It is important to mention that the velocity at the boundary is embedded in the POD-Galerkin PINN formulation. In fact, the first coefficient of the reduced velocity vector $\bm{a}$, namely $a_l$ corresponds to a normalized version of the velocity at the inlet. The reduced approximation of the velocity in this example read as in \eqref{eq:uROM_Lift}.

The next step is to compute the reduced operators which appear in the following POD-Galerkin system that defines the reduced equations:
\begin{subequations}\label{eq:dae1}
\begin{align}
& \nu (\bm{B}+\bm{B_T})\bm{a}-\bm{c}-\bm{H}\bm{b} = \bm{0},
 \\
&\bm{P} \bm{a}= \bm{0}.
\end{align}
\end{subequations} 

The input of the PINN is the physical viscosity $\nu$ and its output is formed by the reduced velocity (except the lifting coefficient $a_l$), reduced pressure and the reduced convective term. The number of the PINN output variables is $N_o = 3N_u + 3N_S + N_p + 1$.\par

To solve the inverse problem involved and to approximate the relation between the physical viscosity and the reduced velocity and pressure, we put forward a physics informed neural network which has $10$ layers with each layer containing $100$ neurons with tangent hyperbolic activation. This PINN takes as output the reduced variables mentioned above and uses the data available from the offline snapshots together with the knowledge given by the system in \eqref{eq:dae1} to approximate the unknown coefficient $a_l$ (or the velocity at the inlet). This is done by training the PINN with a loss function that takes into consideration the data given by the coefficients of the $L^2$ projections and also by constraining the output to follow the POD-Galerkin reduced equations. The output data is obtained by performing the projections in equations \eqref{eq:vec_coeff}, \eqref{eq:pressure_coeff} and \eqref{eq:conv_coeff}. We would like to remark that both input and output values have been standardized to range of $[0,1]$ in order to make the neural networks learning task easier. The PINN loss function is written as
\begin{equation}
E(\bm{w}) = E_\textrm{data}(\bm{w}) + \alpha_1 E_1(\bm{w}) + \alpha_2 E_2(\bm{w}),
\end{equation}
where
\begin{equation}
E_\textrm{data}(\bm{w}) = \sum_{n=1}^{N_s} \frac{1}{N_o}  \sum_{k=1}^{N_o} \{ y_k(\bm{l}^n,\bm{w}) - r_k^n \}^2,
\end{equation}

\begin{equation}
E_1(\bm{w}) = \sum_{n=1}^{N_s} \frac{1}{N_u}  \sum_{k=1}^{N_u} \{ {\bm{R}^a_k(\bm{l}^n,\bm{y},\bm{w})} \}^2,
\end{equation}

\begin{equation}
E_2(\bm{w}) = \sum_{n=1}^{N_s} \frac{1}{N_p}  \sum_{k=1}^{N_p} \{ {\bm{R}^b_k(\bm{l}^n,\bm{y},\bm{w})} \}^2,
\end{equation}
and
\[
\bm{R}^a = \nu (\bm{B}+\bm{B_T})[a_l, \bm{a}_\textrm{PINN}]-\bm{c}_\textrm{PINN}-\bm{H}\bm{b}_\textrm{PINN} \in \mathbb{R}^{N_u}, \quad \bm{R}^b = \bm{P} [a_l, \bm{a}_\textrm{PINN}] \in \mathbb{R}^{N_p},
\]
\[
\bm{y} = [ \bm{a}_\textrm{PINN}, \bm{b}_\textrm{PINN}, \bm{c}_\textrm{PINN}] \in \mathbb{R}^{N_o}.
\]
In the above formulation, the weights and biases vector $\bm{w}$ contains all trainable parameters which include the scalar coefficient $a_l$ which is then learned during the training procedure of the PINN. As for the coefficients of the equations losses $\alpha_1$ and $\alpha_2$, they are determined in a heuristic fashion or they can also be trained like the other weights of the network.\par

The total number of trainable parameters in the PINN for the first test is $83627$. The PINN is run for $30000$ epochs with a learning rate of $1 * 10^{-3}$ and with one batch per epoch (batch size is equal to the number of data points $N_s$). The two weighting coefficients $\alpha_1$ and $\alpha_2$ are set to $0.01$. The physical parameter $a_l$ is initialized with zero value.\par 

\begin{figure}[htbp]
\centering
\begin{subfigure}[b]{1\textwidth} 
\includegraphics[width=0.98\linewidth]{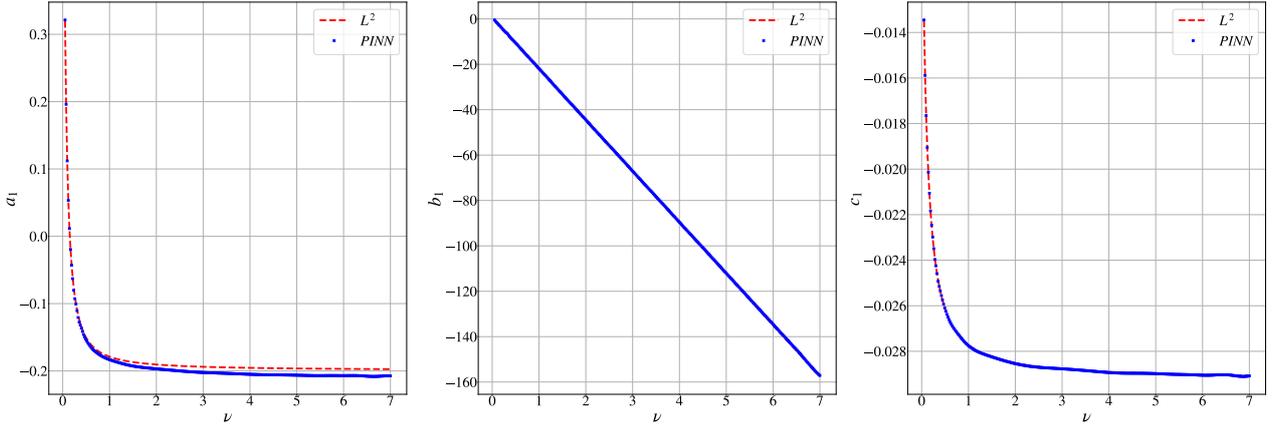} 
\caption{The first reduced coefficients for velocity, pressure and convective terms compared to the ones obtained by the $L^2$ projection.}
\label{fig:BackStep_1stCase_a1b1} 
\end{subfigure}\vspace{5pt}
\begin{subfigure}[b]{1\textwidth} 
\centering
\includegraphics[width=0.98\linewidth]{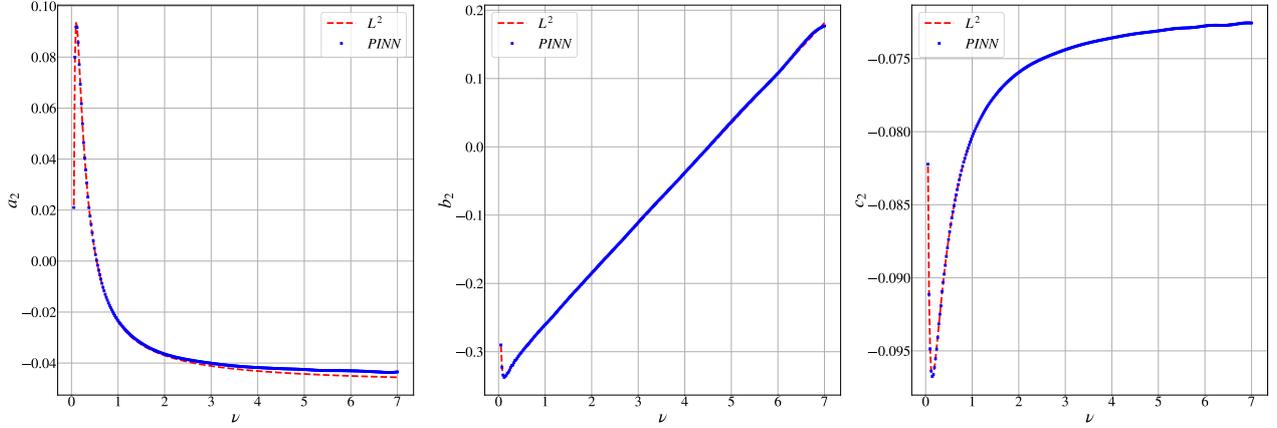} 
\caption{The second reduced coefficients for velocity, pressure and convective terms compared to the ones obtained by the $L^2$ projection.}
\label{fig:BackStep_1stCase_a2b2}
\end{subfigure}\vspace{5pt}
\begin{subfigure}[b]{1\textwidth} 
\centering
\includegraphics[width=0.98\linewidth]{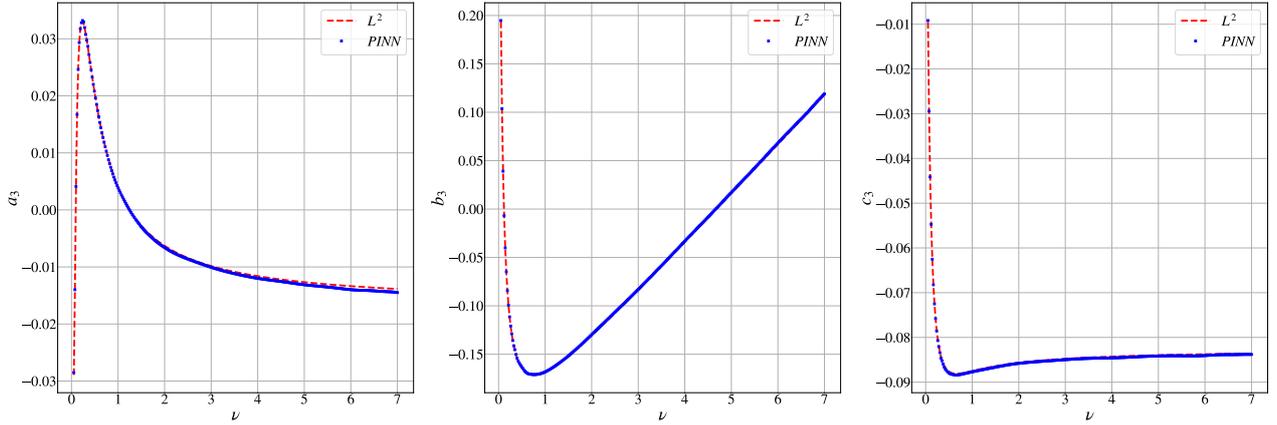} 
\caption{The third reduced coefficients for velocity, pressure and convective terms compared to the ones obtained by the $L^2$ projection.}
\label{fig:BackStep_1stCase_a2b2}
\end{subfigure}\vspace{5pt}
\caption{The results of the POD-Galerkin PINN predictions for the $1$st, $2$nd and $3$rd components of the velocity, pressure and convective reduced coefficients for the first numerical test. The plots compare the reduced coefficients with the $L^2$ projection coefficients of the test velocity and pressure fields onto the corresponding POD modes. The red-dashed lines refers to the $L^2$ projection coefficients, while the blue-dots correspond to the reduced coefficients obtained by the PINNs. The coefficients are plotted versus the physical viscosity values at which the test data was generated.}\label{fig:BackStep_1stCase_NN} 
\end{figure}

The first results are shown for the following number of modes $N_u = N_p = N_S = 5$. This number of modes gives a total number of PINN outputs of $N_o = 26$. The true value of $a_l$ is $2.7302$, while the PINN has identified the value of $a_l$ to be $2.7291$. The relative error in approximating $a_l$ is about $0.0409$ $\%$. As mentioned above $a_l$ has been initialized with zero value, however, we show also the results for different initial values of $a_l$ in \autoref{tab:pinn}, one can see that the PINN is not sensitive to the initial values of the unknown parameter $a_l$. As for the forward task in the first test, we have generated $300$ samples for $\nu$ which are equidistant samples in the range $[0.05, 7]$. Then another simulation campaign is launched for these viscosity samples in order to validate the PINN model. After the training of the PINN for the identification of $a_l$, the PINN is used for approximating the output for the newly created set of input $\nu$.  \autoref{fig:BackStep_1stCase_NN} show the results of the forward task, where one can see the validation results for the first, second and third components of the reduced variables of the velocity, pressure and convective terms versus the value of the viscosity. We recall that the values of the physical viscosity for which the latter figure is depicted were not used in the training procedure of the POD-Galerkin ROM or the PINN. The error committed in approximating the reduced variables in the mean squared sense is about $9.8921* 10^{-6}$, we remark that the latter error is computed on the standardized variables. As for the operators errors for the test data we have $\tilde{E}_1(\bm{w}) = 0.001027$ and $\tilde{E}_2(\bm{w}) = 2.6525* 10^{-6}$. The last quantitative values of the errors show that the PINN was able to generalize for unseen values of the parameter and at the same time constraining the results to satisfy the algebraic system. As for the training time of the PINNs, it ranges from $6$ to $7.402$ minutes using "Intel(R) Core(TM) i7-10610U CPU @ 1.80GHz".\par

\begin{table}[htp]
\centering
{
\begin{tabular}{ c | c | c | c | c  }
Initial $a_l$ & $E_\textrm{data}(\bm{w})$ & $E_1(\bm{w})$ & $E_2(\bm{w})$ & PINN $a_l$\\
\hline  			
$0$ & $9.152303* 10^{-6}$ & $0.00045994046$ & $3.0014705* 10^{-6}$ & $2.7291148$ \\ 
$5$ & $9.2302425* 10^{-6}$ & $0.0009832102$ & $3.0044891* 10^{-6}$ & $2.7288582$ \\ 
$10$ & $0.00020264629$ & $0.00036979085$ & $2.9998373* 10^{-6}$ & $2.7270014$ \\ 
$20$ & $1.4284992* 10^{-5}$ & $0.0016940477$ & $3.0059625* 10^{-6}$ & $2.7295387$ \\ 
\end{tabular}}\caption{{The PINNs results for the data mean squared error and the mean squared errors corresponding to the residual functions defining the POD-Galerkin ROM DAE. The values of the errors are reported for different initial values of the added weight $a_l$, the PINN identified value of $a_l$ is reported in the last column. The PINNs are run for $30000$ epochs with a learning rate of $1 * 10^{-3}$.}}
\label{tab:pinn}
\end{table}

As a final result for the first case, we report an assessment of the approximation accuracy of the POD-Galerkin PINN ROM. Namely, we compute the relative $L^2$ error for velocity and pressure which, respectively, read as:

\begin{equation}\label{eq:l2_error}
\epsilon_u = \frac{{\left\lVert \bm{u} - \bm{u}_r \right\rVert}_{L^2(\Omega)}}{{\left\lVert \bm{u}\right\rVert}_{L^2(\Omega)}} \times 100 \%,\\
\epsilon_p = \frac{{\left\lVert p - p_r \right\rVert}_{L^2(\Omega)}}{{\left\lVert p \right\rVert}_{L^2(\Omega)}} \times 100 \%,
\end{equation}
where $\bm{u}_r$ and $p_r$ are the reduced order velocity and pressure fields, respectively. The values of the relative errors for the velocity and the pressure are computed for the $300$ test parameter values. The mean value of the $\epsilon_u$ reduced velocity error is $0.3081$ $\%$, while the one of the pressure is $0.3459$ $\%$.\par

In the second test, we assume that the input values for the test data in the latter test are unknown. We aim at solving the inverse problems involved with the test data. To this end, we put forward a PINN based on the POD-Galerkin ROM with the following loss function:

\begin{equation}
E(\bm{w}) = E_\textrm{data}(\bm{w}) + \alpha_1 E_1(\bm{w}) + \alpha_2 E_2(\bm{w}) + \alpha_3 \tilde{E}_1(\bm{w}),
\end{equation}

where 
\begin{equation}
\tilde{E}_1 = \sum_{n=1}^{300} \frac{1}{N_u}  \sum_{k=1}^{N_u} \{ {\tilde{\bm{R}}^a_k(\tilde{l}^n,\bm{y},\bm{w})} \}^2,
\end{equation}

and

\begin{equation}
\tilde{\bm{R}}^a = \tilde{l}^n (\bm{B}+\bm{B_T})[a_l, \bm{\tilde{a}}]-\bm{\tilde{c}}-\bm{H}\bm{\tilde{b}} \in \mathbb{R}^{N_u},
\end{equation}

where $\tilde{l}^n$ is the $n$-th value of the unknown input physical viscosity. The PINN used in this second test has the same structure as the one utilized in the first test. The training parameters are also the same, where the PINN is run for $30000$ epochs with learning rate of $1 * 10^{-3}$. The vector of unknown viscosity values $\bm{\tilde{l}}$ is considered as additional trainable weight of the PINN and is embedded into $\bm{w}$. The identified values of the physical viscosity match to high degree of accuracy the true values. The mean squared error of the difference of the true input vector and the PINN-identified one is $0.00055$.

\subsection{The Flow Around a Circular Cylinder}\label{sec:cube}

\begin{figure}[htp]
\centering
\includegraphics[width=15cm]{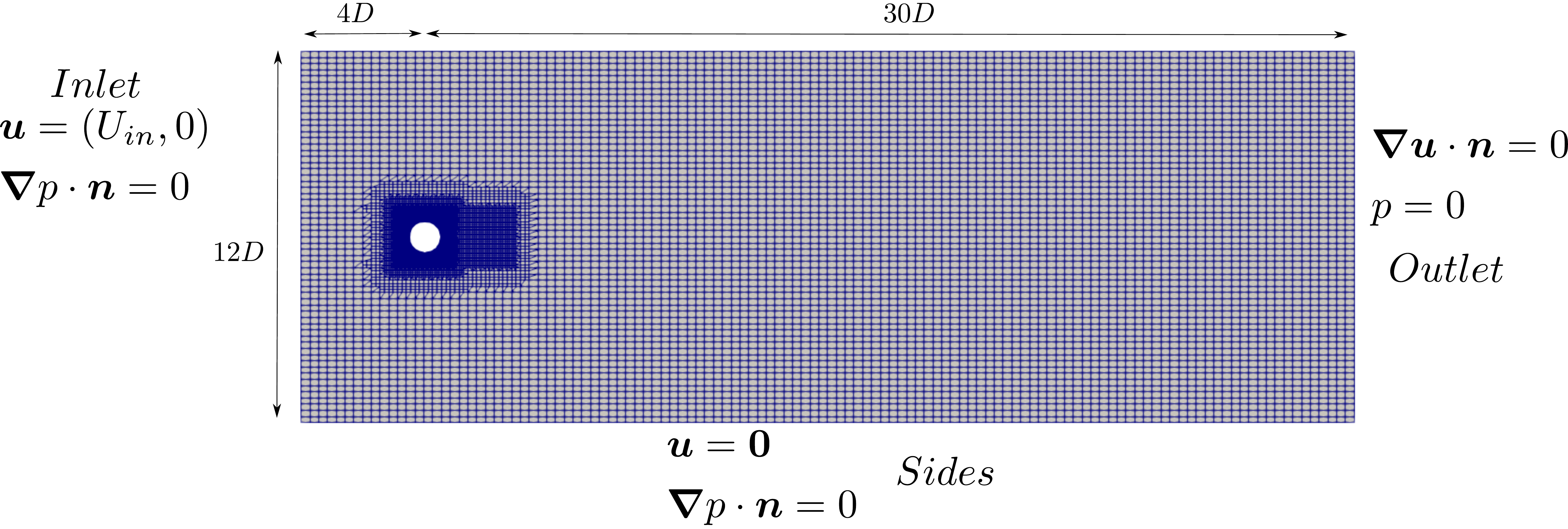} 
\caption{The computational grid of the problem of the flow around a circular cylinder.} 
\label{fig:comp_domain_cy} 
\end{figure}

In this subsection we address the case of having incomplete data for different input configurations/settings. The computational problem considered is the one of the flow around a circular cylinder. The problem is $2$D turbulent one, where turbulence is modeled using the RANS approach. The domain of the problem is $\Omega := [-4D,30D] \times [-6D,6D] \setminus B_D(0,0)$, where $D = 1 \si{m}$ is the diameter of the cylinder. \autoref{fig:comp_domain_cy} shows the grid used for simulating the problem using OpenFOAM, it also reports the boundary conditions for the velocity and pressure fields. The grid has around $18000$ cells, the physical viscosity is $2.5 \times 10^{-4}$ \si{m^2 \per s}.\par 

We assume in this test that we are presented with an incomplete set of data for the fluid dynamics fields. This set of data contains full information about the fluid dynamics fields for some parameter values and contains a set of partial output data for an unknown input/parameter. In particular, we consider the parameter in this test to be the velocity at the inlet $U_{in}$. The fluid dynamics fields for velocity, pressure and the eddy viscosity are available for three different known values of the parameter $U_{in}$ which are $\{ 1, 1.5, 2 \}$ \si{m \per s}. Another set of fluid data for unknown parameter value is presented, the latter set contains partial information as it lacks the pressure fields. The inference tasks in this example are (i) to approximate the unknown velocity at the inlet $U^\star_{in}$, (ii) to recover the missing pressure field history and finally (iii) to compute the lift and drag forces acting on the surface of the cylinder which are dependent on both $U^\star_{in}$ and the missing pressure data.\par 

The snapshots coming from the fluid simulations are covering at least $2$ periodic cycles of the developed regime. For each parameter value, the regime frequency known as the vortex shedding frequency is different, resulting in various snapshots time windows for the three parameter data samples. \autoref{tab:parOffline} shows the snapshots time windows for each parameter sample and the corresponding time period. The number of snapshots per parameter sample is $201$ giving a total of $603$ snapshots. The POD is done on the velocity and pressure snapshots which yields the velocity and pressure POD basis, then an enrichment procedure is carried out by solving the supremizer problems. The ROM is then obtained via a Galerkin projection, in this example the penalty method is used for the enforcement of the inlet velocity at the reduced level.\par

\begin{table}[htp]
\centering
\begin{tabular}{ c | c | c | c | c }
  \textbf{Parameter sample : $U_{in}$ in $\si{m\per s}$}  & \textbf{Time period} $T_p$ in $\si{s}$ & \textbf{Snapshot time window} in $\si{s}$ \\
    \hline  			
  $1$ & $4.255$ & $[0, 11]$ \\
  \hline
  $1.5$ & $2.830$ & $[0, 7.5]$ \\
  \hline  
  $2$ & $2.1127$ & $[0, 5.5]$ \\
  \hline  
\end{tabular}\caption{Offline parameter samples and the corresponding vortex shedding frequency and snapshots time window}
\label{tab:parOffline}
\end{table}

At this stage, one may compute the input and output data matrices which will be used to train the PINN. The input of the PINN is the combination of time and the parameter. However, a non-dimensionalization conversion for time is needed in order to give meaningful information for the PINN. To this end, the non-dimensionalized time $t^\star$ is defined as $t^\star = \frac{tU_{in}}{D} = t U_{in}$. As for the output data, it consist in the following reduced coefficients:

\begin{itemize}
\item The $L^2$ projection coefficients of the velocity snapshots onto the velocity POD modes, see \autoref{eq:vec_coeff}.

\item The $L^2$ projection coefficients of the pressure snapshots onto the pressure POD modes, see \autoref{eq:pressure_coeff}.

\item The $L^2$ projection coefficients of the convective terms snapshots onto the velocity POD modes, see \autoref{eq:conv_coeff}.

\item The $L^2$ projection coefficients of the turbulent terms snapshots onto the velocity POD modes, see \autoref{eq:turb_coeff}.

\end{itemize}

The output of the PINN is the stacked vector of $[\bm{a},\bm{b},\bm{h},\bm{c}]$. The POD-Galerkin DAE that models this problem is the one reported in \autoref{eq:penalty_DAE}. The matrices and vectors which appear in the latter DAE are computed during the offline stage. The partially known output fields for the unknown parameter are also projected onto the velocity POD space and the resulting $L^2$ coefficients are also used for training the PINN.\par

The set of original weights and biases of the PINN denoted by $\bm{w}$ is then enlarged. This is done by introducing an additional set of weights denoted by $\bm{w}^\star$ which contains trainable weights that correspond to the unknown quantities to be approximated. In more details, $\bm{w}^\star$ contains the scalar weight $w_{U^\star_{\textrm{in}}}$ which is introduced for the approximation of the unknown velocity at the inlet. Also $\bm{w}^\star$ encapsulates a matrix of weights denoted by $\bm{w}^{b^\star}$ whose $i$-th column $\bm{w}^{b^\star}_{i}$ is the reduced pressure vector for the unknown pressure output data corresponding to $U^\star_{in}$ at a fixed time instant. Hence, the number of additional weights is $201*N_p+1$ (we assume that the number of data points for the unknown input is $N^\star_T =201$).\par

The PINN loss function $E(\bm{w},\bm{w}^\star)$ is defined as follows:

\begin{equation}
E(\bm{w},\bm{w}^\star) = E_\textrm{data}(\bm{w},\bm{w}^\star) + \sum_{i=1}^3  E_i(\bm{w}) + \sum_{i=1}^3 E^\star_i(\bm{w},\bm{w}^\star) + E^\star_b(\bm{w},\bm{w}^\star),
\end{equation}

as one can notice, the loss function incorporates four different types of loss, the first one $E_\textrm{data}(\bm{w},\bm{w}^\star)$ is the fitting data loss for both the fully known input-output data and the unknown input and partially known output data. The second loss $\sum_{i=1}^3  E_i(\bm{w})$ corresponds to the reduced equation losses evaluated only at the known input samples. The third loss is the same as the second one but computed at the unknown input data points. The final loss is defined as $E^\star_b(\bm{w},\bm{w}^\star) = \frac{1}{N_p} \sum_{i=1}^{N^\star_T} \norm{ \bm{w}^{b^\star}_{i} - \bm{b}_{\textrm{PINN}}(t_i^\star,w_{U^\star_{\textrm{in}}}) }_{\mathbb{R}^{N_p}}^2,$ which penalizes the difference between the reduced pressure output of the PINN computed at the unknown input data points and the weights $\bm{w}^{b^\star}$. The last loss component will ensure that the reduced missing pressure will be recovered through the additional weights. It is worth mentioning that initial values of $w_{U^\star_{in}}$ is set to be the media of $U_{in}$ of the known data, while a full zero matrix is used as a starting point for $\bm{w}^{b^\star}$.\par 

\begin{figure}[htp]
  \centering
  \begin{minipage}[b]{0.8\linewidth}
    \centering
    \includegraphics[width=0.8\linewidth]{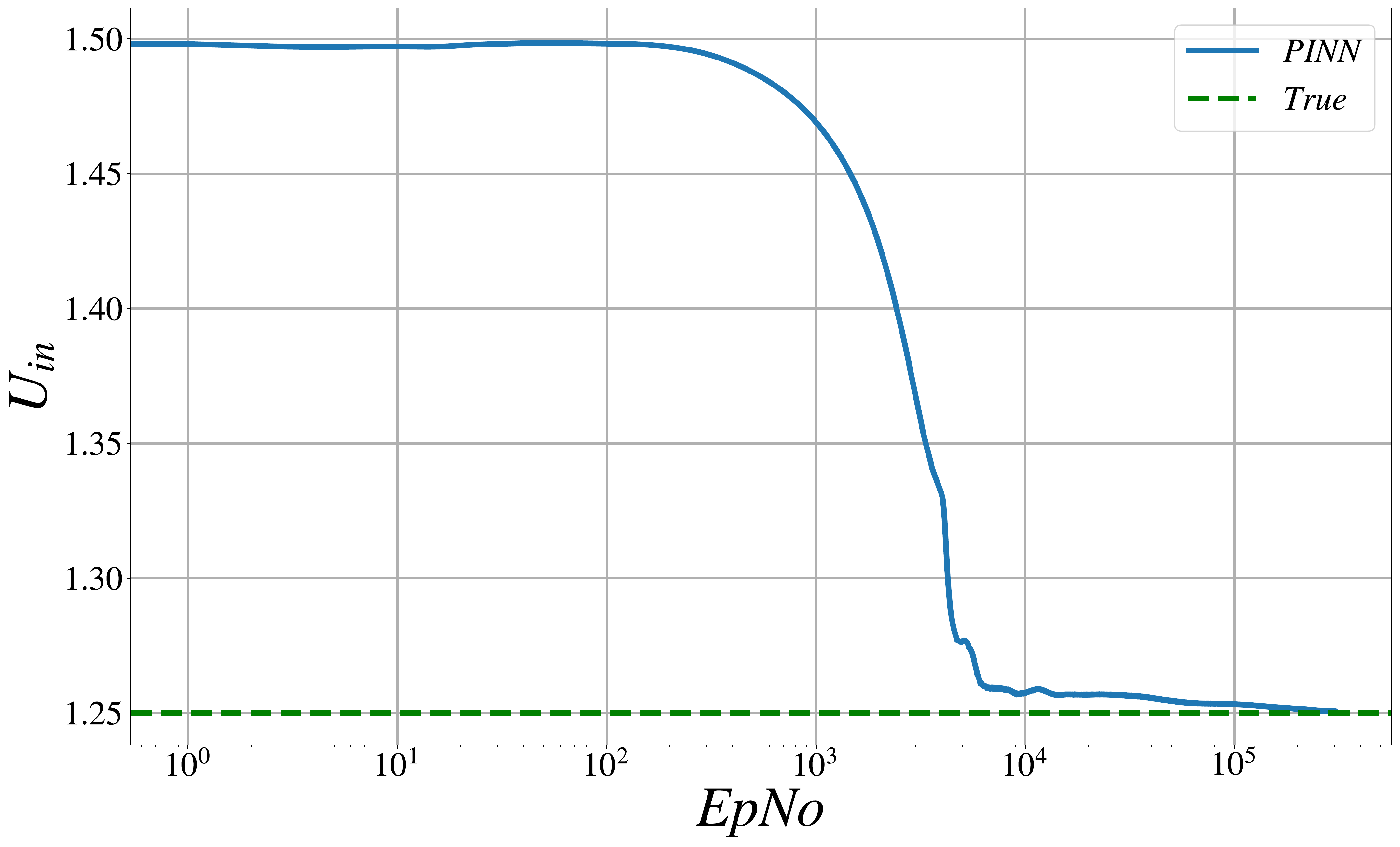} 
  \end{minipage}
  \caption{The PINN approximation of the $U^\star_{in}$.}
  \label{fig:PINN_w_uBC_History} 
\end{figure}

The PINN used in this numerical test has $5$ layers and $64$ neurons per layer with mixed activations (tanh and SIREN). The Adam optimizer is used for solving the optimization problem. The PINN is run for $3 \times 10^5$ epochs, we show in \autoref{fig:PINN_w_uBC_History} the evolution history of the inlet velocity weight during training. One can see that the PINN approximated inlet velocity has converged to a value which is close to the true value of $1.25$ \si{m \per s}. In fact the weight $w_{U^\star_{\textrm{in}}}$ at the end of the training was $1.25029$ which implies that the approximated Reynolds number is $5001.178$ while the real one is $5000$. The reduced order settings for the last result are $N_u = N_p = N_S = 15$.\par

\begin{figure}[htp]
  \centering
  \begin{minipage}[b]{1\linewidth}
    \centering
    \includegraphics[width=1\linewidth]{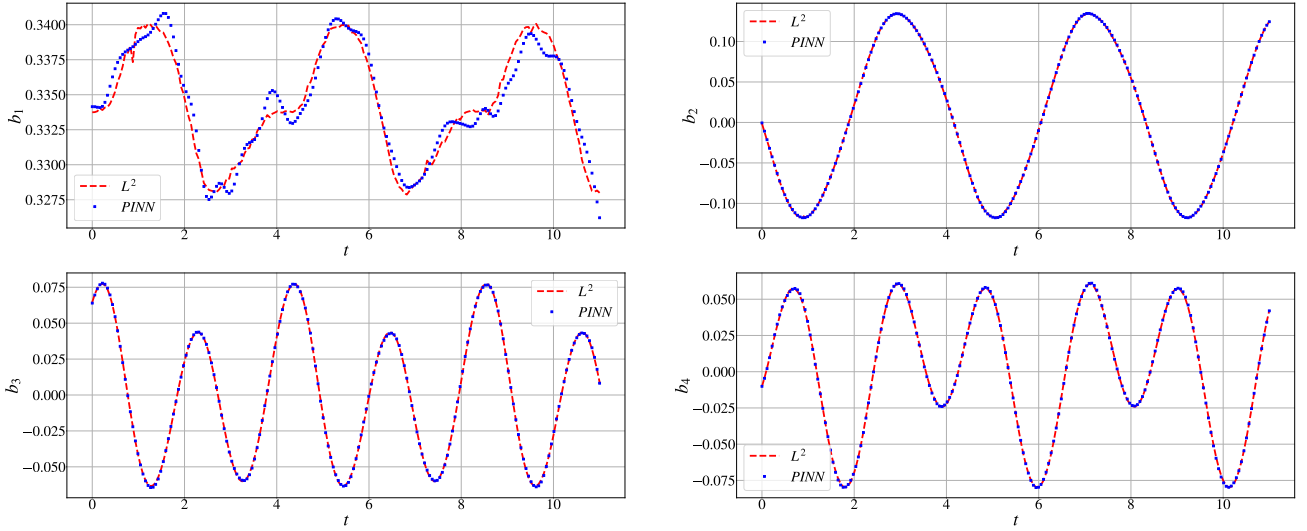} 
  \end{minipage}
  \caption{The first four pressure reduced coefficients for the unknown input data.}
  \label{fig:PINN_w_b} 
\end{figure}

As for the reconstruction of the pressure fields, the additional matrix of weights $\bm{w}^{b^\star}$ which correspond to the reduced pressure of the unknown input data has been optimized in the PINN training process. \autoref{fig:PINN_w_b} depicts the time history of the first four reduced coefficients in the ROM approximation of the missing pressure fields and it shows also the corresponding four coefficients obtained by the $L^2$ projection of the pressure data onto the pressure POD modes.\par

To assess the accuracy of the approximation of the missing pressure fields, we compute the error $\epsilon_p$ in \autoref{eq:l2_error} between the inferred PINN pressure fields and the FOM ones for the $201$ different snapshots. For the reduced model with $N_u = N_p = N_S = 15$, the values of the maximum and mean relative pressure error are $3.1396$ $\%$ and $1.4457$ $\%$, respectively.

Besides solving the inverse problem, we are interested in having an accurate reconstruction of the time-history of the cylinder  drag and lift coefficients. These coefficients come from the fluid dynamics forces $\bm{F}$ acting on the surface of the cylinder which depend locally on the pressure and velocity fields as follows:
\begin{equation}\label{eq:forces}
\bm{F} = \int_{\partial \Gamma_{cy}} (2\mu \bm{\nabla} \bm{u} - p \bm{I}) \bm{n}ds.
\end{equation}

\begin{figure}[htp]
  \centering
  \begin{minipage}[b]{1\linewidth}
    \centering
    \includegraphics[width=1\linewidth]{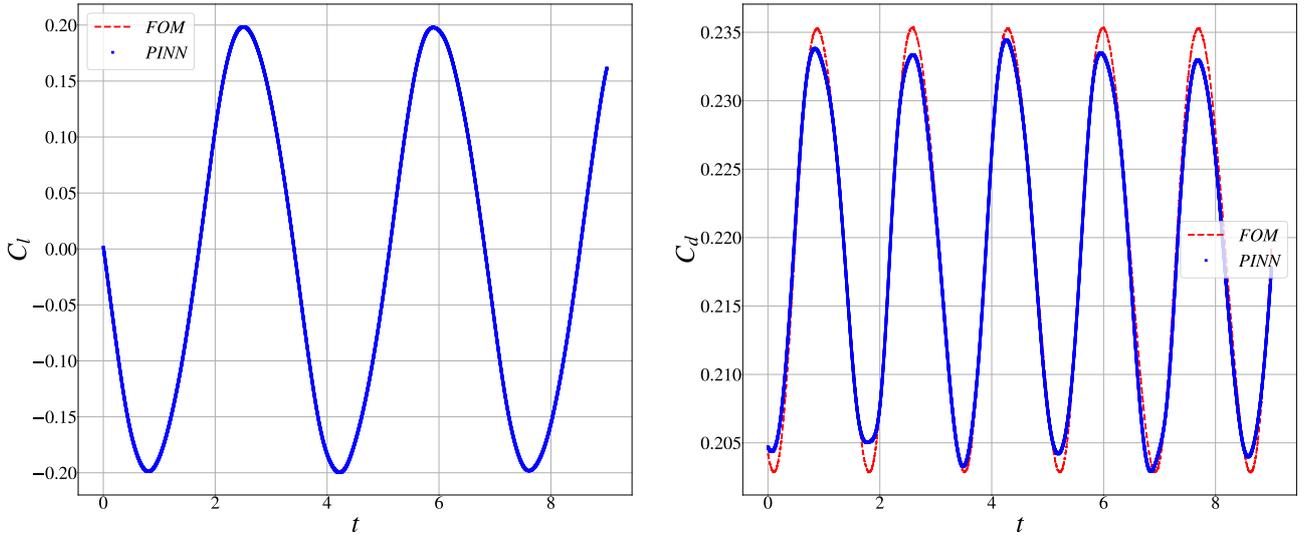} 
  \end{minipage}
  \caption{The lift and drag coefficients curve for the unknown input data ($N_u = N_p = N_S = 15$).}
  \label{fig:PINN_forces} 
\end{figure}

If $F_l$ and $F_d$ are the forces components acting on the surface of the cylinder in the lift and drag direction (the lift direction is the one perpendicular to the flow, while the drag direction is horizontal to the flow), respectively, then the non-dimensionalized drag and lift forces coefficients denoted by $C_d$ and $C_l$, respectively, are given by:
\begin{equation}\label{eq:forcesCoeffs}
C_d=\frac{F_d}{\frac{1}{2}\rho U_{in}^2 A_{ref}}, \quad C_l=\frac{F_l}{\frac{1}{2}\rho U_{in}^2 A_{ref}},
\end{equation}
where $\rho$ is the fluid density and $A_{ref}$ is the reference area. The evaluation of the forces at the reduced level is done in a way that respects the full decoupling of the offline and the online stages, for more details we refer the reader to section $2.8$ in \cite{PhDHijazi}. The PINN is used to recover the lift and drag coefficients by performing forward test for all time values at which we had recorded the forces during the FOM simulation. Then the PINN forces approximation is used together with the PINN-inferred inlet velocity to obtain the reduced lift and drag coefficients signals. The results of this test are shown in \autoref{fig:PINN_forces}.

In order to have a quantitative evaluation of the accuracy of the lift and drag coefficients approximation, we compute the following $L^2$ relative errors

\begin{equation}\label{eq:l2error_cd_cl}
\epsilon_{C_l} = \frac{{\left\lVert C_l(t) - {C_l}^r(t) \right\rVert}_{L_2(T_1,T_2)}}{{\left\lVert C_l(t) \right\rVert}_{L_2(T_1,T_2)}} \times 100 \%, \quad \epsilon_{C_d} = \frac{{\left\lVert C_d(t) - {C_d}^r(t) \right\rVert}_{L_2(T_1,T_2)}}{{\left\lVert C_d(t) \right\rVert}_{L_2(T_1,T_2)}} \times 100 \%,
\end{equation}
where $C_l(t)$ and $C_d(t)$ are the signal functions corresponding to the FOM lift and drag coefficients, respectively. As for ${C_l}^r(t)$ and ${C_d}^r(t)$ they are the corresponding ROM signals, and $[T_1,T_2]$ is the time interval in which the error is sought. The error values for the reduced model with $N_u = N_p = N_S = 15$ are $0.8551$ $\%$ and $0.5699$ $\%$ for lift and drag, respectively. This shows that the POD-Galerkin PINN ROM has been able to reconstruct important CFD performance indicators such as the lift and drag coefficients despite the uncertainty in presence.\par 

\begin{figure}[htp]
  \centering
  \begin{minipage}[b]{0.8\linewidth}
    \centering
    \includegraphics[width=0.8\linewidth]{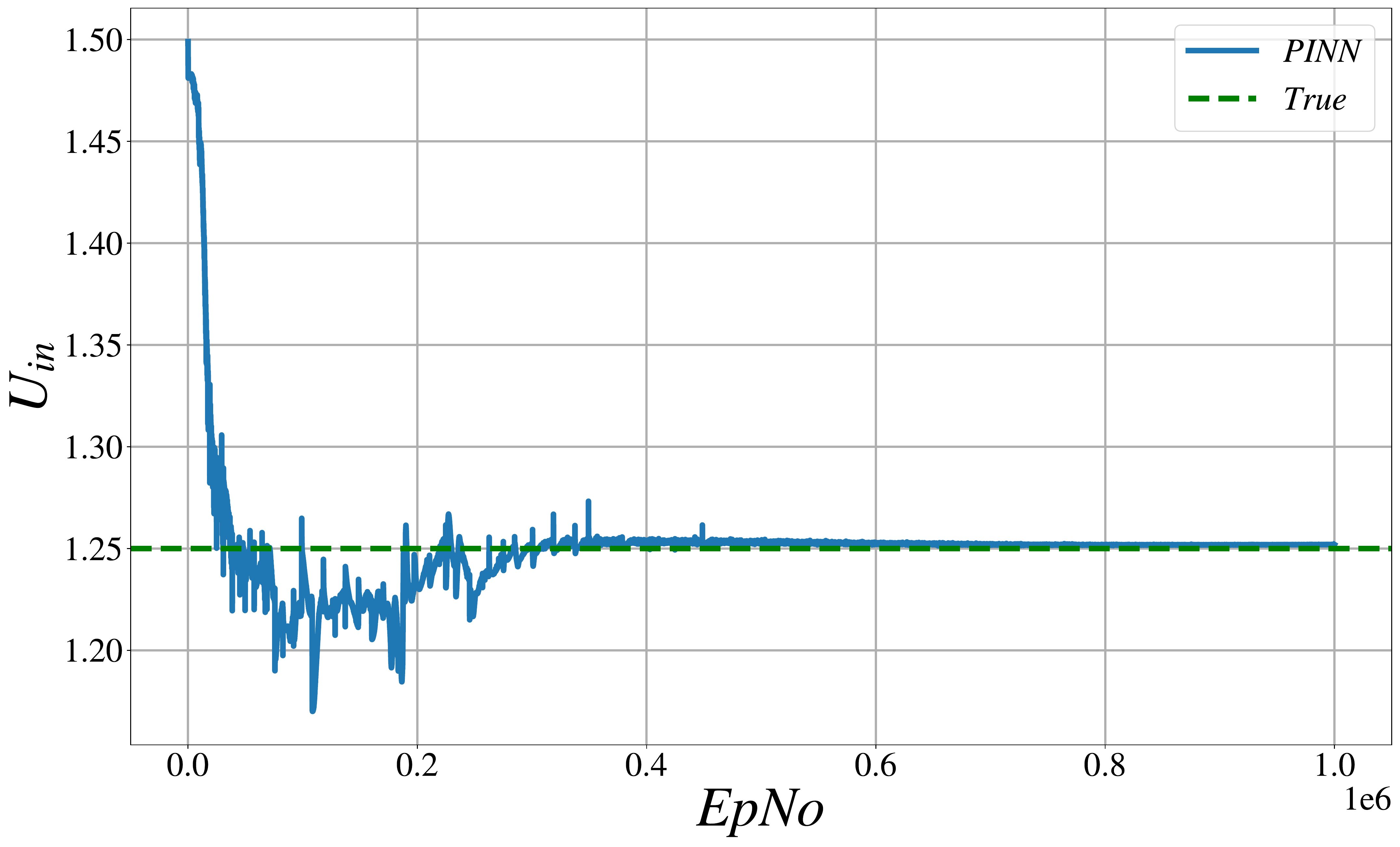} 
  \end{minipage}
  \caption{The PINN approximation of the $U^\star_{in}$ in the case of forces data.}
  \label{fig:PINN_w_uBC_History_ForcesData} 
\end{figure}

In the last results, we have considered the presence of incomplete output data, nevertheless the set of output data was known on the whole internal domain. Now we assume that we have only local data points given by the vector forces acting on the cylinder in \autoref{eq:forces}. This makes the inference problem more difficult to solve because of the locality of the data presented. In spite of that, the POD-Galerkin PINN ROM could be used to solve the inference problem using the data points of the forces. This can be carried out thanks to the physical (reduced) equations which correspond to \autoref{eq:forces} and which are then incorporated in the PINN loss function. We show the evolution of the PINN approximation of the $U_{in}$ in \autoref{fig:PINN_w_uBC_History_ForcesData} for this ultimate test.

\subsection{The Flow Around a Surface Mounted Cube}\label{sec:cube}

\begin{figure}[htp]
\begin{subfigure}[b]{\linewidth} 
\centering\includegraphics[width=15cm]{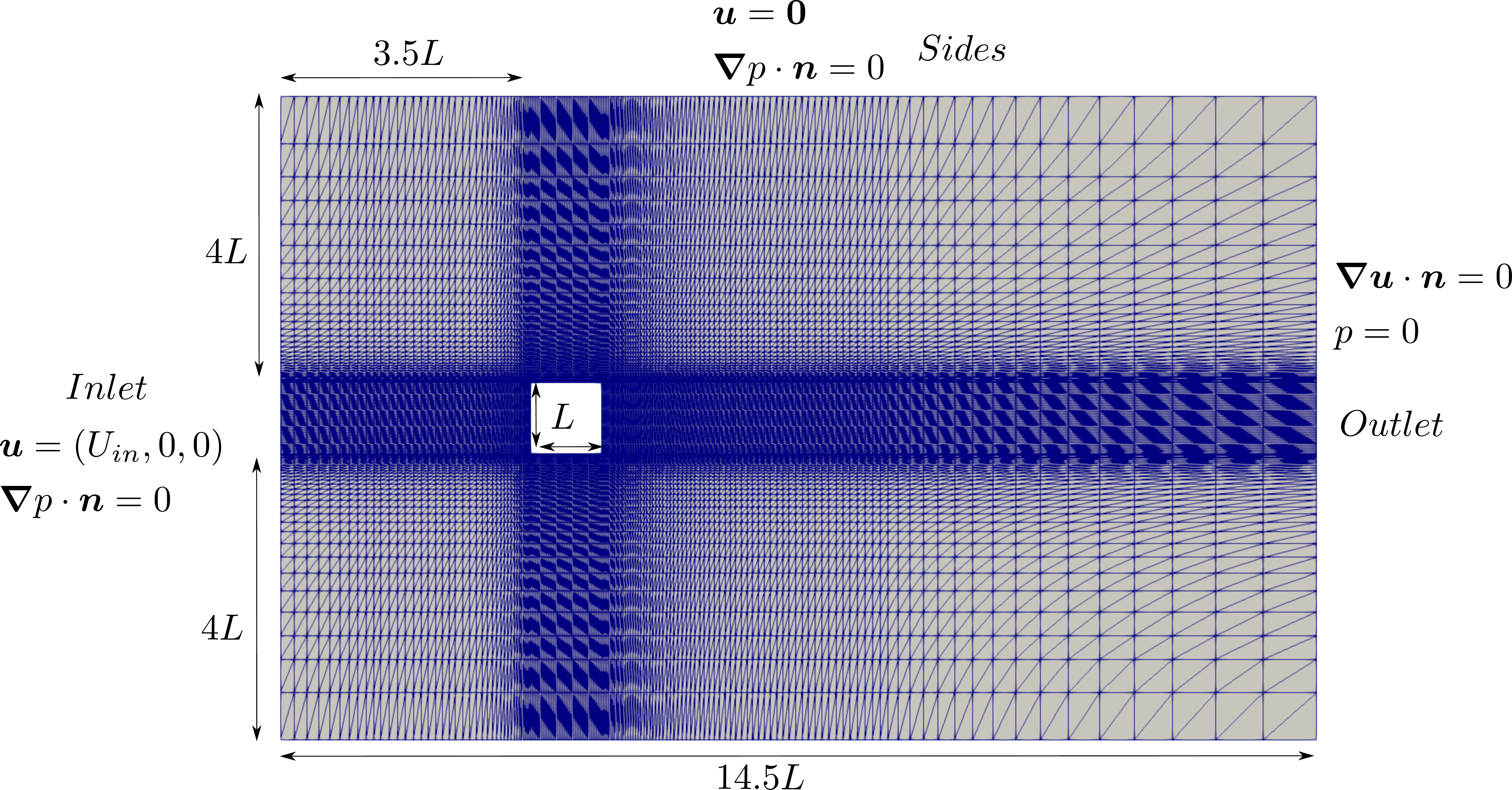} 
\caption{\label{fig:meshZ} A cross-sectional view of the grid at $x_3 = L$, the boundary conditions for the velocity and pressure are reported for the inlet, outlet and the sides.} 
\end{subfigure} 

\centering\begin{subfigure}[b]{0.5\linewidth} 
\centering\includegraphics[width=9cm]{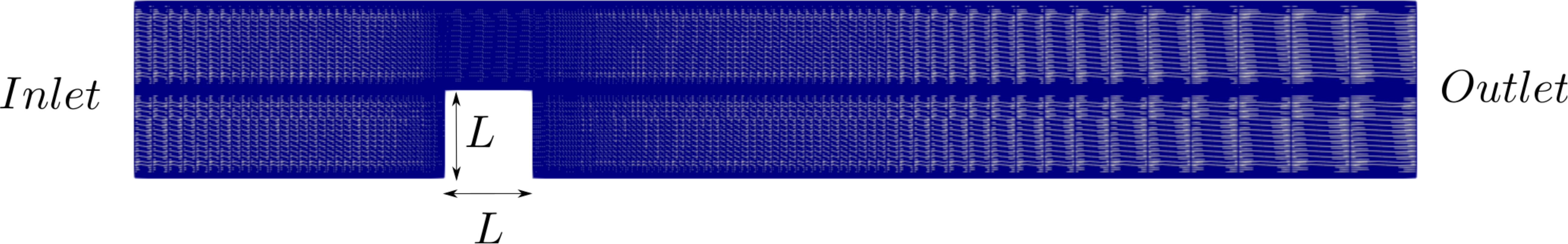} 
\caption{\label{fig:meshY} A cross-sectional view of the grid at $x_2 = 4.5L$} 
\end{subfigure}\hfill
\begin{subfigure}[b]{0.5\linewidth} 
\centering\includegraphics[width=.69\linewidth]{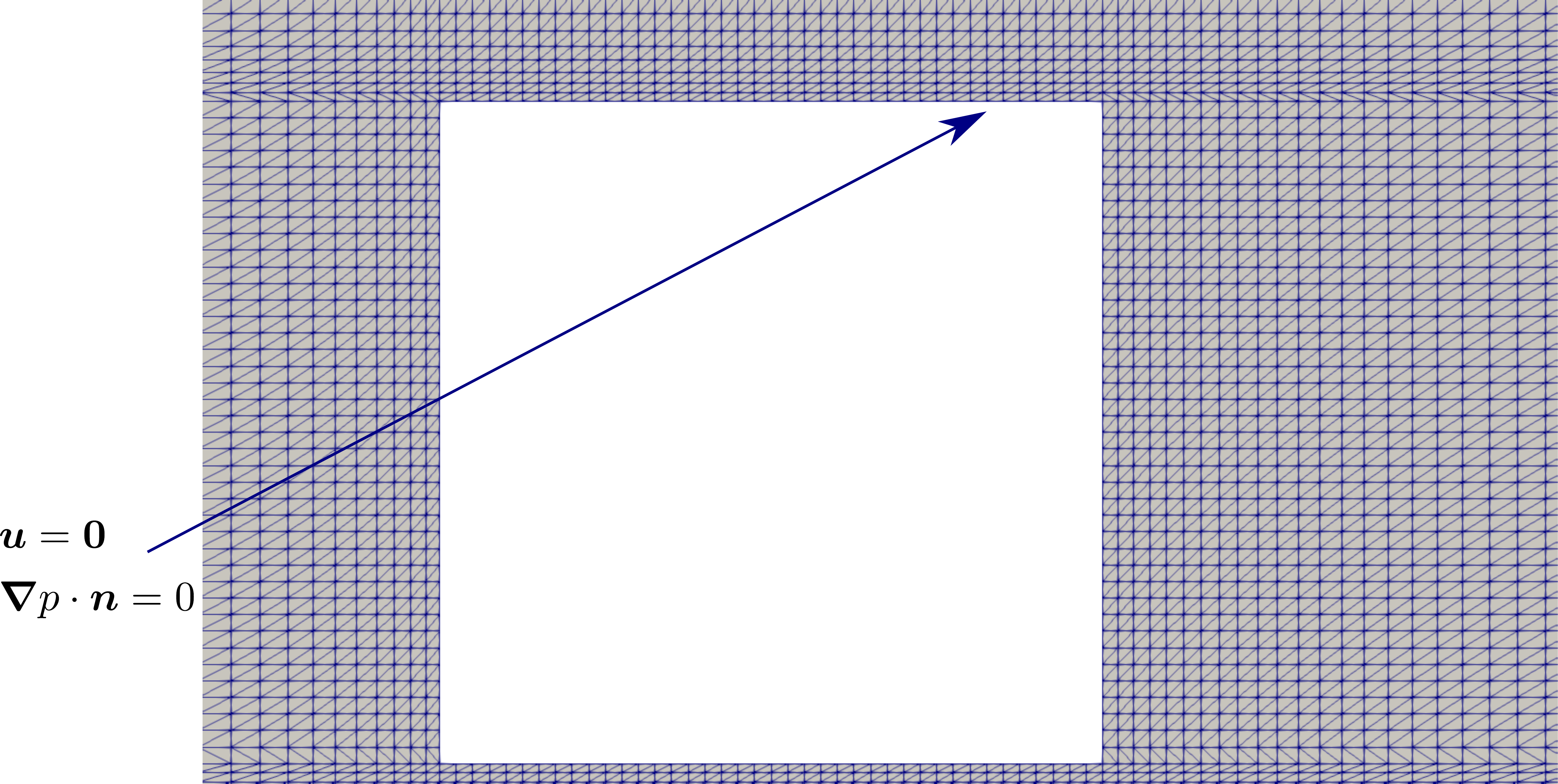} 
\caption{\label{fig:meshZ_zoom} A zoomed picture near the box with boundary conditions imposed on it} 
\end{subfigure}\vspace{10pt}

\caption{The OpenFOAM mesh used in the simulations for the case of the flow around a surface mounted cube.} 
\label{fig:comp_domain} 
\end{figure}

The case considered in this subsection is the one of a flow past a cubic obstacle. The problem is considered in a turbulent setting in three dimensions. Turbulence is modeled using the LES strategy, in particular, the SGS turbulence model used is the one-equation eddy viscosity model named "dynamicKEqn" in OpenFOAM. This model is proposed in \cite{Kim1995} as a continuation of the SGS model presented in \cite{Germano1991}. The fluid domain is $\Omega := [0,14.5L] \times [0,9L] \times [0,2L]$, where $L = 1 \si{m}$ is the length of the cube. The boundary of the domain $\Gamma$ is formed by the three parts which are the inlet $\Gamma_\textrm{inlet} = \{ 0 \} \times [0,9L] \times [0,2L]$ , the walls (the sides and the cube) $\Gamma_0$, and the outlet $\Gamma_\textrm{outlet} := \{14.5L\} \times [0,9L] \times [0,2L]$. The computational grid used in the simulations is depicted in \autoref{fig:comp_domain}, where one can see in image \ref{fig:meshZ} a cross-sectional view of the domain at $x_3 = L$. The cube is mounted on the ground surface with a distance of $3.5L$ from the inlet, where the velocity of the flow is horizontal with magnitude $U_{in}$. In \autoref{fig:meshY}, a similar image is shown for the cross-sectional view at $x_2 = 4.5L$ and a zoomed picture of the cube is viewed in \autoref{fig:meshZ_zoom}. The boundary conditions for the velocity and pressure at each part of the boundary are reported in the latter figures. The finite volume mesh features around $1.2$ millions cells. The physical viscosity $\nu$ is equal to $2.5 \times 10^{-5}$ \si{m^2 \per s}. The velocity at the inlet $U_{in}$ is $1$ \si{m\per s}. This gives a Reynolds number (based on the cube length) of $40000$.\par 

In this numerical test, the physical viscosity $\nu$ is assumed to be unknown, the goal is to identify its value with a high degree of accuracy and to approximate the non-dimensionalized forces coefficients coming from the lift and drag forces acting on the surface of the cubic obstacle.\par 

The fluid problem is simulated for a timespan which is long enough to observe stable values of the time-average of certain output quantities. These quantities include the mean and the Root Mean Squared (RMS) values of the non-dimensionalized forces coefficients coming from the lift and drag forces acting on the surface of the cubic obstacle (see \autoref{eq:forces} and \autoref{eq:forcesCoeffs}).\par 

\begin{figure}[htbp]
\centering
\begin{subfigure}[b]{0.49\textwidth} 
\includegraphics[width=0.98\linewidth]{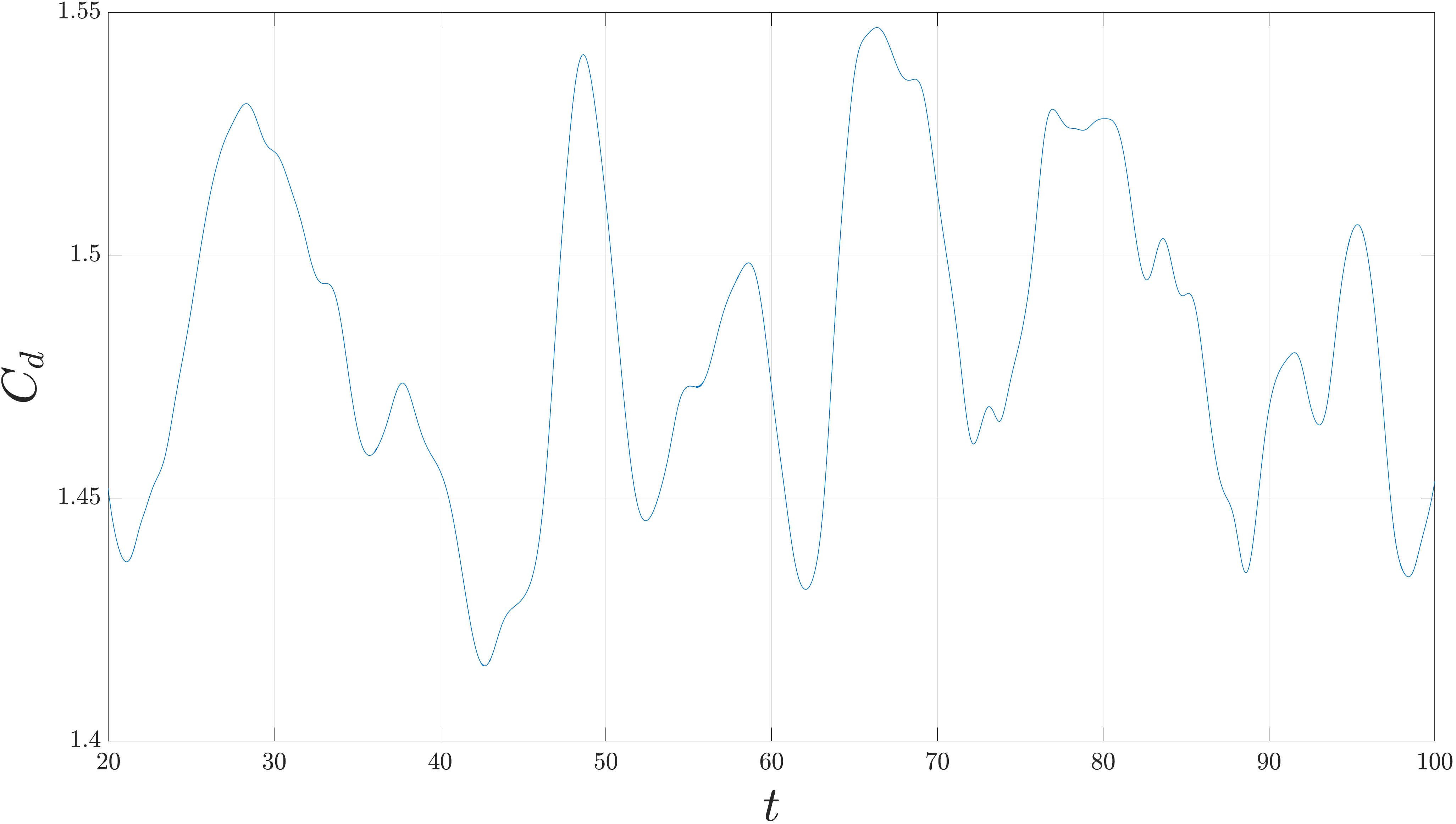} 
\caption{The time-history of the drag coefficient $C_d$ during the build-up phase in the time interval $[20,100]$.}
\label{fig:cdBuildUp} 
\end{subfigure}
\begin{subfigure}[b]{0.49\textwidth} 
\includegraphics[width=0.98\linewidth]{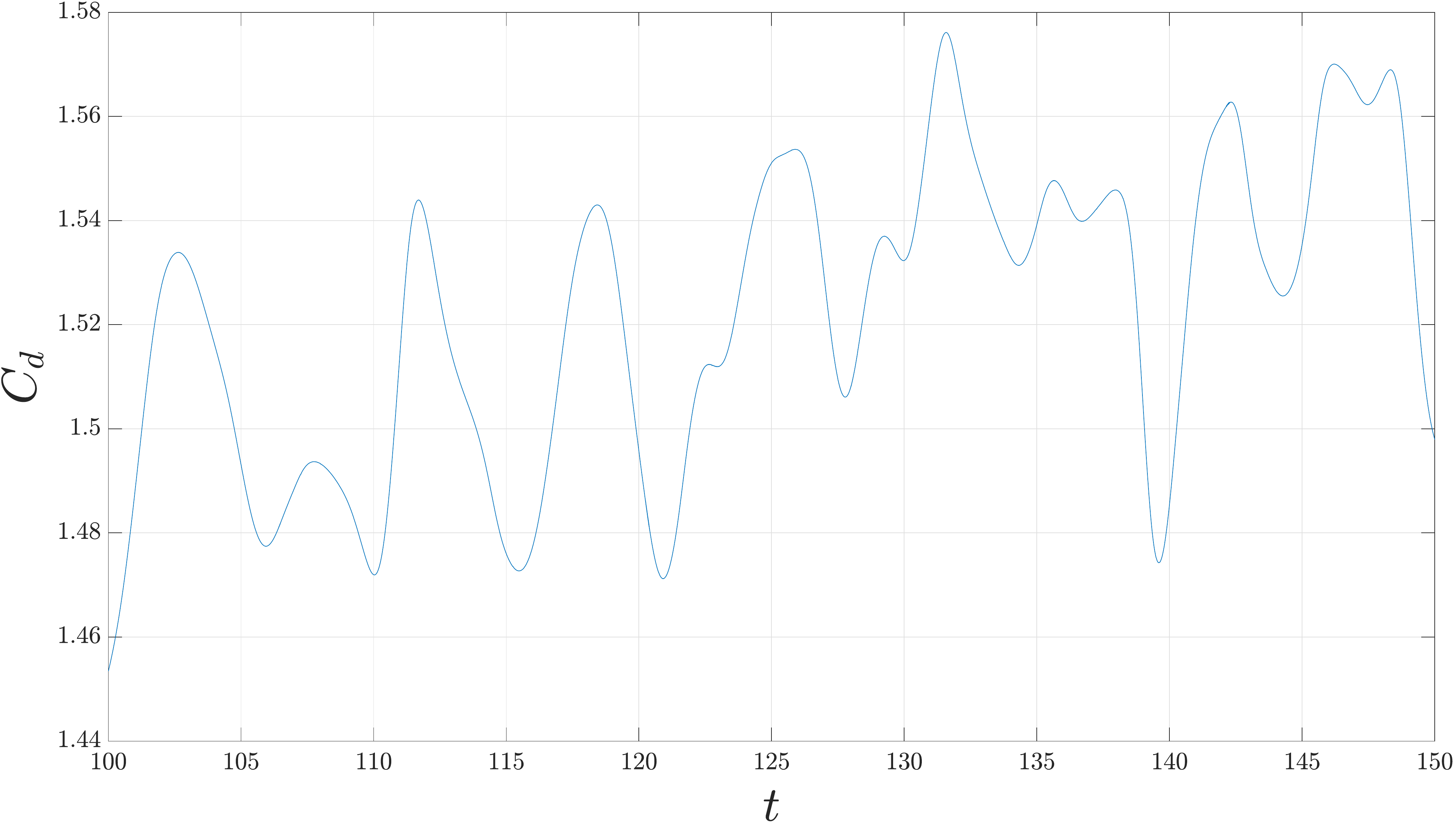}
\caption{The time-history of the drag coefficient $C_d$ in the snapshots time-window $[100,150]$.}
\label{fig:cdOff} 
\end{subfigure}
\caption{The drag coefficient of the cubic obstacle for the build-up phase and the offline snapshots time-window.}\label{fig:drag} 
\end{figure}

This problem has been studied for the same value of the Reynolds number mentioned above in \cite{Shah1997,Breuer1996,Krajnovic2002}. In the latter studies, RANS and LES simulation were carried of for the approximation of the values of the lift and drag coefficients of the cubic box. The nature of this problem is characterized by having chaotic turbulent response for the velocity and pressure field profiles. Thus, in order to have an accurate approximation of the mean drag and lift coefficients, at least $100$ non-dimensionalized time units $t^\star$ were simulated, where $t^\star = \frac{tU_{in}}{L}$. Hence, the NSE are simulated for $100$ \si{s}. The resulted graph for the drag coefficient time-history in the build-up phase is shown in \autoref{fig:cdBuildUp}. The mean value of the drag coefficient across the build-up phase is $1.4829$. The RMS values of the mean subtracted drag coefficients signal is $0.0648$.\par

After completing the build-up phase, snapshots are taken for the construction of the reduced order model, where the simulation is resumed for another $50$ \si{s}. Snapshots are acquired each $0.25$ \si{s} which results in a total of $201$ snapshots. The time-history of the drag coefficient during the offline snapshots time-window is depicted in \autoref{fig:cdOff}.\par

\begin{figure}[htbp]
\centering
\begin{subfigure}[b]{0.49\textwidth} 
\includegraphics[width=0.98\linewidth]{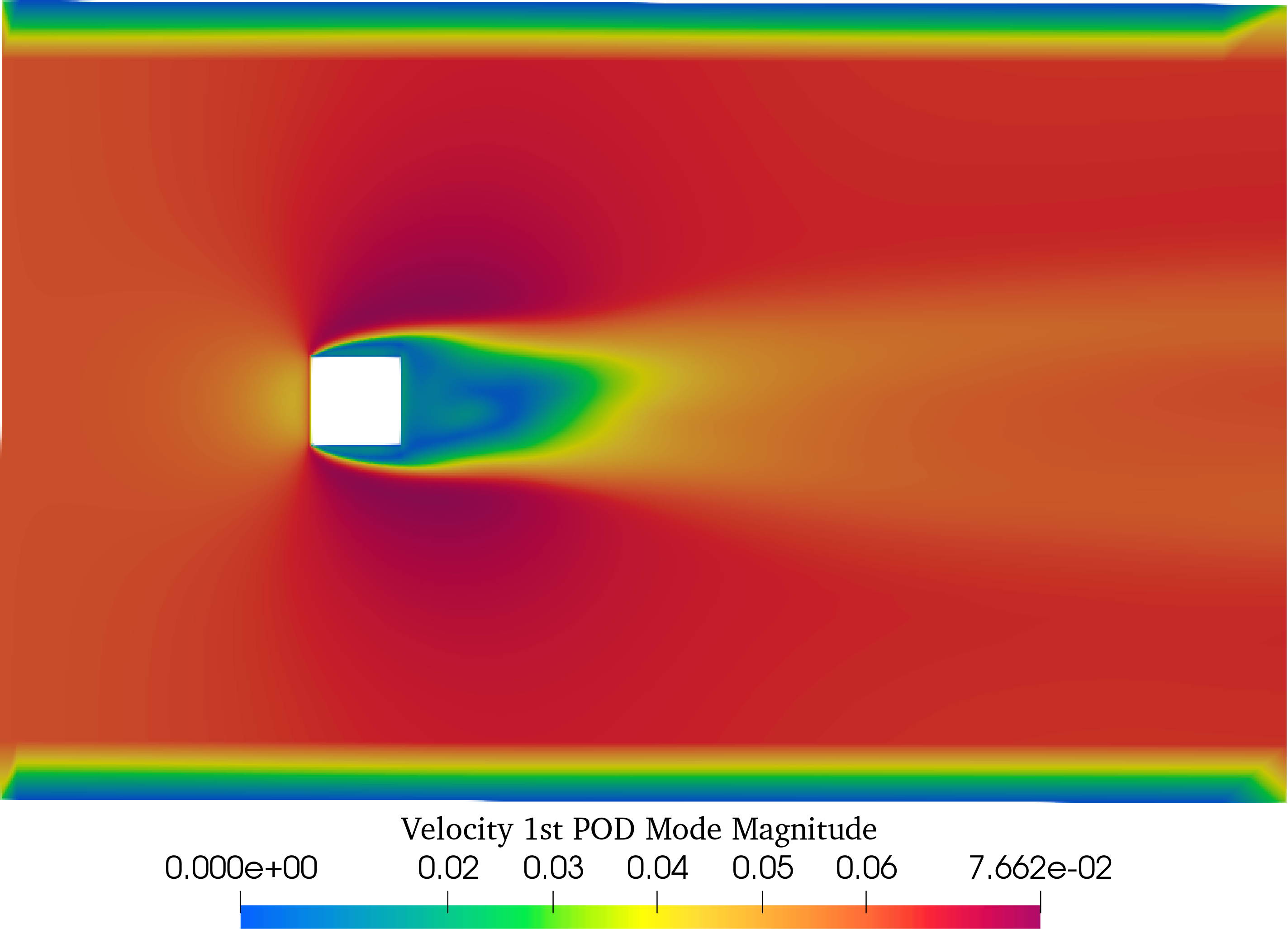} 
\caption{The first velocity POD mode.}
\label{fig:uPOD1} 
\end{subfigure}
\begin{subfigure}[b]{0.49\textwidth} 
\includegraphics[width=0.98\linewidth]{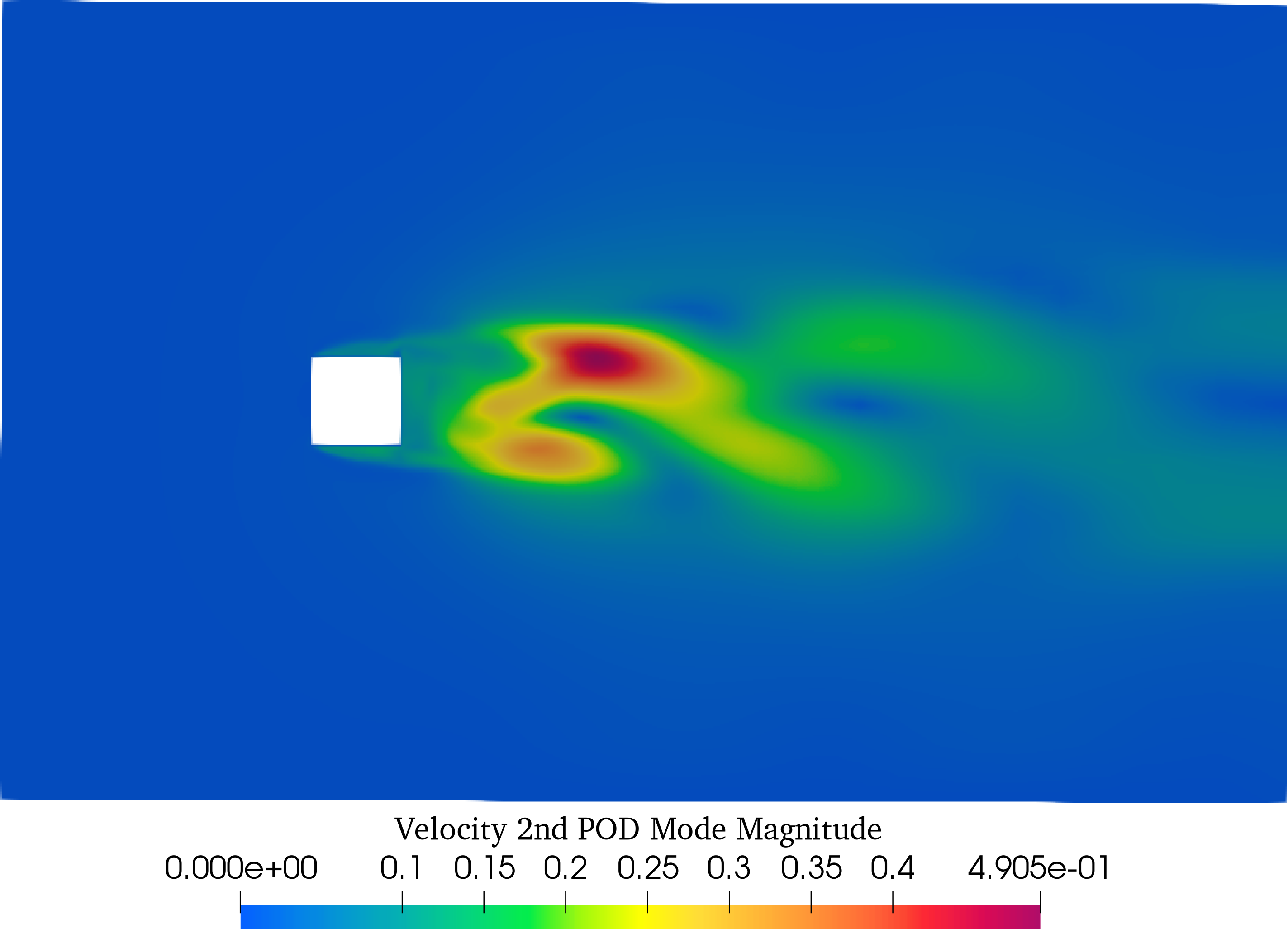}
\caption{The second velocity POD mode.}
\label{fig:uPOD2} 
\end{subfigure}
\begin{subfigure}[b]{0.49\textwidth} 
\centering
\includegraphics[width=0.98\linewidth]{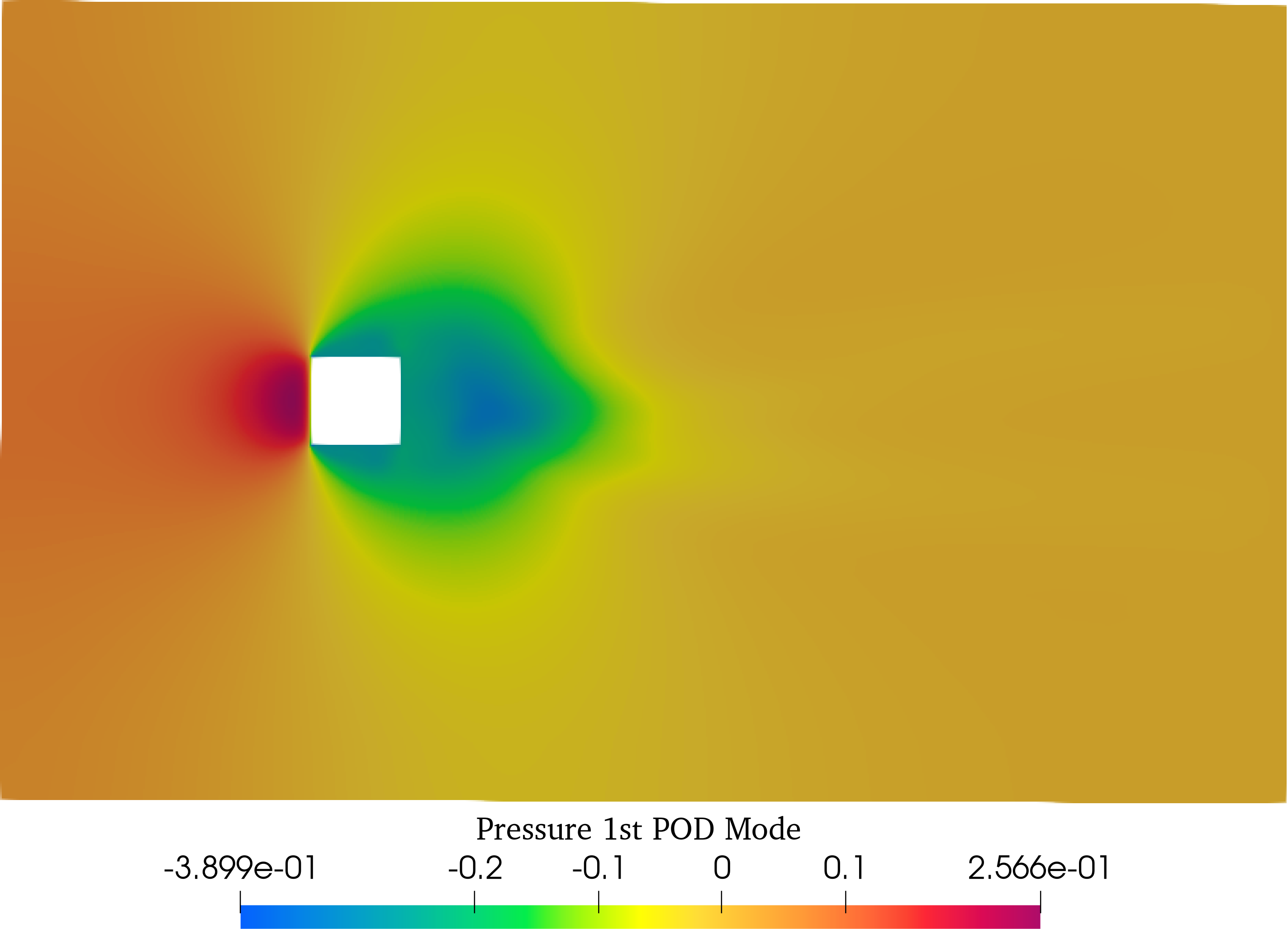} 
\caption{The first pressure POD mode.}
\label{fig:pPOD1}
\end{subfigure}
\begin{subfigure}[b]{0.49\textwidth} 
\centering
\includegraphics[width=0.98\linewidth]{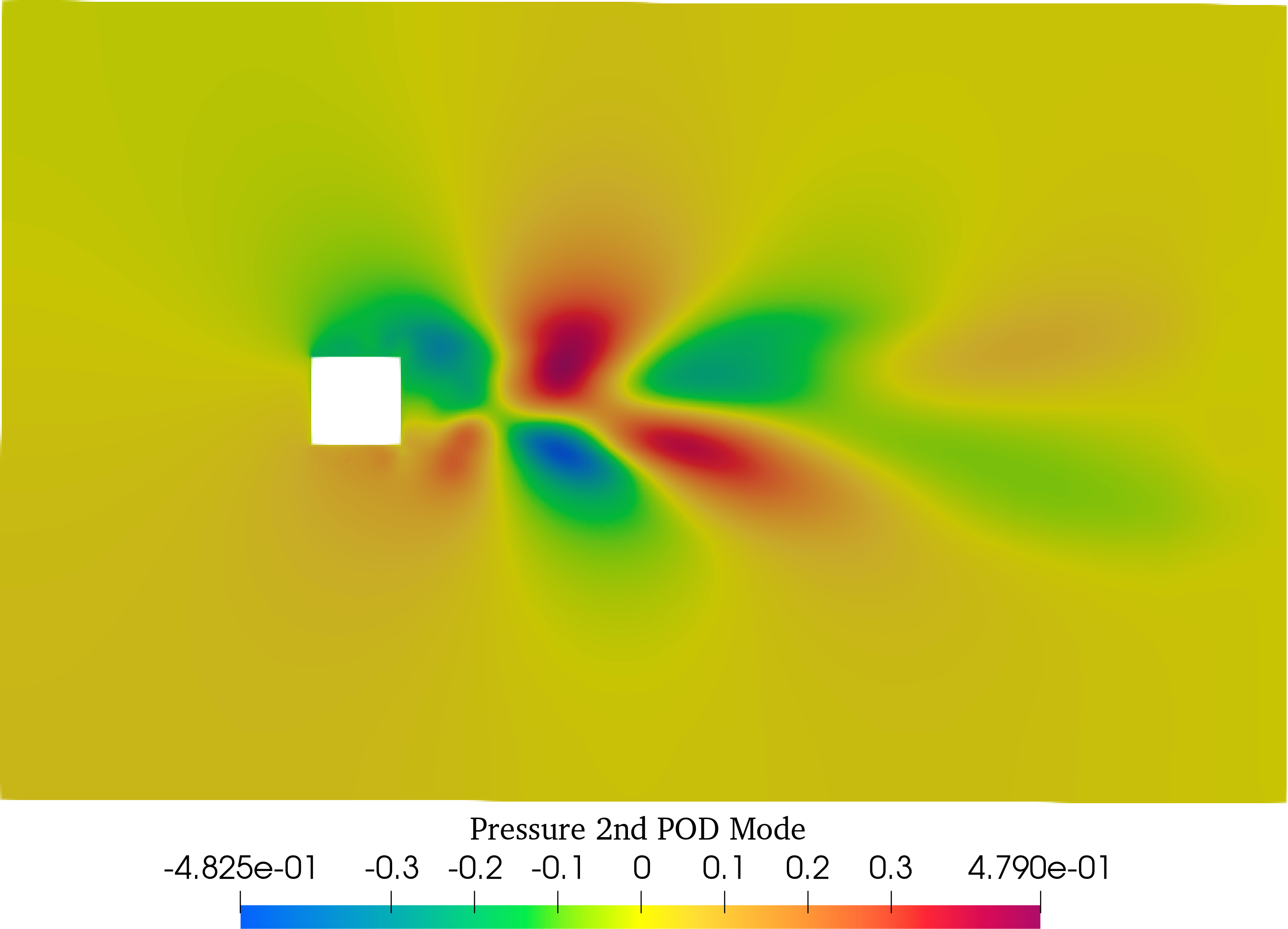} 
\caption{The second pressure POD mode.}
\label{fig:pPOD2}
\end{subfigure}\vspace{10pt}
\caption{The first two POD modes of the velocity and pressure, a $2D$ cross-sectional view of the domain at $x_3 = L$ is shown.}\label{fig:pod_1st_case} 
\end{figure}

The POD method is then applied on the snapshots matrices of velocity and pressure. \autoref{fig:pod_1st_case} shows the first two POD modes of the velocity and the pressure. After the computation of the POD modes of the velocity and pressure, one may solve the supremizer problems in order to obtain the supremizer modes which are then added to the original velocity POD basis.\par


At this point, the computation of the output data of the PINNs is carried out. The penalty method is used for the enforcement of the inhomogeneous Dirichlet boundary condition at the inlet, therefore, the POD-Galerkin DAE that models this problem is the one reported in \autoref{eq:penalty_DAE}. The matrices and vectors which appear in the latter DAE are computed during the offline stage.\par

In order to approximate the value of the unknown physical viscosity, we put forth deep neural networks which have time as the input and the stacked vector of $[\bm{a},\bm{b},\bm{h},\bm{c}]$ as their output. The neural networks used in this test have $7$ layers with each layer containing $200$ neurons. The activation function used at each neuron is the tangent hyperbolic function. The Adam optimizer is used for training the neural networks with a decaying learning rate of initial value of $1 \times 10^{-4}$, the optimization is run for $5 \times 10^5$ training epochs. The loss function which has to be minimized during the training procedure of the PINNs is the following:

\begin{equation}
E(\bm{w}) = E_\textrm{data}(\bm{w}) + E_1(\bm{w}) + E_2(\bm{w}) + E_3(\bm{w}),
\end{equation}
where $E_\textrm{data}(\bm{w})$, $E_1(\bm{w})$, $E_2(\bm{w})$ and $E_3(\bm{w})$ are defined as in \autoref{eq:dataLoss}, \autoref{eq:eq1Loss} , \autoref{eq:eq2Loss} and \autoref{eq:eq3Loss}, respectively. An additional trainable variable which corresponds to the physical viscosity and denoted by $\nu_\textrm{PINN}$ is added to the set of the PINNs tunable weights $\bm{w}$. This additional trainable parameter of the neural network is present in the loss function through $E_1(\bm{w})$. Therefore, the Adam optimizer will be able to compute the gradient of the total loss function $E(\bm{w})$ with respect to $\nu_\textrm{PINN}$ and as a result optimize its value. The initial value of $\nu_\textrm{PINN}$ is assumed to be $10^{-4}$. \par

\begin{figure}[htbp]
\centering
\begin{subfigure}[b]{0.48\textwidth} 
\includegraphics[width=0.98\linewidth]{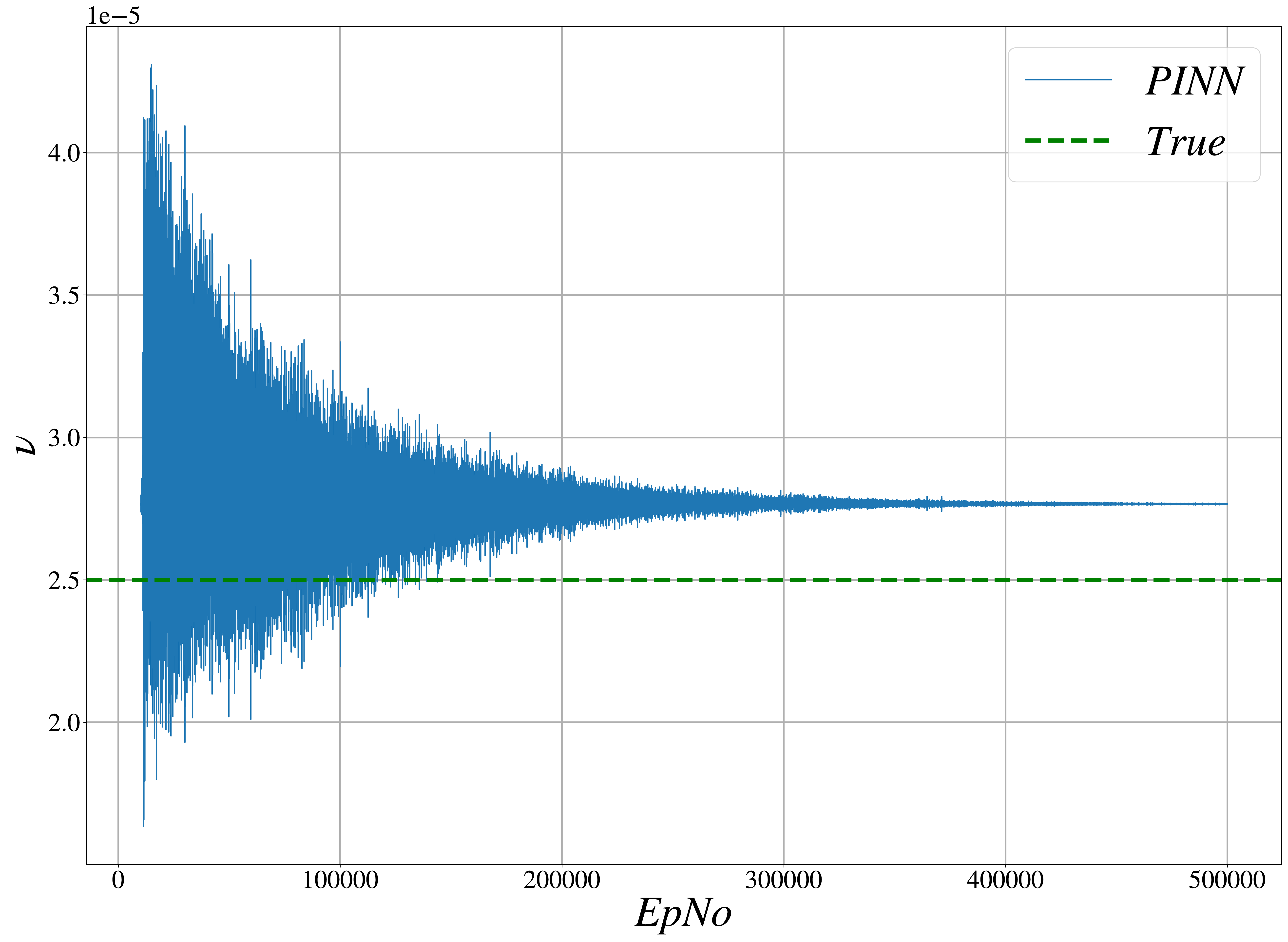} 
\caption{Reduced setting : $N_u = N_p = N_S = 10$}
\end{subfigure}
\begin{subfigure}[b]{0.48\textwidth} 
\includegraphics[width=0.98\linewidth]{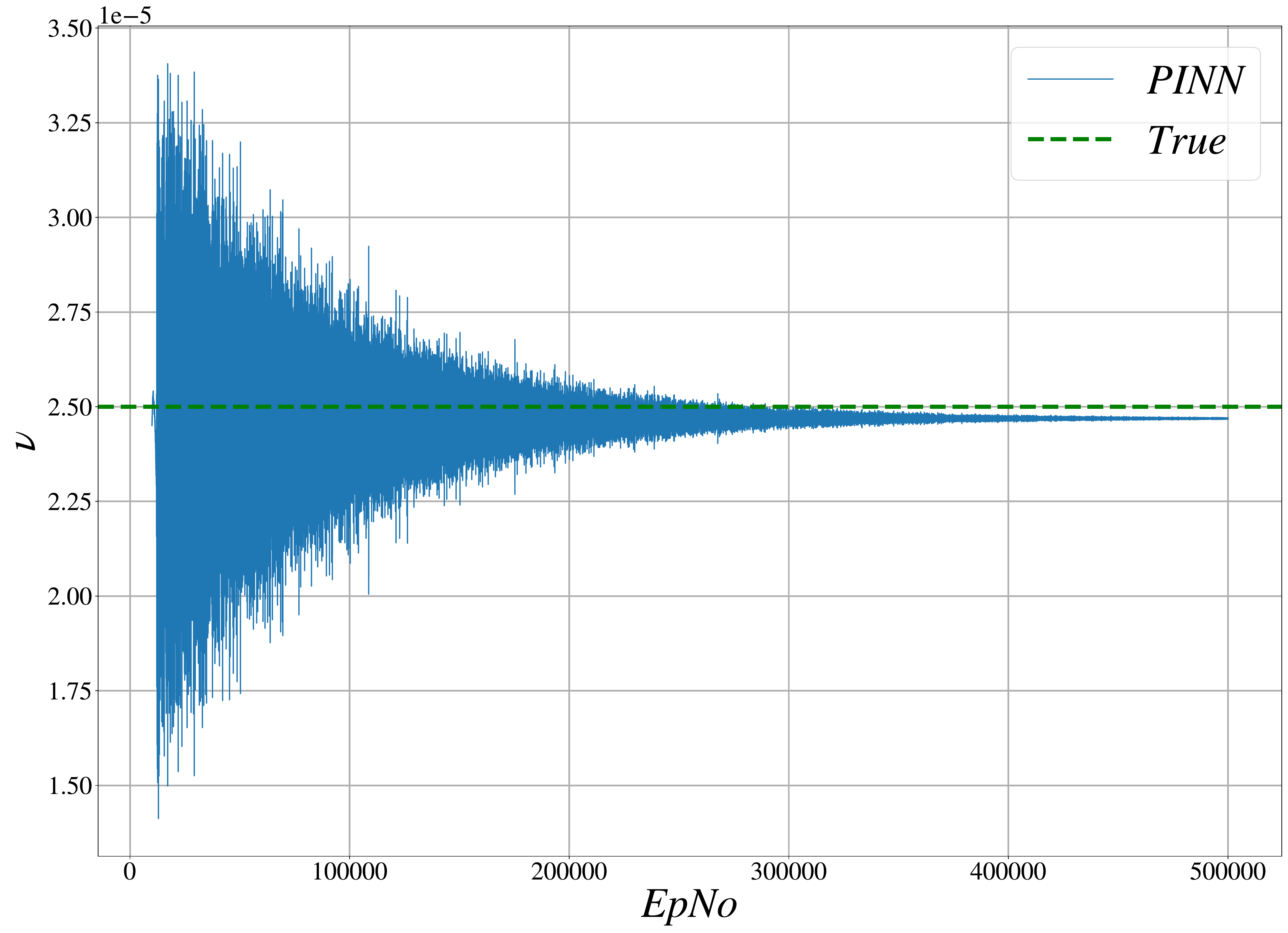}
\caption{Reduced setting : $N_u = N_p = N_S = 30$}
\end{subfigure}
\begin{subfigure}[b]{0.48\textwidth} 
\centering
\includegraphics[width=0.98\linewidth]{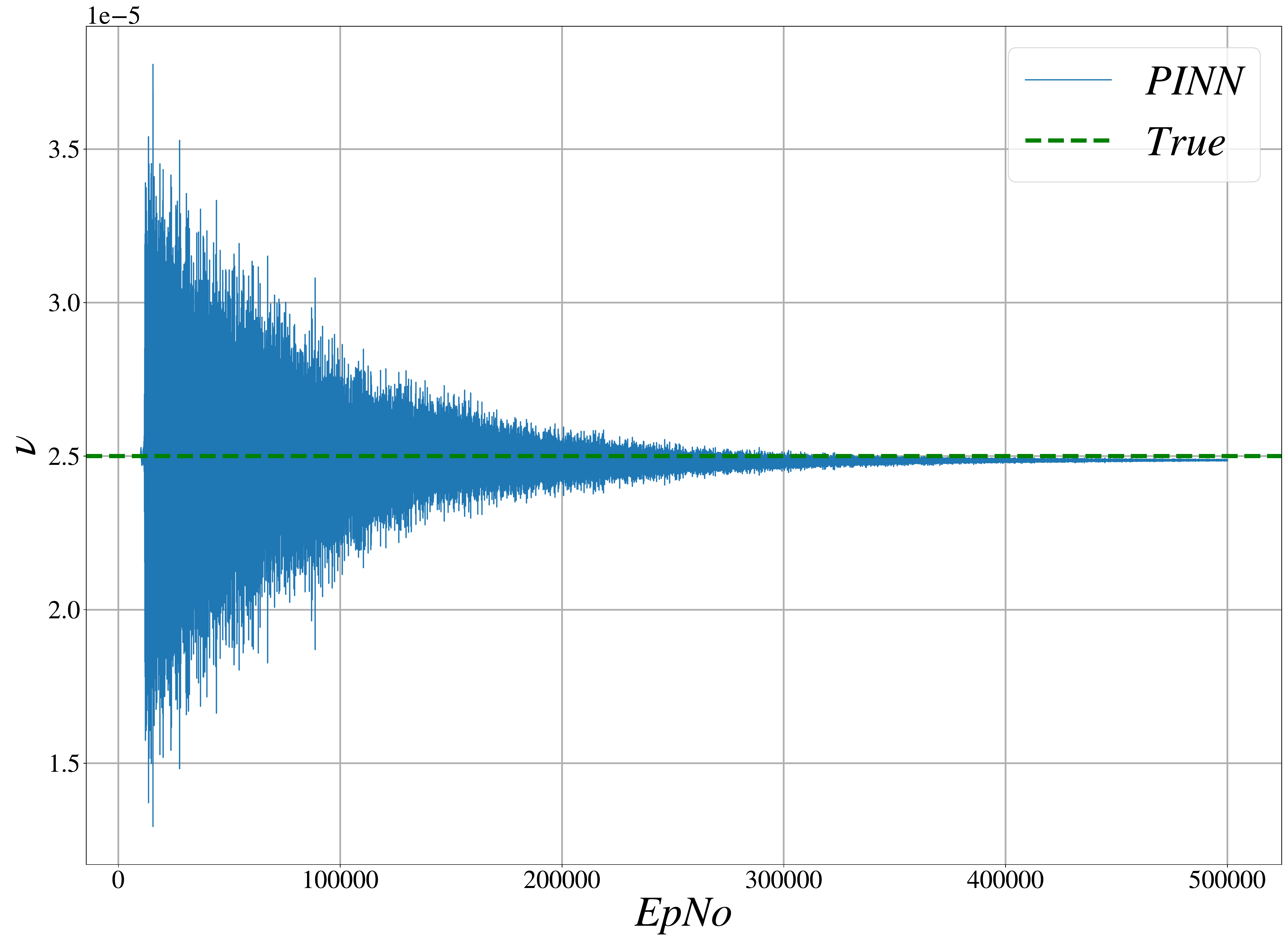} 
\caption{Reduced setting : $N_u = N_p = N_S = 60$}
\end{subfigure}
\caption{The approximation of the physical viscosity by the PINN at each training epoch for different reduced number of modes.}\label{fig:nuPINN_history} 
\end{figure}

The resulted approximation of the identified PINN physical viscosity is shown in \autoref{fig:nuPINN_history}. The last figure shows that PINN has obtained accurate approximation when $60$ modes were used in the construction of the ROM for each reduced variable, where the relative error in approximating $\nu$ is as low as $1$ $\%$.\par

\begin{figure}[htbp]
\centering
\begin{subfigure}[b]{1\textwidth} 
\includegraphics[width=0.98\linewidth]{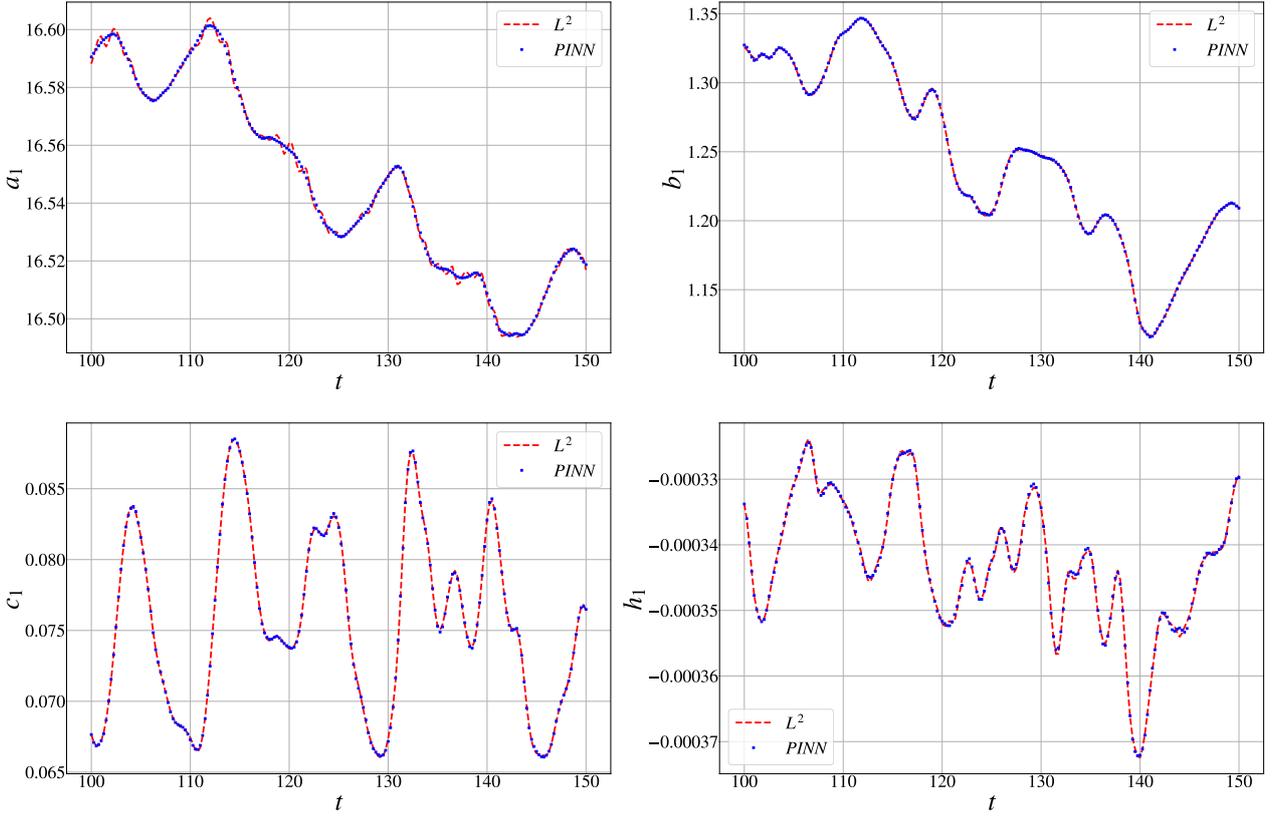} 
\caption{The first reduced coefficients for velocity, pressure, turbulent and convective terms compared to the ones obtained by the $L^2$ projection.}
\label{fig:LES_Box_a1b1} 
\end{subfigure}\vspace{10pt}
\begin{subfigure}[b]{1\textwidth} 
\centering
\includegraphics[width=0.98\linewidth]{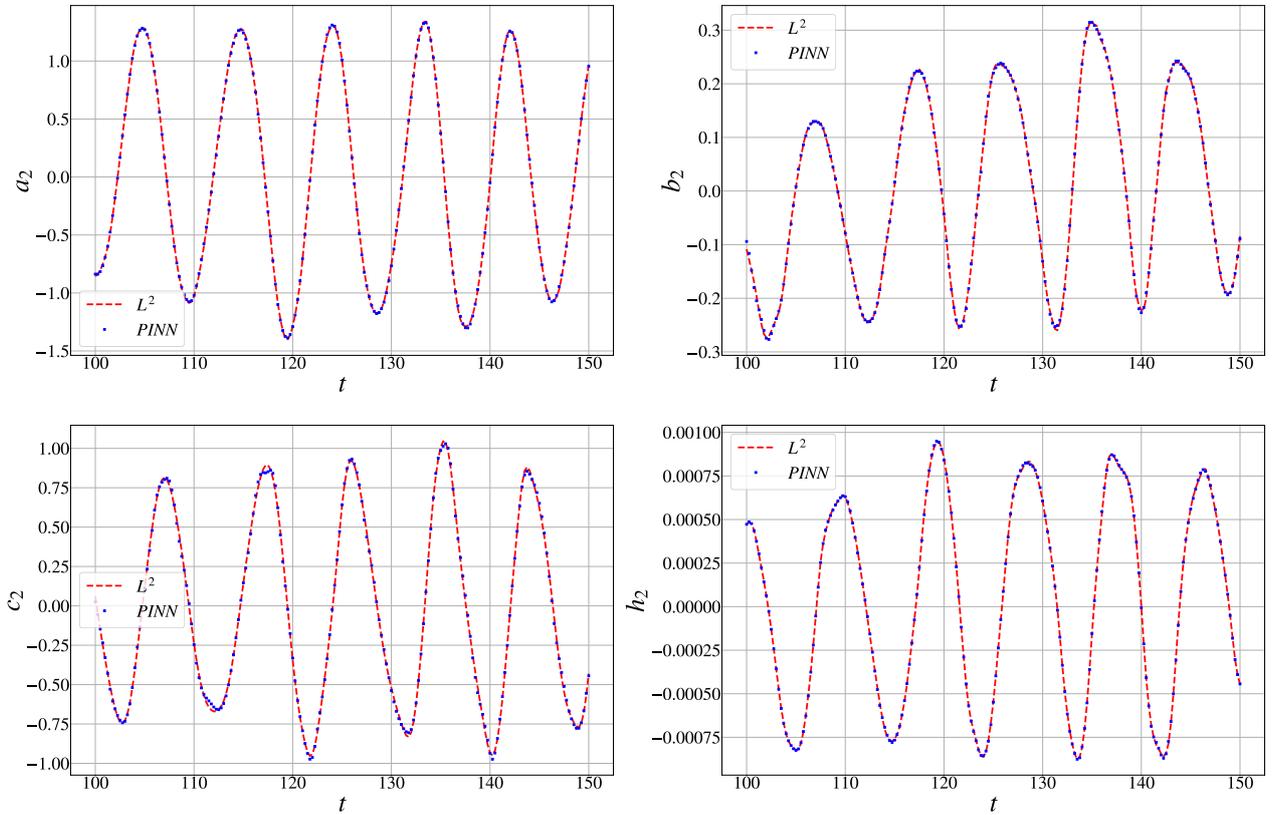} 
\caption{The second reduced coefficients for velocity, pressure, turbulent and convective terms compared to the ones obtained by the $L^2$ projection.}
\label{fig:LES_Box_a2b2}
\end{subfigure}\vspace{10pt}
\caption{The results of the PINNs predictions for all reduced variables, the reduced order spaces are constructed with $N_u=N_p=N_S=60$. The plots compare the reduced coefficients with the $L^2$ projection coefficients. The red-dashed lines refers to the $L^2$ projection coefficients, while the blue-dots correspond to the reduced coefficients obtained by the PINNs.}\label{fig:LES_Box_PINN} 
\end{figure}

The POD-Galerkin PINN ROM is also used for the approximation of the lift and drag coefficients of the cubic obstacle. The reduced forces coefficients are then compared to the FOM ones which are recorded during the full order simulation. In order to obtain the reduced forces, we perform a set of forward computations using the trained PINNs for the time values at which the FOM has recorded the forces. The number of time steps performed by the FOM solver during the acquisition of the offline snapshots is $28370$. The results of the PINNs forward computations are the reduced velocity and reduced pressure vectors at the aforementioned $28370$ time instants. We remark that these computations are carried out in a computational time which is significantly low, yielding high speed-up factors. The results of the forward application of the PINNs are shown in \autoref{fig:LES_Box_PINN}, where the first and second coefficients of each reduced variable are plotted, the figure depicts both the $L^2$ projection coefficients and the ROM coefficients. The last figure shows that the PINNs forward predictions are matching the original $L^2$ projection coefficients curves to a high degree despite the presence of uncertain parameter in the ROM formulation. The reduced velocity and pressure outputs will be used together with the reduced forces matrices which were computed during the offline stage to yield the reduced three dimensional forces acting on the surface of the box. Then, the reduced lift and drag coefficients are computed.\par

\begin{figure}[htbp]
\centering
\begin{subfigure}[b]{0.49\textwidth} 
\includegraphics[width=0.98\linewidth]{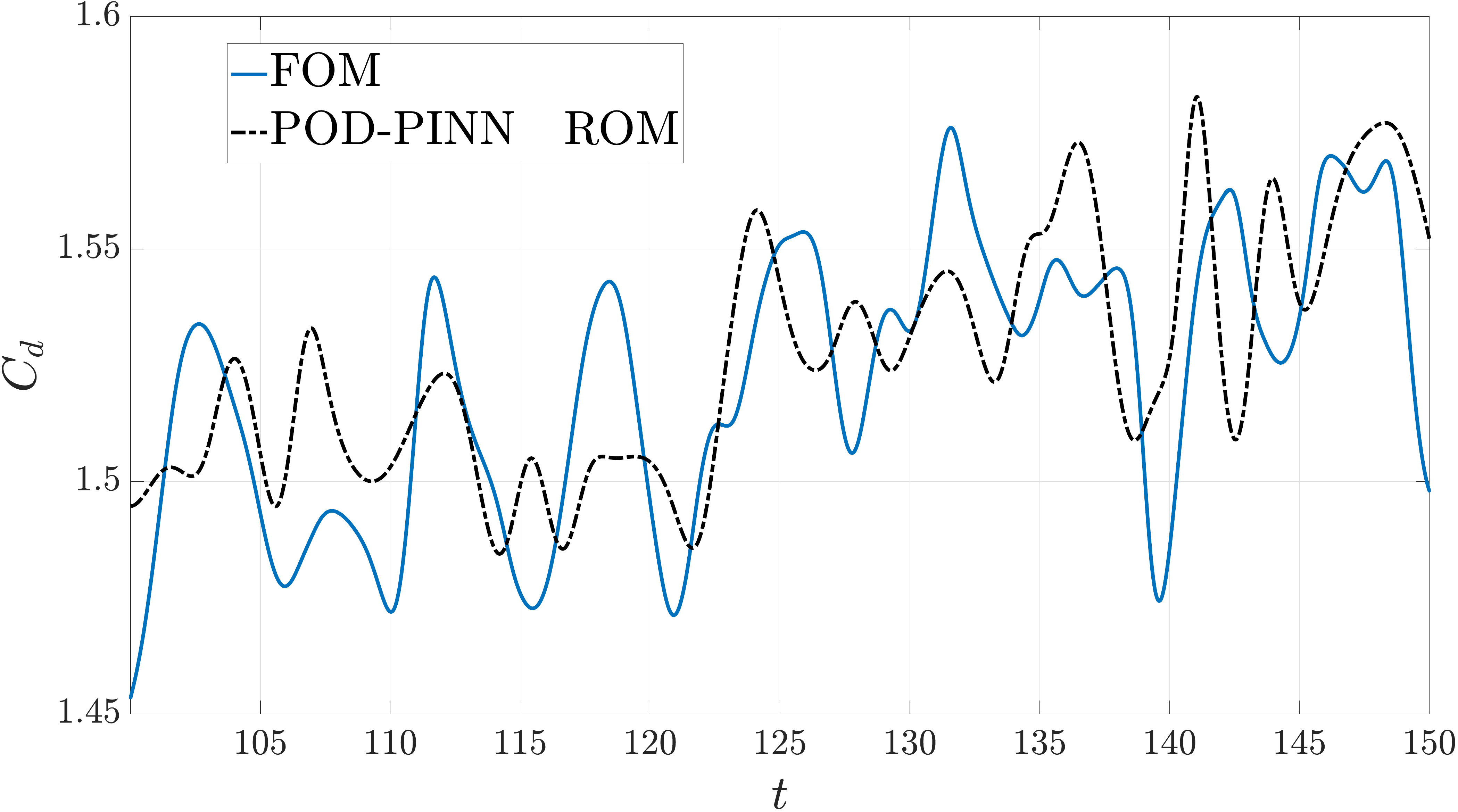} 
\caption{The drag coefficients curve reconstructed by the ROM using $N_u=N_p=N_S=10$ compared to the FOM one.}
\label{fig:Drag_10modes} 
\end{subfigure}
\begin{subfigure}[b]{0.49\textwidth} 
\includegraphics[width=0.98\linewidth]{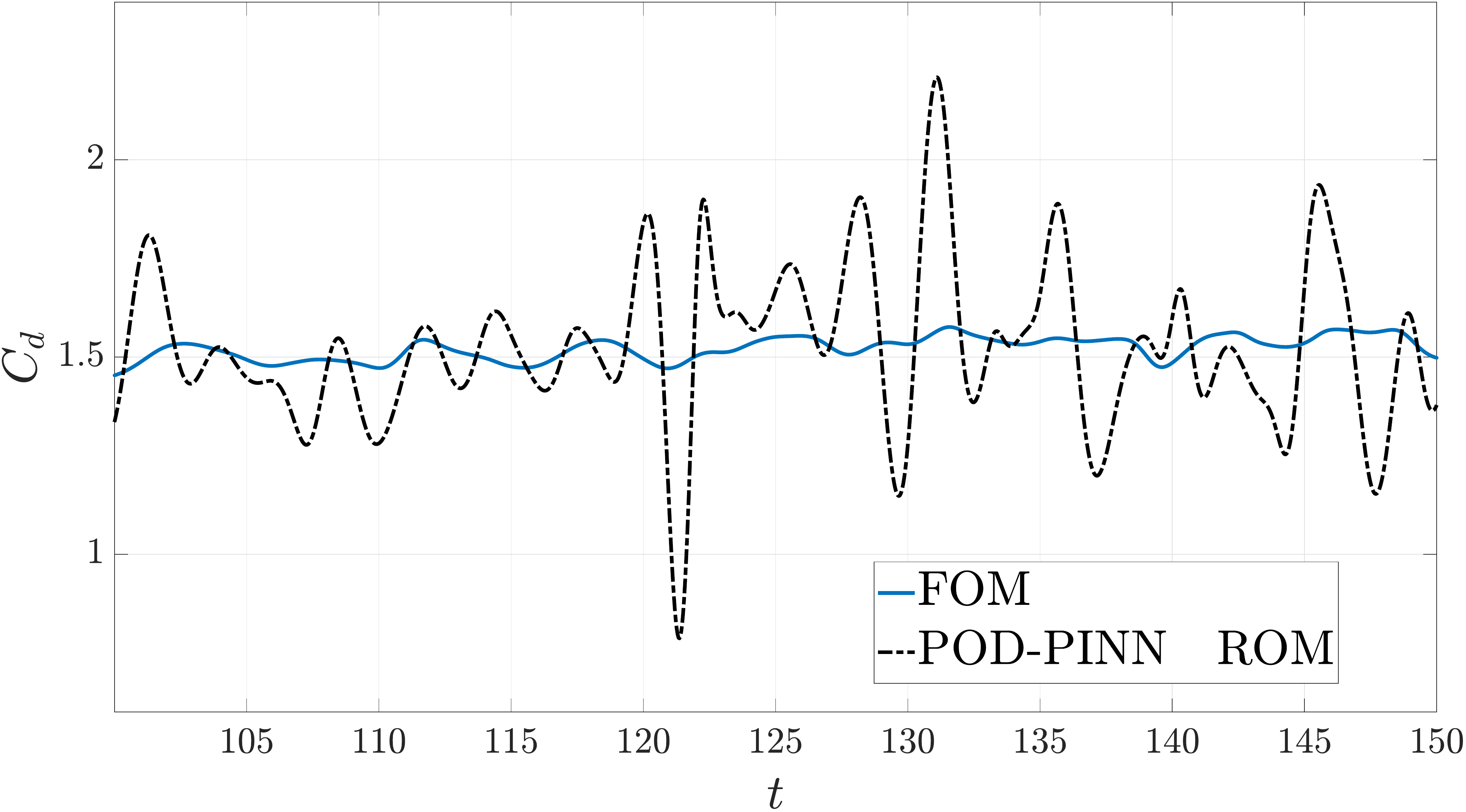}
\caption{The drag coefficients curve reconstructed by the ROM using $N_u=N_p=N_S=20$ compared to the FOM one.}
\label{fig:Drag_20modes} 
\end{subfigure}
\begin{subfigure}[b]{0.49\textwidth} 
\centering
\includegraphics[width=0.98\linewidth]{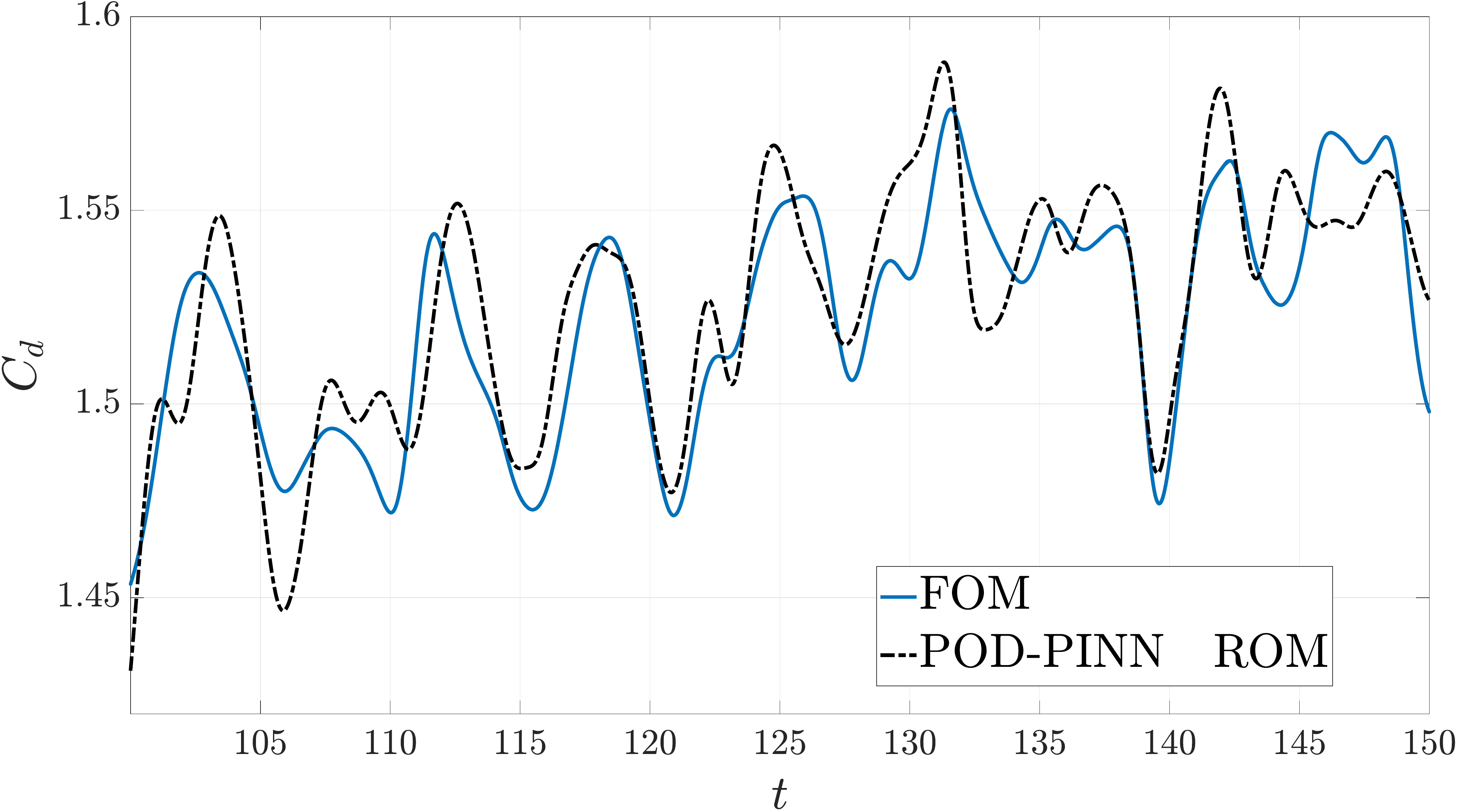} 
\caption{The drag coefficients curve reconstructed by the ROM using $N_u=N_p=N_S=30$ compared to the FOM one.}
\label{fig:Drag_30modes}
\end{subfigure}
\begin{subfigure}[b]{0.49\textwidth} 
\centering
\includegraphics[width=0.98\linewidth]{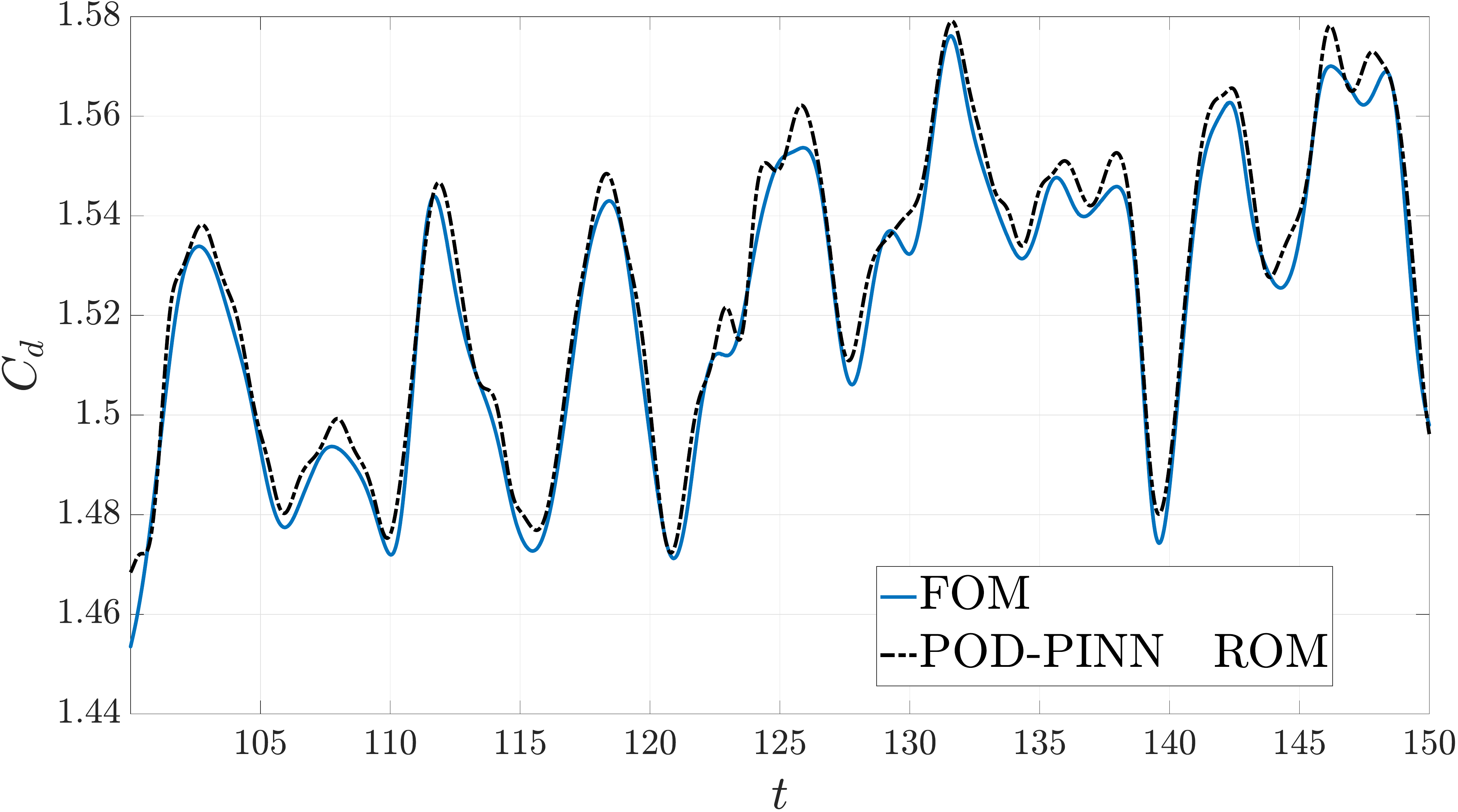} 
\caption{The drag coefficients curve reconstructed by the ROM using $N_u=N_p=N_S=60$ compared to the FOM one.}
\label{fig:Drag_60modes}
\end{subfigure}\vspace{10pt}
\caption{The drag coefficients curve over the time range $[100,150]$ \si{s}. The figures show the one obtained by the FOM solver and the ones approximated by the ROM for different values of and $N_u$, $N_p$ and $N_S$.}\label{fig:DragCoeff} 
\end{figure}

\begin{figure}[htbp]
\centering
\begin{subfigure}[b]{0.49\textwidth} 
\includegraphics[width=0.98\linewidth]{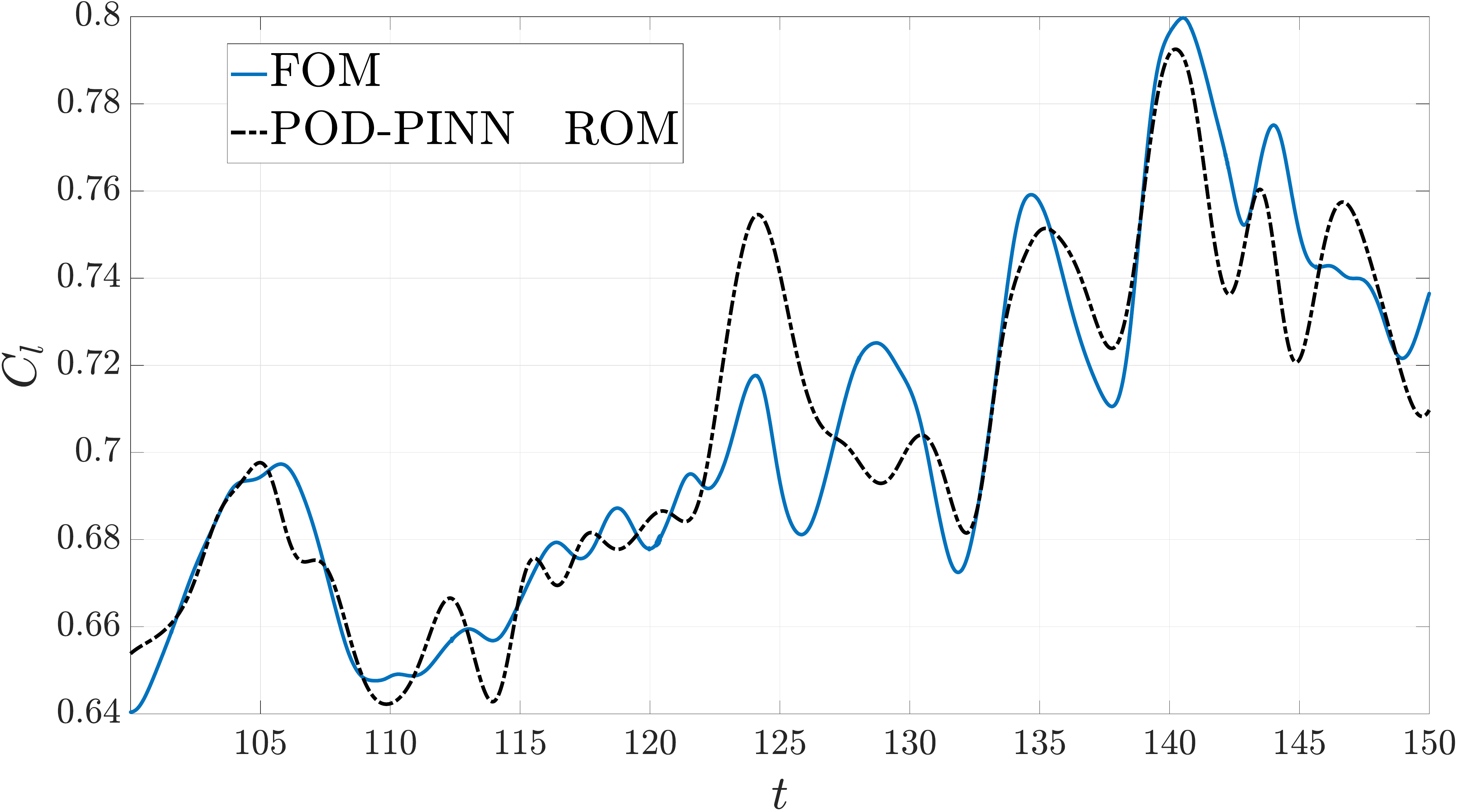} 
\caption{The Lift coefficients curve reconstructed by the ROM using $N_u=N_p=N_S=10$ compared to the FOM one.}
\label{fig:Lift_10modes} 
\end{subfigure}
\begin{subfigure}[b]{0.49\textwidth} 
\includegraphics[width=0.98\linewidth]{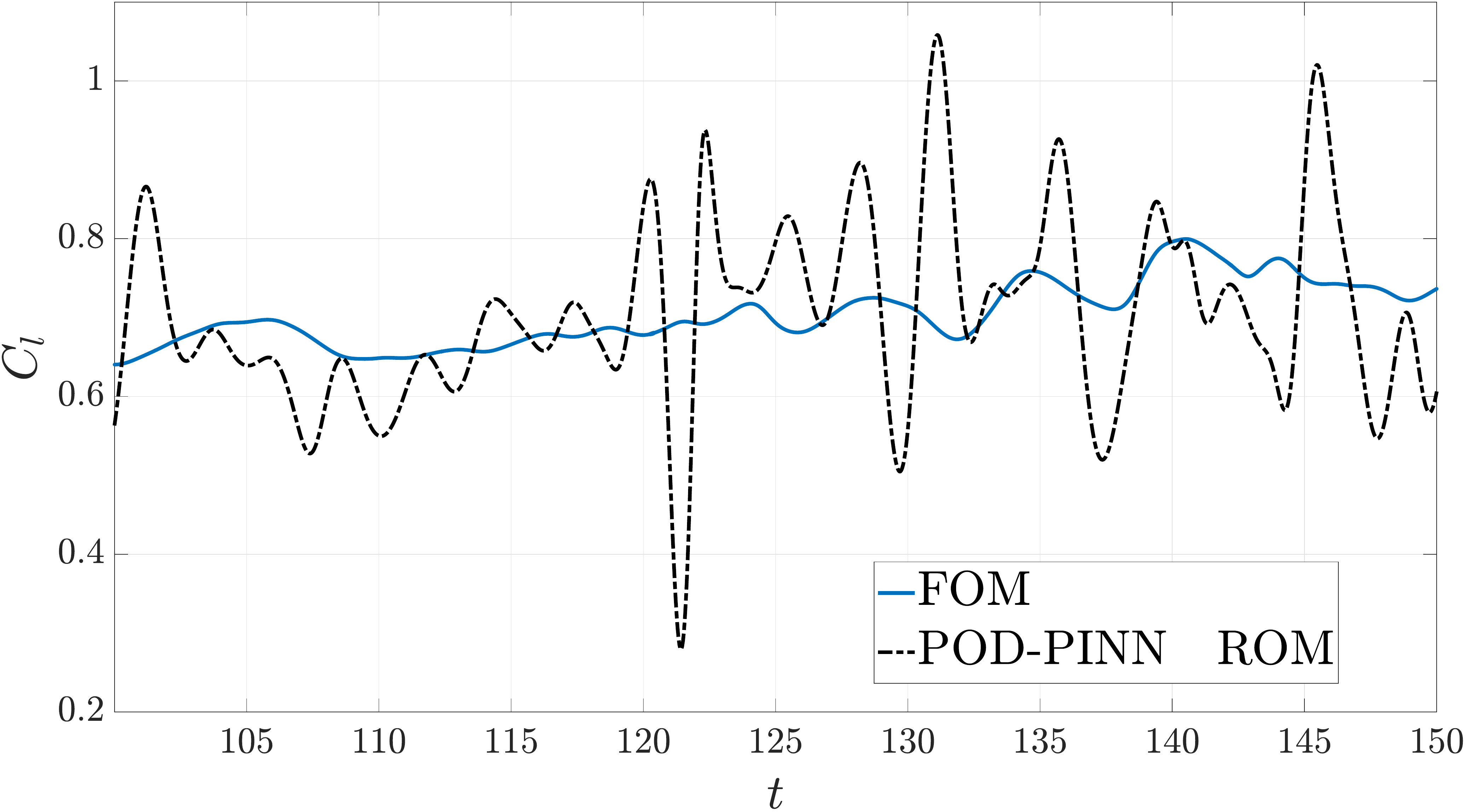}
\caption{The Lift coefficients curve reconstructed by the ROM using $N_u=N_p=N_S=20$ compared to the FOM one.}
\label{fig:Lift_20modes} 
\end{subfigure}
\begin{subfigure}[b]{0.49\textwidth} 
\centering
\includegraphics[width=0.98\linewidth]{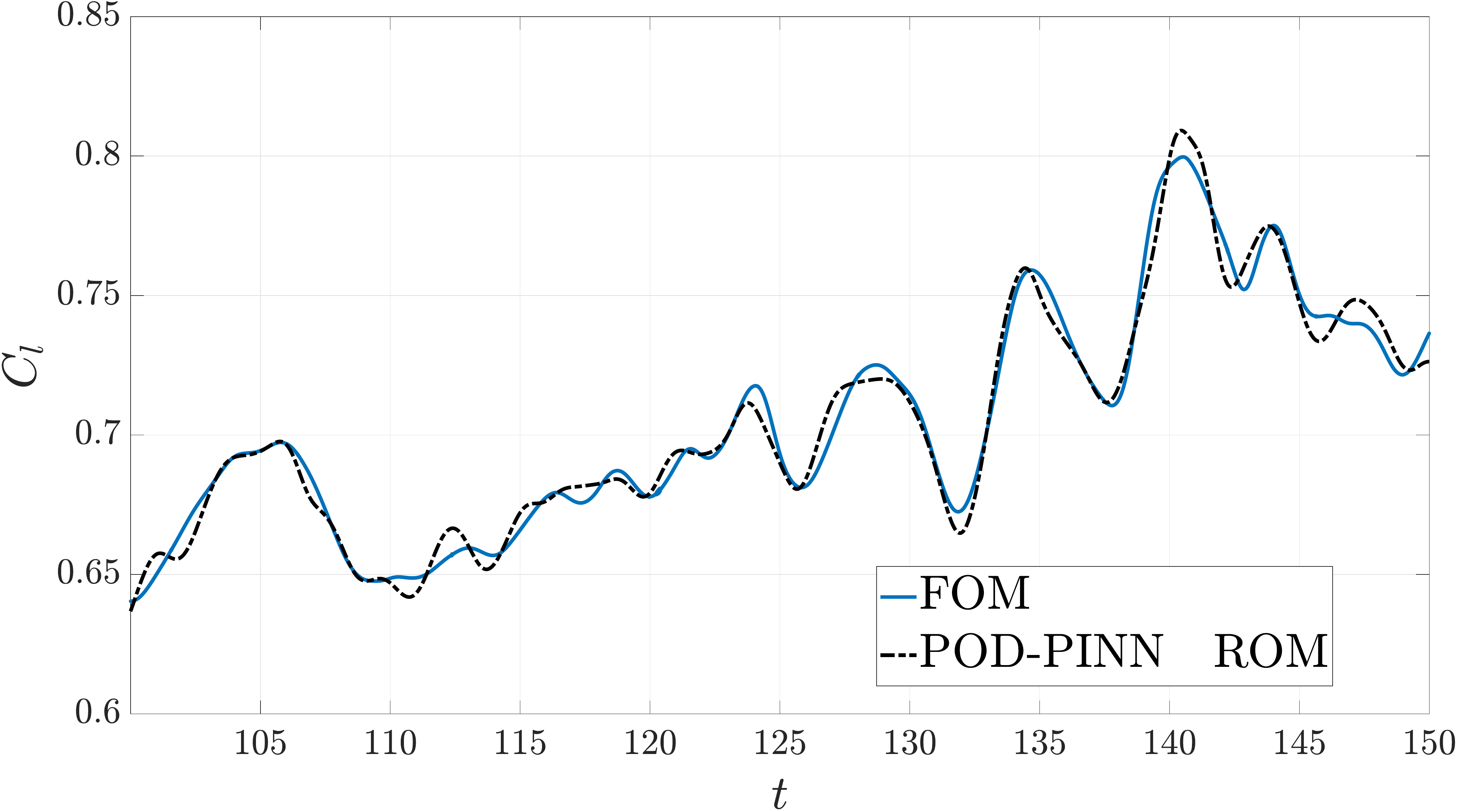} 
\caption{The Lift coefficients curve reconstructed by the ROM using $N_u=N_p=N_S=30$ compared to the FOM one.}
\label{fig:Lift_30modes}
\end{subfigure}
\begin{subfigure}[b]{0.49\textwidth} 
\centering
\includegraphics[width=0.98\linewidth]{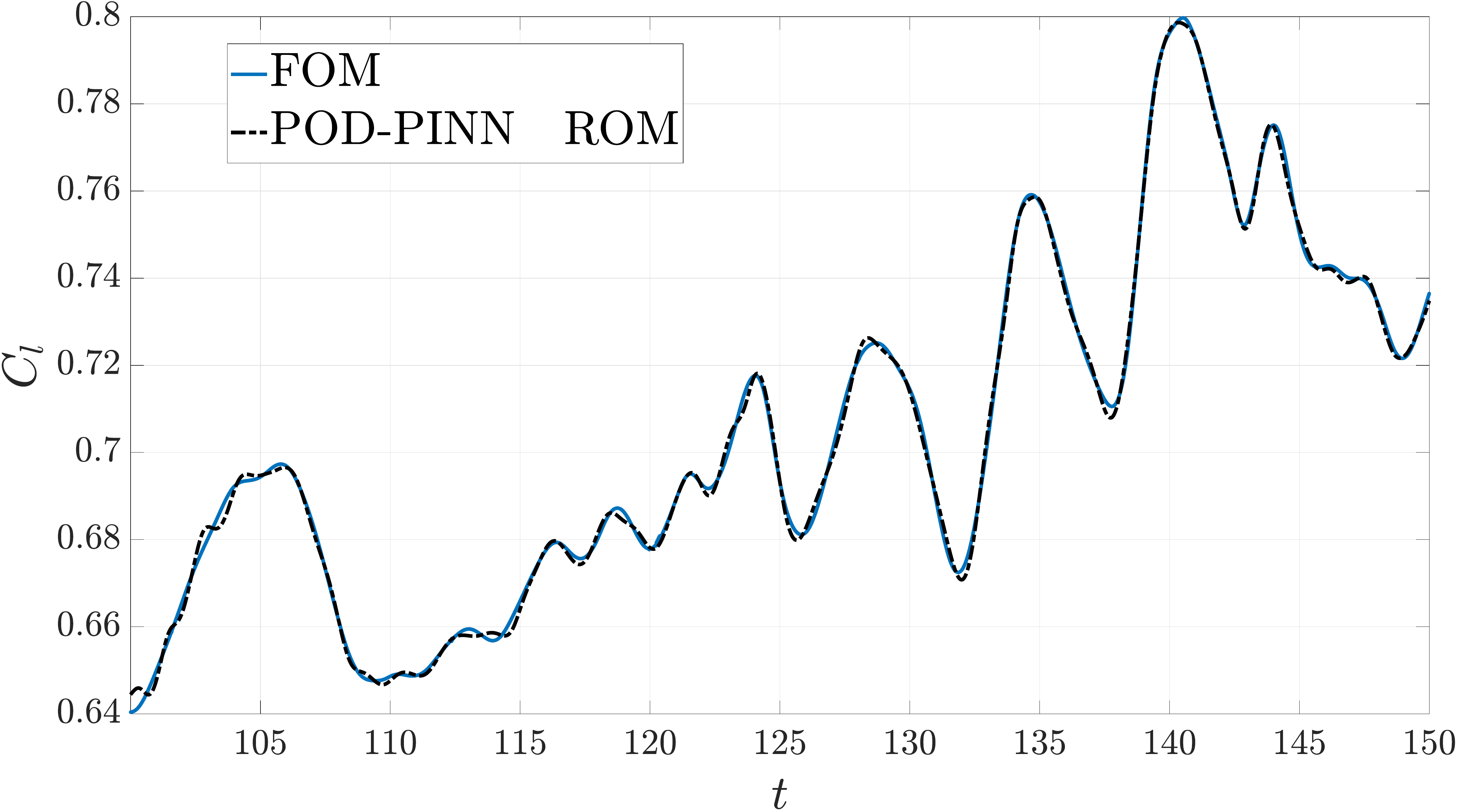} 
\caption{The Lift coefficients curve reconstructed by the ROM using $N_u=N_p=N_S=60$ compared to the FOM one.}
\label{fig:Lift_60modes}
\end{subfigure}\vspace{10pt}
\caption{The Lift coefficients curve over the time range $[100,150]$ \si{s}. The figures show the one obtained by the FOM solver and the ones approximated by the ROM for different values of and $N_u$, $N_p$ and $N_S$.}\label{fig:LiftCoeff} 
\end{figure}

The results of this test for different values of the reduced modes are shown in \autoref{fig:DragCoeff} and \autoref{fig:LiftCoeff}. The latter figures depict the time history of the FOM and the ROM forces coefficients for different reduced spaces dimensions. It is clear from the previous results that the ROM results are not matching their FOM counterparts when only $10-20$ modes for each of $N_u$, $N_p$ and $N_S$ are used for the ROM construction. However, \autoref{fig:Drag_60modes} and \autoref{fig:Lift_60modes} demonstrate that the level of qualitative reproduction of the FOM $C_d$ and $C_l$ curves by the surrogate model becomes substantially better when at least $60$ modes are used for each reduced variable.\par

\begin{figure}[htbp]
\centering
\begin{subfigure}[b]{0.49\textwidth} 
\includegraphics[width=0.98\linewidth]{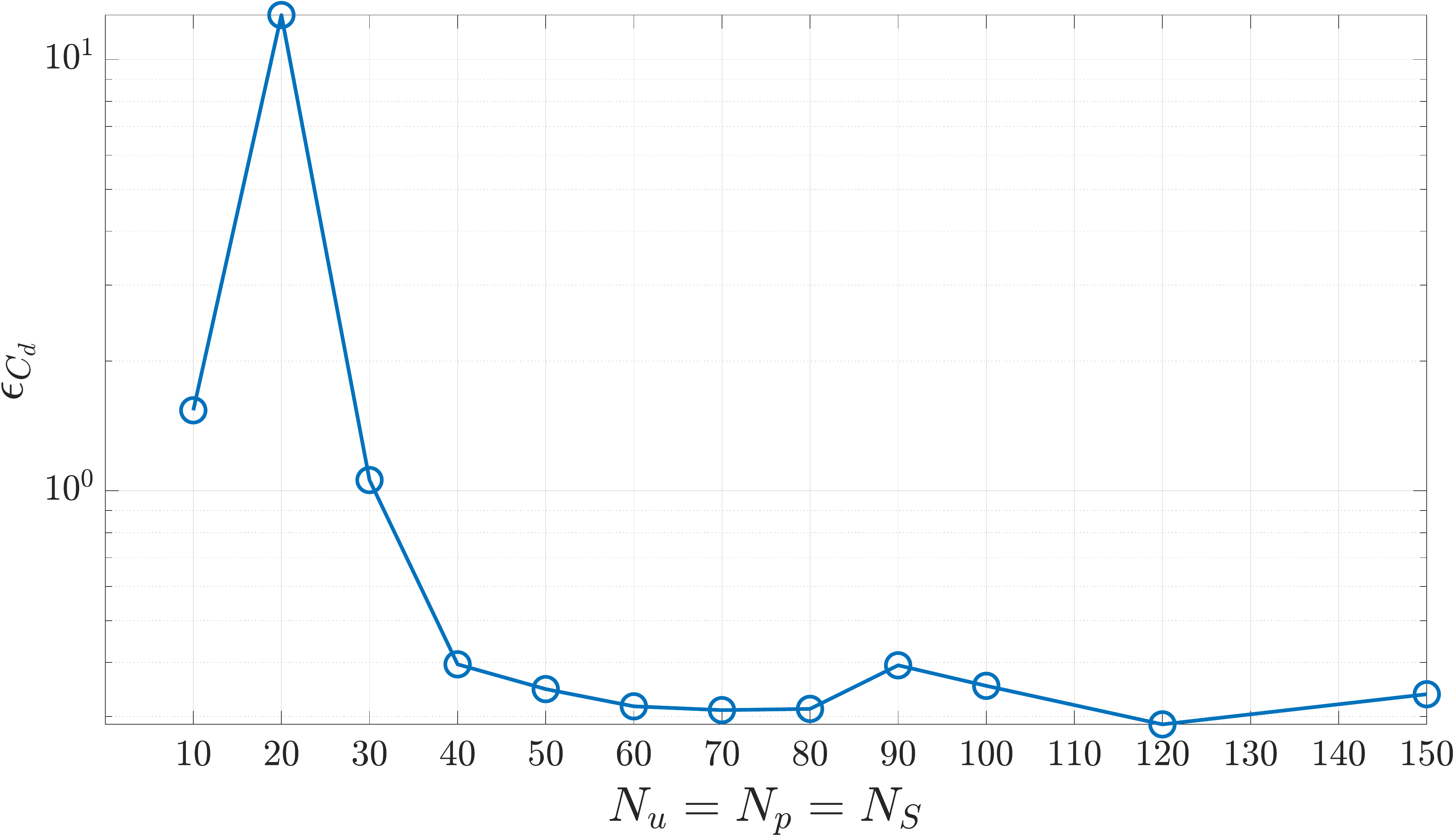} 
\caption{The $L^2$ relative error in approximating the drag coefficient $C_d$ over the time span $[100,150]$ as function of the number of modes.}
\label{fig:Lift_10modes} 
\end{subfigure}
\begin{subfigure}[b]{0.49\textwidth} 
\includegraphics[width=0.98\linewidth]{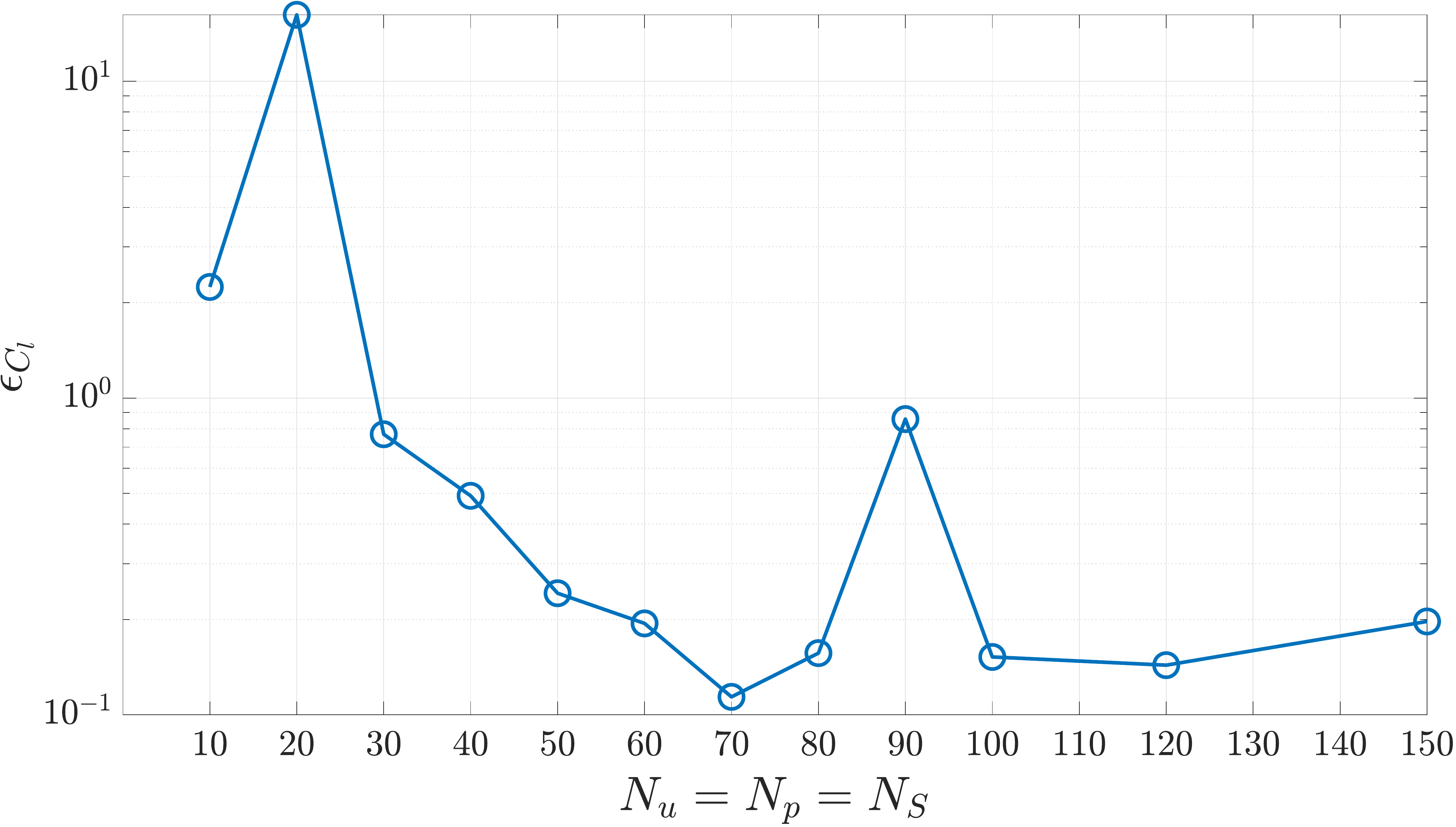}
\caption{The $L^2$ relative error in approximating the lift coefficient $C_l$ over the time span $[100,150]$ as function of the number of modes.}
\label{fig:Lift_20modes} 
\end{subfigure}
\vspace{10pt}
\caption{The $L^2$ relative errors for the approximation of the drag and lift coefficients in the time span $[100,150]$ as function of the number of modes. A base-$10$ logarithmic scale on the $y$-axis is used in both plots, the error values are already in percentages.}\label{fig:errForces} 
\end{figure}

In order to have a quantitative evaluation of the accuracy of the lift and drag coefficients approximation, we compute the $L^2$ relative errors (see \autoref{eq:l2error_cd_cl}). The errors are computed for different number of reduced modes, \autoref{fig:errForces} plots the error values versus the number of modes used for the ROM construction. The last plot shows that the error in approximating the drag coefficients reaches $0.3531$ $\%$ when $N_u=N_p=N_S=100$ modes, while the lift coefficients error is $0.1523$ $\%$ for the same setting.\par

We report a study of the computational time needed for the completion of each task or step involved in the POD-Galerkin PINN ROM. This study is conducted in \autoref{tab:pinn_gpuTime} for different numbers of the reduced modes. Firstly, we mention the computational time taken by the FOM solver for running the problem on $24$ CPUs (CPU model "AMD EPYC 7302 16-Core Processor @ 1498MHz") that is $T_\textrm{Off} = 11460$ \si{s} or approximately $3.1$ hours. In the second column of \autoref{tab:pinn_gpuTime}, we report $T_\textrm{proj,DAE}$ which is the time required for projecting the equations and computing the DAE reduced vectors and matrices. The third column lists the time taken for the computation of the PINNs outputs data, this time is denoted by $T_\textrm{proj,data}$. We remark that the computational cost corresponding to $T_\textrm{proj,DAE}$ is present also in the case of the intrusive POD-Galerkin ROM, while the cost that corresponds to $T_\textrm{proj,data}$ is only present in non-intrusive or hybrid ROMs such as the POD-Galerkin PINN ROM. The most significant cost is the one reported in the fourth column, where one can see the time taken for training the PINNs for $5 \times 10^5$ epochs on the Graphics Processing Unit (GPU). \autoref{tab:pinn_gpuTime} also details in the fifth column the cost for the forward computations carried out by the PINNs for the approximation of the forces denoted by $T_\textrm{PINN,F}$ (GPU cost). Finally the speed-up $\textrm{SU}$ is recorded in the last column, where this value is calculated as follows $\textrm{SU} = \frac{T_\textrm{Off}}{T_\textrm{PINN,F}+T_\textrm{Onl,Forces}}$, with $T_\textrm{Onl,Forces}$ being the time required for assembling the reduced forces in the online stage and whose value is about $7-10 * 10^{-5} $ \si{s}. As it can be seen from the latter study, the use of deep neural networks in reduced order modeling results in a substantial increase in the offline cost given by the time needed for the PINNs training in this setting $T_\textrm{PINNs,T}$. However, the nature of the inverse problem at hand requires this training procedure for the approximation of the unknown parameter. In addition, the POD-Galerkin PINN ROM compensates the high offline-cost by giving in return high speed-up ($\textrm{SU}$) value which reaches as high as $10^6$. These speed-up factors are not easily attainable in fully intrusive POD-Galerkin ROMs.

\begin{table}[htp]
\centering
{
\begin{tabular}{ c | c | c | c | c | c  }
$N_u=N_p=N_S$ & $T_\textrm{proj,DAE}$ in \si{s} & $T_\textrm{proj,data}$ in \si{s} & $T_\textrm{PINNs,T}$ in \si{\minute} & $T_\textrm{PINN,F}$ in \si{s} & $\textrm{Speedup}$ \\
\hline
$10$ & $8.269$    & $37.6084$ &  $58.01$ & $0.004612$   & $2.4438 * 10^6$ \\   
$20$ & $25.219$   & $59.5178$ &  $59.94$ & $0.004715$   & $2.3711 * 10^6$ \\   
$30$ & $55.0379$  & $77.1995$ & $61.78$ & $0.004778$    & $2.3557 * 10^6$ \\   
$40$ & $96.2729$  & $100.27687$ & $58.01$ & $0.004873$    & $2.2907 * 10^6$ \\   
$50$ & $145.838$  & $126.1021$ &  $71.7$ & $0.006929$  & $1.6261 * 10^6$  \\   						
$60$ & $207.629$  & $159.0223$ &  $70.63$  & $0.006481$ & $1.7393 * 10^6$  \\
$70$ & $289.609$  & $187.9712$ &  $59.91$  & $0.006036$ & $1.8584 * 10^6$  \\ 
$80$ & $387.812$  & $233.1214$ &  $72.06$ & $0.005258$  & $2.1340 * 10^6$  \\
$90$ & $462.827$  & $270.7364$ &  $65.49$ & $0.005462$  & $2.0480 * 10^6$  \\  
$100$ & $649.555$ & $322.3108$ &  $89.91$ & $0.005543$  & $2.0314 * 10^6$  \\ 
$120$ & $850.28$  & $428.9114$ &  $101.0$ & $0.005463$  & $2.0368 * 10^6$  \\ 
$150$ & $1289.12$ & $625.9247$ &  $136.8$ & $0.006089$  & $1.8480 * 10^6$  \\

\end{tabular}}\caption{{The computational time taken by different tasks needed for the implementation of the POD-Galerkin PINN ROM (i) the second column reports the time taken for performing the projection of the equations (computing the reduced quantities in the reduced DAE system) in parallel setting using $24$ processors, (ii) the third column details the CPU time (also using $24$ processors in parallel) consumed for the computation of the PINNs output data represented by the $L^2$ projection coefficients of the different variables. (iii) The fourth column shows the time required for running the PINNs on the Graphics Processing Unit (GPU), (iii) the fifth column indicates the time taken for carrying out the forward online computations of the PINNs for the approximation of the reduced variables. (iv) The last column reports the speedup achieved by the reduced order model. The speedup is calculated as follows $\textrm{Speedup} = \frac{T_\textrm{Off}}{T_\textrm{PINN,F}+T_\textrm{Onl,Forces}}$, where $T_\textrm{Off} =  11460$ \si{s} is the CPU time needed for simulating the FOM for the snapshots time window using $24$ processors in parallel, and $T_\textrm{Onl,Forces}$ is the time consumed for assembling the reduced forces in the online stage which has taken around $7-10 * 10^{-5} $ \si{s} in the current experiments. The table shows the computational costs for different sizes of reduced modes.}}
\label{tab:pinn_gpuTime}
\end{table}

To conclude, it is evident that the POD-Galerkin PINN ROM has provided accurate approximation for parameter estimation problems. At the same time, the presence of unknown parameter has not affected the quality of the approximation of important outputs such as the lift and drag forces acting on the surface of the box. However, the results for the study of the quality of the PINNs approximation for the forward and inverse cases illustrate that the quality of the inverse approximation is not so much damaged when only $10-20$ modes are used for each reduced variable. Unlike the inverse case, the forward approximation of the lift and drag coefficients needs substantially larger number of reduced modes.\par

\section{Conclusions} 
We have presented a reduced order model which is designed to learn unknown input parameters or physical quantities for fluid problems governed by the Navier--Stokes equations.

The proposed model employs the POD for the generation of the reduced order space and then utilizes Galerkin projection for the construction of the reduced order system. The solution of the reduced order system is obtained by the use of physics-informed neural networks (PINNs). The PINNs have as input time and/or parameters, while their output is the combined vector of the reduced velocity, pressure, turbulent and convective terms. The training procedure of the PINNs is carried out by minimizing a loss function which is a combination of the data loss and the reduced equations losses. Unknown physical parameters which appear in the POD-Galerkin DAE could be approximated using the PINNs by exploiting their feature of introducing additional trainable parameters. The PINNs optimizer was then used to compute the gradient of the loss function with respect to the additional trainable weight and consequently  optimize its value.\par 

The proposed POD-Galerkin PINN ROM has proven being accurate in solving inverse problems involving unknown physical quantities such as the physical viscosity. At the same time, this ROM is able to reconstruct fluid dynamics outputs with high degree of accuracy despite having input uncertainty. Three test cases have been used for the validation of the proposed model. The first case is the steady flow past a backward step, the second case is the one of a circular cylinder immersed in horizontal flow, while the last one is the unsteady flow past a surface mounted cube. The steady case was considered in laminar setting while the unsteady cases are turbulent ones with Reynolds number up to $Re=40000$. The POD-Galerkin PINN ROM has given very promising results when it comes to the approximation of the unknown parameters and also for the prediction of the fluid dynamics outputs, both for the non-turbulent and the turbulent case.\par

\section*{Acknowledgements} 
The research of S. Hijazi has been supported by a Jacobi Fellowship at the University of Potsdam. The research of M. Freitag is partially funded by the Deutsche Forschungsgemeinschaft (DFG)- Project-ID 318763901 - SFB1294.

\bibliographystyle{amsplain_mod}
\bibliography{bib/bibfile} 

\providecommand{\bysame}{\leavevmode\hbox to3em{\hrulefill}\thinspace}
\providecommand{\MR}{\relax\ifhmode\unskip\space\fi MR }
\providecommand{\MRhref}[2]{%
  \href{http://www.ams.org/mathscinet-getitem?mr=#1}{#2}
}
\providecommand{\href}[2]{#2}
\begin{thebibliography}{100}

\bibitem{TensorFlow}
M.~Abadi, P.~Barham, J.~Chen, Z.~Chen, A.~Davis, J.~Dean, M.~Devin,
  S.~Ghemawat, G.~Irving, M.~Isard, M.~Kudlur, J.~Levenberg, R.~Monga,
  S.~Moore, D.~Murray, B.~Steiner, P.~Tucker, V.~Vasudevan, P.~Warden, and
  X.~Zhang, \emph{Tensorflow: A system for large-scale machine learning},
  (2016).

\bibitem{Akhtar2009}
I.~Akhtar, A.~H. Nayfeh, and C.~J. Ribbens, \emph{{On the stability and
  extension of reduced-order Galerkin models in incompressible flows}},
  Theoretical and Computational Fluid Dynamics \textbf{23} (2009), no.~3,
  213--237.

\bibitem{Amsallem2012}
D.~Amsallem and C.~Farhat, \emph{Stabilization of projection-based
  reduced-order models}, International Journal for Numerical Methods in
  Engineering \textbf{91} (2012), no.~4, 358--377.

\bibitem{Babuka1973}
I.~Babu{\v{s}}ka, \emph{The finite element method with penalty}, Mathematics of
  Computation \textbf{27} (1973), no.~122, 221--221.

\bibitem{Bader2016}
E.~Bader, M.~K\"{a}rcher, M.~A. Grepl, and K.~Veroy, \emph{{Certified Reduced
  Basis Methods for Parametrized Distributed Elliptic Optimal Control Problems
  with Control Constraints}}, {SIAM} Journal on Scientific Computing
  \textbf{38} (2016), no.~6, A3921--A3946.

\bibitem{Baiges2014189}
J.~Baiges, R.~Codina, and S.~Idelsohn, \emph{{Reduced-order modelling
  strategies for the finite element approximation of the incompressible
  Navier-Stokes equations}}, Computational Methods in Applied Sciences
  \textbf{33} (2014), 189--216.

\bibitem{Balajewicz2012}
M.~Balajewicz and E.~H. Dowell, \emph{{Stabilization of projection-based
  reduced order models of the Navier{\textendash}Stokes}}, Nonlinear Dynamics
  \textbf{70} (2012), no.~2, 1619--1632.

\bibitem{Ballarin2014}
F.~Ballarin, A.~Manzoni, A.~Quarteroni, and G.~Rozza, \emph{{Supremizer
  stabilization of {POD}-Galerkin approximation of parametrized steady
  incompressible Navier-Stokes equations}}, International Journal for Numerical
  Methods in Engineering \textbf{102} (2014), no.~5, 1136--1161.

\bibitem{Ballarin2016}
F.~Ballarin and G.~Rozza, \emph{{{POD}-Galerkin monolithic reduced order models
  for parametrized fluid-structure interaction problems}}, International
  Journal for Numerical Methods in Fluids \textbf{82} (2016), no.~12,
  1010--1034.

\bibitem{Banks1989}
H.~T. Banks and K.~Kunisch, \emph{Estimation techniques for distributed
  parameter systems}, Birkh\"{a}user Boston, 1989.

\bibitem{Barrault2004}
M.~Barrault, Y.~Maday, N.~C. Nguyen, and A.~T. Patera, \emph{An `empirical
  interpolation' method: application to efficient reduced-basis discretization
  of partial differential equations}, Comptes Rendus Mathematique \textbf{339}
  (2004), no.~9, 667--672.

\bibitem{Barrett1986}
J.~W. Barrett and C.~M. Elliott, \emph{Finite element approximation of the
  dirichlet problem using the boundary penalty method}, Numerische Mathematik
  \textbf{49} (1986), no.~4, 343--366.

\bibitem{Baydin2017}
A.~G. Baydin, B.~A. Pearlmutter, A.~A. Radul, and J.~M. Siskind,
  \emph{Automatic differentiation in machine learning: A survey}, J. Mach.
  Learn. Res. \textbf{18} (2017), no.~1, 5595--5637.

\bibitem{Benner2015}
P.~Benner, S.~Gugercin, and K.~Willcox, \emph{{A Survey of Projection-Based
  Model Reduction Methods for Parametric Dynamical Systems}}, {SIAM} Review
  \textbf{57} (2015), no.~4, 483--531.

\bibitem{bennerParSys}
P.~Benner, M.~Ohlberger, A.~Pater, G.~Rozza, and K.~Urban, \emph{{Model
  Reduction of Parametrized Systems.}}, vol. 1st ed. 2017, MS\&A series, no.
  Vol. 17, Springer, 2017.

\bibitem{Bergmann2009516}
M.~Bergmann, C.-H. Bruneau, and A.~Iollo, \emph{{Enablers for robust POD
  models}}, Journal of Computational Physics \textbf{228} (2009), no.~2,
  516--538.

\bibitem{berselli2005mathematics}
L.~C. Berselli, T.~Iliescu, and W.~J. Layton, \emph{Mathematics of large eddy
  simulation of turbulent flows}, Springer Science \& Business Media, 2005.

\bibitem{Binev2011}
P.~Binev, A.~Cohen, W.~Dahmen, R.~DeVore, G.~Petrova, and P.~Wojtaszczyk,
  \emph{Convergence rates for greedy algorithms in reduced basis methods},
  {SIAM} Journal on Mathematical Analysis \textbf{43} (2011), no.~3,
  1457--1472.

\bibitem{Bizon2012}
K.~Bizon and G.~Continillo, \emph{Reduced order modelling of chemical reactors
  with recycle by means of {POD}-penalty method}, Computers {\&} Chemical
  Engineering \textbf{39} (2012), 22--32.

\bibitem{Bonomi2017}
D.~Bonomi, A.~Manzoni, and A.~Quarteroni, \emph{A matrix {DEIM} technique for
  model reduction of nonlinear parametrized problems in cardiac mechanics},
  Computer Methods in Applied Mechanics and Engineering \textbf{324} (2017),
  300--326.

\bibitem{boussinesq1877essa}
J.~Boussinesq, \emph{Essa sur latheories des eaux courantes. memoires presentes
  par divers savants a l’academic des sciences de l’institut national de
  france}, Tome XXIII (1877), no.~1.

\bibitem{Breuer1996}
M.~Breuer, D.~Lakehal, and W.~Rodi, \emph{Flow around a surface mounted cubical
  obstacle: Comparison of {LES} and {RANS}-results}, Notes on Numerical Fluid
  Mechanics ({NNFM}), Vieweg$\mathplus$Teubner Verlag, 1996, pp.~22--30.

\bibitem{Burkardt2006}
J.~Burkardt, M.~Gunzburger, and H.-C. Lee, \emph{{POD} and {CVT}-based
  reduced-order modeling of navier{\textendash}stokes flows}, Computer Methods
  in Applied Mechanics and Engineering \textbf{196} (2006), no.~1-3, 337--355.

\bibitem{Carlberg2013}
K.~Carlberg, C.~Farhat, J.~Cortial, and D.~Amsallem, \emph{{The {GNAT} method
  for nonlinear model reduction: Effective implementation and application to
  computational fluid dynamics and turbulent flows}}, Journal of Computational
  Physics \textbf{242} (2013), 623--647.

\bibitem{Chen2021}
W.~Chen, Q.~Wang, J.~S. Hesthaven, and C.~Zhang, \emph{Physics-informed machine
  learning for reduced-order modeling of nonlinear problems}, Journal of
  Computational Physics \textbf{446} (2021), 110666.

\bibitem{Chinesta2011}
F.~Chinesta, P.~Ladeveze, and E.~Cueto, \emph{{A Short Review on Model Order
  Reduction Based on Proper Generalized Decomposition}}, Archives of
  Computational Methods in Engineering \textbf{18} (2011), no.~4, 395.

\bibitem{Cotter2010}
S.~L. Cotter, M.~Dashti, and A.~M. Stuart, \emph{Approximation of {Bayesian}
  inverse problems for {PDEs}},  \textbf{48} (2010), no.~1, 322--345.

\bibitem{Couplet2005}
M.~Couplet, C.~Basdevant, and P.~Sagaut, \emph{{Calibrated reduced-order
  {POD}-Galerkin system for fluid flow modelling}}, Journal of Computational
  Physics \textbf{207} (2005), no.~1, 192--220.

\bibitem{Tiangang2014}
T.~Cui, Y.~M. Marzouk, and K.~E. Willcox, \emph{Data-driven model reduction for
  the {Bayesian} solution of inverse problems}, International Journal for
  Numerical Methods in Engineering \textbf{102} (2014), no.~5.

\bibitem{Dashti2017}
M.~Dashti and A.~Stuart, \emph{The {Bayesian} approach to inverse problems},
  2017.

\bibitem{DeVore2013}
R.~DeVore, G.~Petrova, and P.~Wojtaszczyk, \emph{Greedy algorithms for reduced
  bases in banach spaces}, Constructive Approximation \textbf{37} (2013),
  no.~3, 455--466.

\bibitem{Dumon20111387}
A.~Dumon, C.~Allery, and A.~Ammar, \emph{{Proper General Decomposition (PGD)
  for the resolution of Navier-Stokes equations}}, Journal of Computational
  Physics \textbf{230} (2011), no.~4, 1387--1407.

\bibitem{Fresca_2021}
S.~Fresca, L.~Dede', and A.~Manzoni, \emph{A comprehensive deep learning-based
  approach to reduced order modeling of nonlinear time-dependent parametrized
  {PDEs}}, Journal of Scientific Computing \textbf{87} (2021), no.~2.

\bibitem{Fresca2022}
S.~Fresca and A.~Manzoni, \emph{{POD}-{DL}-{ROM}: Enhancing deep learning-based
  reduced order models for nonlinear parametrized {PDEs} by proper orthogonal
  decomposition}, Computer Methods in Applied Mechanics and Engineering
  \textbf{388} (2022), 114181.

\bibitem{Galbally2009}
D.~Galbally, K.~Fidkowski, K.~Willcox, and O.~Ghattas, \emph{Non-linear model
  reduction for uncertainty quantification in large-scale inverse problems},
  International Journal for Numerical Methods in Engineering (2009),
  1581--1608.

\bibitem{GALLETTI2004}
B.~Galletti, C.~H. Bruneau, L.~Zannetti, and A.~Iollo, \emph{Low-order
  modelling of laminar flow regimes past a confined square cylinder}, Journal
  of Fluid Mechanics \textbf{503} (2004), 161--170.

\bibitem{Garmatter2016}
D.~Garmatter, B.~Haasdonk, and B.~Harrach, \emph{A reduced basis {Landweber}
  method for nonlinear inverse problems},  \textbf{32} (2016), no.~3, 035001.

\bibitem{Germano1991}
M.~Germano, U.~Piomelli, P.~Moin, and W.~H. Cabot, \emph{A dynamic
  subgrid-scale eddy viscosity model}, Physics of Fluids A: Fluid Dynamics
  \textbf{3} (1991), no.~7, 1760--1765.

\bibitem{Graham1999}
W.~R. Graham, J.~Peraire, and K.~Y. Tang, \emph{Optimal control of vortex
  shedding using low-order models. {Part I}--open-loop model development},
  International Journal for Numerical Methods in Engineering \textbf{44}
  (1999), no.~7, 945--972.

\bibitem{Gunzburger2007}
M.~D. Gunzburger, J.~S. Peterson, and J.~N. Shadid, \emph{Reduced-order
  modeling of time-dependent {PDEs} with multiple parameters in the boundary
  data}, Computer Methods in Applied Mechanics and Engineering \textbf{196}
  (2007), no.~4-6, 1030--1047.

\bibitem{Guo2018}
M.~Guo and J.~S. Hesthaven, \emph{Reduced order modeling for nonlinear
  structural analysis using gaussian process regression}, Computer Methods in
  Applied Mechanics and Engineering \textbf{341} (2018), 807--826.

\bibitem{Guo2019}
\bysame, \emph{Data-driven reduced order modeling for time-dependent problems},
  Computer Methods in Applied Mechanics and Engineering \textbf{345} (2019),
  75--99.

\bibitem{Haasdonk2008}
B.~Haasdonk and M.~Ohlberger, \emph{{Reduced basis method for finite volume
  approximations of parametrized linear evolution equations}}, Mathematical
  Modelling and Numerical Analysis \textbf{42} (2008), no.~2, 277--302.

\bibitem{Hanjalic1972}
K.~Hanjalic and B.~E. Launder, \emph{{A Reynolds stress model of turbulence and
  its application to thin shear flows}}, Journal of Fluid Mechanics \textbf{52}
  (1972), no.~04, 609.

\bibitem{hesthaven2015certified}
J.~S. Hesthaven, G.~Rozza, and B.~Stamm, \emph{{Certified Reduced Basis Methods
  for Parametrized Partial Differential Equations}}, Springer International
  Publishing, 2016.

\bibitem{Hesthaven2018}
J.~Hesthaven and S.~Ubbiali, \emph{Non-intrusive reduced order modeling of
  nonlinear problems using neural networks}, Journal of Computational Physics
  \textbf{363} (2018), 55--78.

\bibitem{PhDHijazi}
S.~Hijazi, \emph{{Reduced order methods for laminar and turbulent flows in a
  finite volume setting: projection-based methods and data-driven techniques}},
  dissertation, SISSA, 2020.

\bibitem{Hijazi2020}
S.~Hijazi, S.~Ali, G.~Stabile, F.~Ballarin, and G.~Rozza, \emph{{The Effort of
  Increasing Reynolds Number in Projection-Based Reduced Order Methods: From
  Laminar to Turbulent Flows}}, Lecture Notes in Computational Science and
  Engineering, Springer International Publishing, 2020, pp.~245--264.

\bibitem{Hijazi2020JCP}
S.~Hijazi, G.~Stabile, A.~Mola, and G.~Rozza, \emph{{Data-driven {POD}-Galerkin
  reduced order model for turbulent flows}}, Journal of Computational Physics
  \textbf{416} (2020), 109513.

\bibitem{HijaziStabileMolaRozza2020a}
S.~Hijazi, G.~Stabile, A.~Mola, and G.~Rozza, \emph{Non-intrusive polynomial
  chaos method applied to full-order and reduced problems in computational
  fluid dynamics: A comparison and perspectives}, pp.~217--240, Springer
  International Publishing, Cham, 2020.

\bibitem{Himpe2015}
C.~Himpe and M.~Ohlberger, \emph{Data-driven combined state and parameter
  reduction for inverse problems}, Advances in Computational Mathematics
  \textbf{41} (2015), no.~5, 1343--1364.

\bibitem{Ionita2014}
A.~C. Ionita and A.~C. Antoulas, \emph{Data-driven parametrized model reduction
  in the loewner framework}, {SIAM} Journal on Scientific Computing \textbf{36}
  (2014), no.~3, A984--A1007.

\bibitem{jasak1996error}
H.~Jasak, \emph{{Error analysis and estimation for the finite volume method
  with applications to fluid flows}}, Ph.D. thesis, Imperial College,
  University of London, 1996.

\bibitem{Ji2019}
W.~Ji, Z.~Ren, Y.~Marzouk, and C.~K. Law, \emph{Quantifying kinetic uncertainty
  in turbulent combustion simulations using active subspaces}, Proceedings of
  the Combustion Institute \textbf{37} (2019), no.~2, 2175--2182.

\bibitem{JOHNSTON2004221}
H.~Johnston and J.-G. Liu, \emph{{Accurate, stable and efficient
  Navier–Stokes solvers based on explicit treatment of the pressure term}},
  Journal of Computational Physics \textbf{199} (2004), no.~1, 221 -- 259.

\bibitem{Kaipio2005}
J.~Kaipio and E.~Somersalo, \emph{Statistical and computational inverse
  problems}, Springer-Verlag, 2005.

\bibitem{Kaiser2014}
E.~Kaiser, B.~R. Noack, L.~Cordier, A.~Spohn, M.~Segond, M.~Abel, G.~Daviller,
  J.~\"{O}sth, S.~Krajnovi{\'{c}}, and R.~K. Niven, \emph{Cluster-based
  reduced-order modelling of a mixing layer}, Journal of Fluid Mechanics
  \textbf{754} (2014), 365--414.

\bibitem{Kalashnikova2012}
I.~Kalashnikova and M.~F. Barone, \emph{Efficient non-linear proper orthogonal
  decomposition/{Galerkin} reduced order models with stable penalty enforcement
  of boundary conditions}, International Journal for Numerical Methods in
  Engineering \textbf{90} (2012), no.~11, 1337--1362.

\bibitem{Kim1995}
W.-W. Kim and S.~Menon, \emph{A new dynamic one-equation subgrid-scale model
  for large eddy simulations}, 33rd Aerospace Sciences Meeting and Exhibit,
  American Institute of Aeronautics and Astronautics, January 1995.

\bibitem{Kirsch2011}
A.~Kirsch, \emph{An introduction to the mathematical theory of inverse
  problems}, Springer New York, 2011.

\bibitem{Krajnovic2002}
S.~Krajnovic and L.~Davidson, \emph{Large-eddy simulation of the flow around a
  bluff body}, {AIAA} Journal \textbf{40} (2002), no.~5, 927--936.

\bibitem{Kunisch2002492}
K.~Kunisch and S.~Volkwein, \emph{{Galerkin proper orthogonal decomposition
  methods for a general equation in fluid dynamics}}, SIAM Journal on Numerical
  Analysis \textbf{40} (2002), no.~2, 492--515.

\bibitem{Lagaris1998}
I.~Lagaris, A.~Likas, and D.~Fotiadis, \emph{Artificial neural networks for
  solving ordinary and partial differential equations}, {IEEE} Transactions on
  Neural Networks \textbf{9} (1998), no.~5, 987--1000.

\bibitem{Lee1990}
H.~Lee and I.~S. Kang, \emph{Neural algorithm for solving differential
  equations}, Journal of Computational Physics \textbf{91} (1990), no.~1,
  110--131.

\bibitem{Lee2020}
K.~Lee and K.~T. Carlberg, \emph{Model reduction of dynamical systems on
  nonlinear manifolds using deep convolutional autoencoders}, Journal of
  Computational Physics \textbf{404} (2020), 108973.

\bibitem{Liu2010}
J.-G. Liu, J.~Liu, and R.~L. Pego, \emph{Stable and accurate pressure
  approximation for unsteady incompressible viscous flow}, Journal of
  Computational Physics \textbf{229} (2010), no.~9, 3428--3453.

\bibitem{Lorenzi2016}
S.~Lorenzi, A.~Cammi, L.~Luzzi, and G.~Rozza, \emph{{{POD}-Galerkin method for
  finite volume approximation of Navier{\textendash}Stokes and {RANS}
  equations}}, Computer Methods in Applied Mechanics and Engineering
  \textbf{311} (2016), 151--179.

\bibitem{Marzouk2009}
Y.~M. Marzouk and H.~N. Najm, \emph{Dimensionality reduction and polynomial
  chaos acceleration of {Bayesian} inference in inverse problems}, Journal of
  Computational Physics \textbf{228} (2009), no.~6, 1862--1902.

\bibitem{Matthies2016}
H.~G. Matthies, E.~Zander, B.~V. Rosi{\'{c}}, A.~Litvinenko, and O.~Pajonk,
  \emph{Inverse problems in a {Bayesian} setting}, pp.~245--286, Springer
  International Publishing, Cham, 2016.

\bibitem{Menter1994}
F.~R. Menter, \emph{{Two-equation eddy-viscosity turbulence models for
  engineering applications}}, AIAA Journal \textbf{32} (1994), no.~8,
  1598--1605.

\bibitem{Mou2021}
C.~Mou, B.~Koc, O.~San, L.~G. Rebholz, and T.~Iliescu, \emph{Data-driven
  variational multiscale reduced order models}, Computer Methods in Applied
  Mechanics and Engineering \textbf{373} (2021), 113470.

\bibitem{Moukalled:2015:FVM:2876154}
F.~Moukalled, L.~Mangani, and M.~Darwish, \emph{{The Finite Volume Method in
  Computational Fluid Dynamics: An Advanced Introduction with OpenFOAM and
  Matlab}}, 1st ed., Springer Publishing Company, Incorporated, 2015.

\bibitem{Murata2019}
T.~Murata, K.~Fukami, and K.~Fukagata, \emph{Nonlinear mode decomposition with
  convolutional neural networks for fluid~dynamics}, Journal of Fluid Mechanics
  \textbf{882} (2019).

\bibitem{NOACK2003}
B.~R. Noack, K.~Afanasiev, M.~Morzy{\'{n}}ski, G.~Tadmor, and F.~Thiele,
  \emph{A hierarchy of low-dimensional models for the transient and
  post-transient cylinder wake}, Journal of Fluid Mechanics \textbf{497}
  (2003), 335--363.

\bibitem{Noack1994}
B.~R. Noack and H.~Eckelmann, \emph{A low-dimensional galerkin method for the
  three-dimensional flow around a circular cylinder}, Physics of Fluids
  \textbf{6} (1994), no.~1, 124--143.

\bibitem{noack2005}
B.~R. Noack, P.~Papas, and P.~A. Monkewitz, \emph{{The need for a pressure-term
  representation in empirical Galerkin models of incompressible shear flows}},
  Journal of Fluid Mechanics \textbf{523} (2005), 339--365.

\bibitem{Peherstorfer2015}
B.~Peherstorfer and K.~Willcox, \emph{Dynamic data-driven reduced-order
  models}, Computer Methods in Applied Mechanics and Engineering \textbf{291}
  (2015), 21--41.

\bibitem{quarteroniRB2016}
A.~Quarteroni, A.~Manzoni, and F.~Negri, \emph{{Reduced Basis Methods for
  Partial Differential Equations}}, Springer International Publishing, 2016.

\bibitem{Raissi2019}
M.~Raissi, P.~Perdikaris, and G.~Karniadakis, \emph{Physics-informed neural
  networks: A deep learning framework for solving forward and inverse problems
  involving nonlinear partial differential equations}, Journal of Computational
  Physics \textbf{378} (2019), 686--707.

\bibitem{Raissi2018}
M.~Raissi, Z.~Wang, M.~S. Triantafyllou, and G.~E. Karniadakis, \emph{Deep
  learning of vortex-induced vibrations}, Journal of Fluid Mechanics
  \textbf{861} (2018), 119--137.

\bibitem{Romor2022}
F.~Romor, G.~Stabile, and G.~Rozza, \emph{{Non-linear manifold ROM with
  Convolutional Autoencoders and Reduced Over-Collocation method}}, 2022.

\bibitem{Rozza2007}
G.~Rozza and K.~Veroy, \emph{{On the stability of the reduced basis method for
  Stokes equations in parametrized domains}}, Computer Methods in Applied
  Mechanics and Engineering \textbf{196} (2007), no.~7, 1244 -- 1260.

\bibitem{sagaut}
P.~Sagaut, \emph{Large eddy simulation for incompressible flows}, Springer
  Science \& Business Media, Berlin Heidelberg, 2006.

\bibitem{Schmid2010}
P.~J. Schmid, \emph{Dynamic mode decomposition of numerical and experimental
  data}, Journal of Fluid Mechanics \textbf{656} (2010), 5--28.

\bibitem{Shah1997}
K.~B. Shah and J.~H. Ferziger, \emph{A fluid mechanicians view of wind
  engineering: Large eddy simulation of flow past a cubic obstacle}, Journal of
  Wind Engineering and Industrial Aerodynamics \textbf{67-68} (1997), 211--224.

\bibitem{Sirignano2018}
J.~Sirignano and K.~Spiliopoulos, \emph{{DGM}: A deep learning algorithm for
  solving partial differential equations}, Journal of Computational Physics
  \textbf{375} (2018), 1339--1364.

\bibitem{Sirisup2005}
S.~Sirisup and G.~Karniadakis, \emph{Stability and accuracy of periodic flow
  solutions obtained by a {POD}-penalty method}, Physica D: Nonlinear Phenomena
  \textbf{202} (2005), no.~3-4, 218--237.

\bibitem{sirovich1987snap}
L.~Sirovich, \emph{{Turbulence and the Dynamics of Coherent Structures part I:
  Coherent Structures}}, Quarterly of Applied Mathematics \textbf{45} (1987),
  no.~3, 561--571.

\bibitem{SMAGORINSKY1963}
J.~Smagorinsky, \emph{General circulation experiments with the primitive
  equations i. the basic experiment}, Monthly Weather Review \textbf{91}
  (1963), no.~3, 99--164.

\bibitem{SPALART1992}
P.~Spalart and S.~Allmaras, \emph{A one-equation turbulence model for
  aerodynamic flows}, 30th Aerospace Sciences Meeting and Exhibit, American
  Institute of Aeronautics and Astronautics, jan 1992.

\bibitem{RoSta17}
G.~Stabile and G.~Rozza, \emph{{ITHACA-FV - In real Time Highly Advanced
  Computational Applications for Finite Volumes}},
  \url{http://www.mathlab.sissa.it/ithaca-fv}, Accessed: 2018-01-30.

\bibitem{Stabile2017}
G.~Stabile, S.~Hijazi, A.~Mola, S.~Lorenzi, and G.~Rozza, \emph{{{POD}-Galerkin
  reduced order methods for {CFD} using Finite Volume Discretisation: vortex
  shedding around a circular cylinder}}, Communications in Applied and
  Industrial Mathematics \textbf{8} (2017), no.~1, 210--236.

\bibitem{Stabile2018}
G.~Stabile and G.~Rozza, \emph{{Finite volume {POD}-Galerkin stabilised reduced
  order methods for the parametrised incompressible Navier{\textendash}Stokes
  equations}}, Computers {\&} Fluids \textbf{173} (2018), 273--284.

\bibitem{Stuart2010}
A.~M. Stuart, \emph{Inverse problems: A {Bayesian} perspective}, Acta Numerica
  \textbf{19} (2010), 451--559.

\bibitem{volkwein2013proper}
S.~Volkwein, \emph{Proper orthogonal decomposition: Theory and reduced-order
  modelling}, Lecture Notes, University of Konstanz \textbf{4} (2013), no.~4.

\bibitem{Warner2015}
J.~E. Warner, W.~Aquino, and M.~D. Grigoriu, \emph{Stochastic reduced order
  models for inverse problems under uncertainty}, Computer Methods in Applied
  Mechanics and Engineering \textbf{285} (2015), 488--514.

\bibitem{weller1998tensorial}
H.~G. Weller, G.~Tabor, H.~Jasak, and C.~Fureby, \emph{{A tensorial approach to
  computational continuum mechanics using object-oriented techniques}},
  Computers in physics \textbf{12} (1998), no.~6, 620--631.

\bibitem{wilcox2006turbulence}
D.~Wilcox, \emph{Turbulence modeling for cfd}, Turbulence Modeling for CFD, no.
  v. 1, DCW Industries, La Canada, California, U.S.A, 2006.

\bibitem{Xiao20141}
D.~Xiao, F.~Fang, A.~Buchan, C.~Pain, I.~Navon, J.~Du, and G.~Hu, \emph{{Non
  linear model reduction for the Navier Stokes equations using residual DEIM
  method}}, Journal of Computational Physics \textbf{263} (2014), 1 -- 18.

\bibitem{Xie2018}
X.~Xie, M.~Mohebujjaman, L.~G. Rebholz, and T.~Iliescu, \emph{Data-driven
  filtered reduced order modeling of fluid flows}, {SIAM} Journal on Scientific
  Computing \textbf{40} (2018), no.~3, B834--B857.

\end{thebibliography}
\end{document}